\documentstyle[11pt,epsf]{article}
 1
 1
 1

\def\lsim{\mathrel{\raise.3ex\hbox{$<$\kern-.75em\lower1ex\hbox{$\sim$}}}}
\def\gsim{\mathrel{\raise.3ex\hbox{$>$\kern-.75em\lower1ex\hbox{$\sim$}}}}

\begin{document}
\noindent
\thispagestyle{empty}
\renewcommand{\thefootnote}{\fnsymbol{footnote}}
\begin{flushright}
{\bf DTP/97/68}\\
{\bf UCSD/PTH 97-16}\\
{\bf hep-ph/9707496}\\ 
{\bf July 1997}\\
\end{flushright}
\vspace{0.5cm}
\begin{center}
  \begin{Large}\bf
Analytic Calculation of Two Loop Corrections to Heavy Quark \\[1mm]
Pair Production Vertices Induced by Light Quarks
\\
  \end{Large}
  \vspace{1.5cm}

\begin{large}
 A.H. Hoang$^{a}$,
 T.~Teubner$^{b}$
\end{large}

\vspace{1.5cm}
\begin{it}
${}^a$ Department of Physics, University of California, San Diego,\\
La Jolla, CA 92093-0319, USA\\[0.5cm]
${}^b$ Department of Physics, University of Durham,\\
Durham, DH1 3LE, UK
\end{it}

\vspace{2.5cm}
  {\bf Abstract}\\
\vspace{0.3cm}

\noindent
\begin{minipage}{14.0cm}
\begin{small}
Non-singlet ${\cal{O}}(\alpha_s^2)$ corrections from light quarks are
calculated analytically for the massive quark-antiquark pair
production rates through the vector, axial-vector, scalar and
pseudoscalar currents for all energies above the threshold. 
The results are presented in terms of the moments of the
absorptive part of the vacuum polarization function of the light
quarks, which allows for an immediate determination of
${\cal{O}}(\alpha_s^2)$ corrections coming from other types of light
particles. The kinematic regime close to the threshold is examined,
and a comparison to results based on non-relativistic considerations
is carried out.   
\\[3mm]
\\[1mm]
\end{small}
\end{minipage}
\end{center}
\setcounter{footnote}{0}
\renewcommand{\thefootnote}{\arabic{footnote}}
\vspace{1.2cm}
\newpage
%
%
%
\section{Introduction}
\label{sectionintroduction}
In the past years considerable amount of work has been invested to
obtain higher order QCD radiative corrections to quark pair production
vertices. As far as the quark pair production rates through the
vector and axial-vector currents are
concerned this effort was mainly motivated by the impressive amount of
data on hadronic Z~decay events obtained at the LEP experiments. 
Whereas
the ${\cal{O}}(\alpha_s)$ corrections have been well known since quite a
long time for all energy and mass values~\cite{KallenSabry,Jersak1},
the complete analytic form of the  
${\cal{O}}(\alpha_s^2)$ corrections is known as a high energy
expansion in $M^2/q^2$~\cite{Chetyrkin1}, where $\sqrt{q^2}$ denotes the
c.m. energy and $M$ the mass of the produced quarks. 
For high energies, even the leading leading
${\cal{O}}(\alpha_s^3)$~\cite{Gorishny1} and some mass corrections 
have been calculated~\cite{Chetyrkin2}. The quark pair production
rates through the scalar and pseudoscalar currents have also been
investigated in view of the expected studies 
of hadronic Higgs decays at forthcoming collider experiments. Here, the
${\cal{O}}(\alpha_s)$ corrections are well established for all ratios
$M^2/q^2$~\cite{Braaten1,Drees1}, while the second order corrections are
known as an expansion in $M^2/q^2$~\cite{Higgs1}.
Thus, to ${\cal{O}}(\alpha_s^2)$ the present calculations are
definitely a good 
approximation if the c.m. energy is much larger than the mass of the
produced fermions, which is for example the case for $Z$ or Higgs
decays into bottom quarks, but can also be applied for intermediate
energies if a sufficient number of terms in the high energy expansion
is taken into account (see also~\cite{Multiloop} for a review). 
However, in face of forthcoming experiments ($\tau$-charm-,
B-factory, NLC), where quark pairs
will also be produced in the kinematic regime where the quark masses
and the c.m. energy are of comparable order, i.e. closer to the
threshold, the knowledge of the
entire ${\cal{O}}(\alpha_s^2)$ corrections valid for all values of
$M^2/q^2$ is desirable. Whereas meanwhile there are methods to obtain
numerical approximations for the ${\cal{O}}(\alpha_s^2)$
corrections in the latter kinematic regime based on Pad\'e
approximants~\cite{Pade}, the
entire analytic calculation seems to be an impossible task at the
present stage. However, analytical results are of particular importance
because they provide important cross checks for approximation methods
and serve as a starting point for qualitative considerations.
\par
A first step towards a complete analytical calculation consists of the 
determination of those ${\cal{O}}(\alpha_s^2)$ corrections to the
heavy quark production vertices which are induced by light
quarks. These corrections 
are equal to the sum of all possible cuts of at least two massive
quark lines of current-current correlator diagrams consisting of a
massive quark loop, a massless quark loop and two gluon lines. The
individual contributions can be divided into three classes:
(i) the massive quarks are produced directly by the external current
insertion (primary production), 
(ii) the massive quarks are produced via gluon emission off the
massless quarks (secondary production) and
(iii) the interference between amplitudes of class (i) and (ii). 
Class (iii) corresponds to cuts of the double-triangle or singlet
diagrams. For the vector current correlator the class (iii) diagrams
vanish due to Furry's theorem. The contributions belonging to
class (ii) were determined analytically in~\cite{Kniehl1,HJKT2} for
the vector and axial-vector case.\footnote{
A numerical approach for the contributions belonging to class
(ii) in the vector and axial-vector case was 
presented in~\cite{Soper1}.}
\par
In this work we present analytic expressions for the
${\cal{O}}(\alpha_s^2)$ corrections 
belonging to class (i) for arbitrary values of the quark mass $M$ and
c.m. energy $\sqrt{q^2}$ above threshold. Analytic formulae for
primary production through the
vector and scalar currents were already presented in~\cite{HKT1} and
\cite{Melnikov}, respectively. 
Here we carry out the same program for the  axial-vector and
pseudoscalar case. In addition, we  present the results in terms
of moments of the absorptive part of the vacuum polarization of the
light quark-antiquark pair, proceeding along the lines of~\cite{CHKST1},
which allows for an immediate determination of the
${\cal{O}}(\alpha_s^2)$ corrections coming from light colored scalar
particles or from the gluonic vacuum polarization insertion. Based on the
concept of the moments even a certain class of ${\cal{O}}(\alpha_s^3)$
and ${\cal{O}}(\alpha_s^4)$ corrections can be determined. Thus, for
completeness also the vector and scalar cases are presented in terms
of the moments.
\par
To be definite we consider the light quark induced
${\cal{O}}(\alpha_s^2)$ contributions to the imaginary part of the
massive quark current-current correlators
\begin{eqnarray}
\Pi^{\Theta}_{\mu\nu}(q) & = & 
-\,i \int \!{\rm d}x\,e^{i\,q.x}\,
   \langle 0|\,T\,j^{\Theta}_\mu(x)\,j^{\Theta}_\nu(0)\,|0 \rangle
\nonumber\\ & = &
g_{\mu\nu}\,q^2\,\Pi^{\Theta}(q^2)-q_\mu\,q_\nu\,\Pi_L^{\Theta}(q^2)
\,,\qquad \Theta=V,A,
\label{currentcorrelatorvector}
\end{eqnarray}
for the vector and the axial-vector currents
\begin{equation}
j_\mu^V \, = \, \bar\Psi_Q^0\,\gamma_\mu\,\Psi_Q^0
\,,\qquad\qquad
j_\mu^A \, = \, \bar\Psi_Q^0\,\gamma_\mu\,\gamma_5\,\Psi_Q^0
\,,
\label{defvectorcurrents}
\end{equation}
and
\begin{eqnarray}
q^2\,\Pi^{\Theta}(q^2) & = & 
i \int \!{\rm d}x\,e^{i\,qx}\,
   \langle 0|\,T\,j^{\Theta}(x)\,j^{\Theta}(0)\,|0 \rangle
\,,\qquad \Theta=S,P,
\label{currentcorrelatorscalar}
\end{eqnarray}
for the scalar and pseudoscalar currents
\begin{equation}
j^S \, = \, M^0\,\bar\Psi_Q^0\,\Psi_Q^0
\,,\qquad\qquad
j^P \, = \, i\,M^0\,\bar\Psi_Q^0\,\gamma_5\,\Psi_Q^0
\,,
\label{defscalarcurrents}
\end{equation}
where summation over color degrees of freedom is understood.
$\Psi_Q^0$ denotes the bare  massive quark field 
and $M^0$ its bare mass. In order to maintain vanishing anomalous
dimension for all four currents the scalar and pseudoscalar currents 
are defined including the mass parameter. The longitudinal part of the
vector current correlator $\Pi_L^{V}$ is equal to $\Pi^{V}$ due to
current conservation, whereas $\Pi_L^{A}$ is related to the
pseudoscalar current correlator $\Pi^{P}$ via the axial Ward identity. 
The optical theorem relates the 
renormalization group invariant imaginary parts of $\Pi^{\Theta}$,
$\Theta=V,A,S,P$, to decay rates and cross-sections. The
contribution of ${\rm Im}\Pi^{V,A}$ to the total hadronic Z decay rate
reads 
\begin{eqnarray}
\Delta\,\Gamma_Z^{had} & = &
  \frac{G_F\,M_Z^3}{24\,\sqrt{2}\,\pi}\,\left(\,
  v_Q^2\,R^V(M_Z^2) + a_Q^2\,R^A(M_Z^2)\,
  \right)
\,,\\
a_Q & = & 2\,I_3^Q
\,,\nonumber\\
v_Q & = & 2\,I_3^Q - 4\,q_Q\,\sin^2\theta_W
\,,\nonumber
\end{eqnarray}
where
\begin{equation}
R^{A,V}(q^2) \, \equiv \, 
12\,\pi\,{\rm Im}\Pi^{A,V}(q^2+i\,\epsilon)
\,,
\label{defRVA}
\end{equation}
$q_Q$ is the massive quark electric charge, $\Theta_W$ the Weinberg
angle and $I_3^Q$ the massive quark weak isospin.
Moreover, $R^V$ is equal to the cross-section of $Q\bar Q$ production in
$e^+e^-$ annihilation into a single photon normalized to the point 
cross-section. The contribution of the quantities $R^{V,A}$ to 
Z-mediated $Q\bar Q$ production is straightforward.
The contributions of ${\rm Im}\Pi^{S,P}$ to the hadronic decay widths
of a scalar or pseudoscalar Higgs boson with mass  $m_S$ and $m_P$,
respectively, read 
\begin{equation}
\Delta\,\Gamma_{S,P}^{had} \, = \,
\frac{G_F\,M^2}{4\,\sqrt{2}\,\pi}\,m_{S,P}\,a_{S,P}^F\,R^{S,P}(m_{S,P}^2)
\,,
\end{equation}
where
\begin{equation}
R^{S,P}(q^2) \, \equiv \,
\frac{8\,\pi}{M^2}\,{\rm Im}\Pi^{S,P}(q^2+i\,\epsilon)
\label{defRSP}
\end{equation}
and $M$ denotes the heavy quark pole mass.
The coefficient $a_{S,P}^F$ is the relative weight of the corresponding
partial width with respect to the standard model Higgs decay rate 
($a_H^F=1$). Thus in the framework of the MSSM, where a two doublet 
Higgs sector is realized, 
$a_H^{up}=\sin^2\alpha/\sin^2\beta$,
$a_H^{down}=\cos^2\alpha/\cos^2\beta$,
$a_h^{up}=\cos^2\alpha/\sin^2\beta$,
$a_h^{down}=\sin^2\alpha/\cos^2\beta$,
$a_A^{up}=\cot^2\beta$ and
$a_A^{down}=\tan^2\beta$, where $H,h$ and $A$ denote the heavy and light
scalar and the pseudoscalar Higgs fields, respectively. $\alpha$ describes
the mixing between the two Higgs doublets and $\beta$ is defined via the
ratio of their vacuum expectation values,
$\tan\beta=v_2/v_1$ (see e.g.~\cite{Higgshunters}).
In the following sections we present the results of our calculation
in terms of the quantities $R^\Theta$, $\Theta=V,A,S,P$. 
Their perturbative expansion using the $\overline{\rm MS}$ renormalization
scheme for the strong coupling reads
\begin{equation}
R^\Theta(q^2) \, = \,
r_\Theta^{(0)}(q^2) 
+\left(\frac{\alpha_s^{\overline{\mbox{\tiny
MS}}}(\mu^2)}{\pi}\right)\,r_\Theta^{(1)}(q^2)
+\left(\frac{\alpha_s^{\overline{\mbox{\tiny
MS}}}(\mu^2)}{\pi}\right)^2\, 
        r_\Theta^{(2),{\overline{\mbox{\tiny MS}}}}(q^2,\mu^2)
+\ldots\,,
\label{defR}
\end{equation}
where the light quark ${\cal{O}}(\alpha_s^2)$ corrections determined
in this work will be referred to as
$r_{\Theta,q}^{(2),{\overline{\mbox{\tiny MS}}}}$.
As we use the pole definition for the heavy mass $M$ throughout this
work, only the ${\cal{O}}(\alpha_s^2)$ coefficient of $R_\Theta$ is 
renormalization group $\mu$-dependent. For simplicity we call the quantities
$R^\Theta$, $\Theta=V,A,S,P$, decay rates from now on in this paper.
Because the calculations for the virtual light quark corrections are most
easily performed by using dispersion integration techniques, the results
for all four heavy quark currents 
are derived in the framework of on-shell renormalized QED, where the 
coupling $\alpha$ is defined as the fine structure constant and the
wave function renormalization constant of the quark fields is fixed by
the requirement that the residue of the fermion propagator is one.
In a second separate step the results are then transferred to the QED
$\overline{\rm MS}$ scheme for the coupling and to QCD.
Throughout this paper
the heavy quark will be denoted as $Q$ 
with pole mass $M$ and the light quarks as $q$ with mass $m$. 
It should also be noted that, even in the framework of QED, we will
refer to the fermions as quarks though they may be leptons as well.
For simplicity all quarks are assigned
electric charge one. 
The generalization to different charge assignments is straightforward.
All QED results are marked with a tilde. 
\par 
The program of this work is as follows: in 
Section~\ref{sectionvectorcurrent} we describe in some detail the
calculation of the light quark ${\cal O}(\alpha^2)$ corrections to
massive quark pair production through the vector current in on-shell QED.
Virtual and real
contributions as well as the inclusive sum are discussed separately.
The results are presented in a general form which is common for all
the cases ($V, A, S, P$). The corresponding results for corrections to
axial vector induced quark pair production are then given in
Section~\ref{sectionaxialvectorcurrent}, whereas in
Section~\ref{sectionscalarcurrent} the cases of scalar and
pseudoscalar currents are treated on the same footing. In
Section~\ref{sectiontransitiontoqcd} we carry out the transition from  
on-shell QED to the $\overline{\rm MS}$
scheme and to QCD. As mentioned above our results are presented in
terms of moments of the inserted vacuum polarization from light
quarks. In Section~\ref{sectionmoments} this ``method of the moments''
is discussed in more detail. We present the corresponding moments for
light scalar and gluonic second order corrections.
Section~\ref{sectionthreshold} contains a discussion of our results in
the threshold region and Section~\ref{sectionsummary} our summary.
For the sake of completeness numerous ${\cal{O}}(\alpha)$ formulae 
for the cases of the different currents ($V, A, S, P$) are given in
Appendix~\ref{appendixfirstorder}. Second order virtual corrections in the
case of equal masses $m = M$ are listed in
Appendix~\ref{appendixequalmass}. For the convenience of the reader we
also present expansions for the high energy as well as
for the threshold region in
Appendix~\ref{appendixexpansions}. Phenomenological applications of the
results will be discussed elsewhere. 
\par
\vspace{1cm}
\section{Heavy Quark Production through the Vector Current}
\label{sectionvectorcurrent}
\subsection{Virtual Corrections}
\label{subsectionvectorvirtualcorrections}
\begin{figure}[htb]
\begin{center}
\leavevmode
\epsfxsize=3.0cm
\epsffile[170 270 420 520]{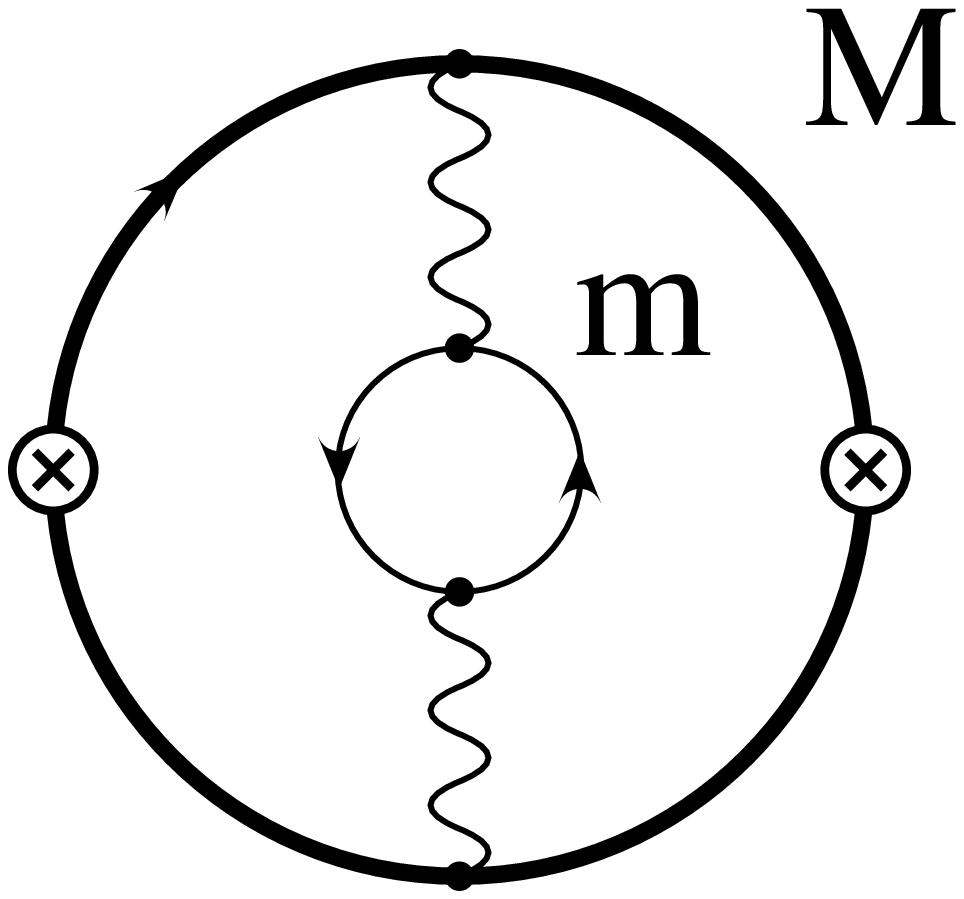}
\hspace{10ex}
\leavevmode
\epsfxsize=3.0cm
\epsffile[170 270 420 520]{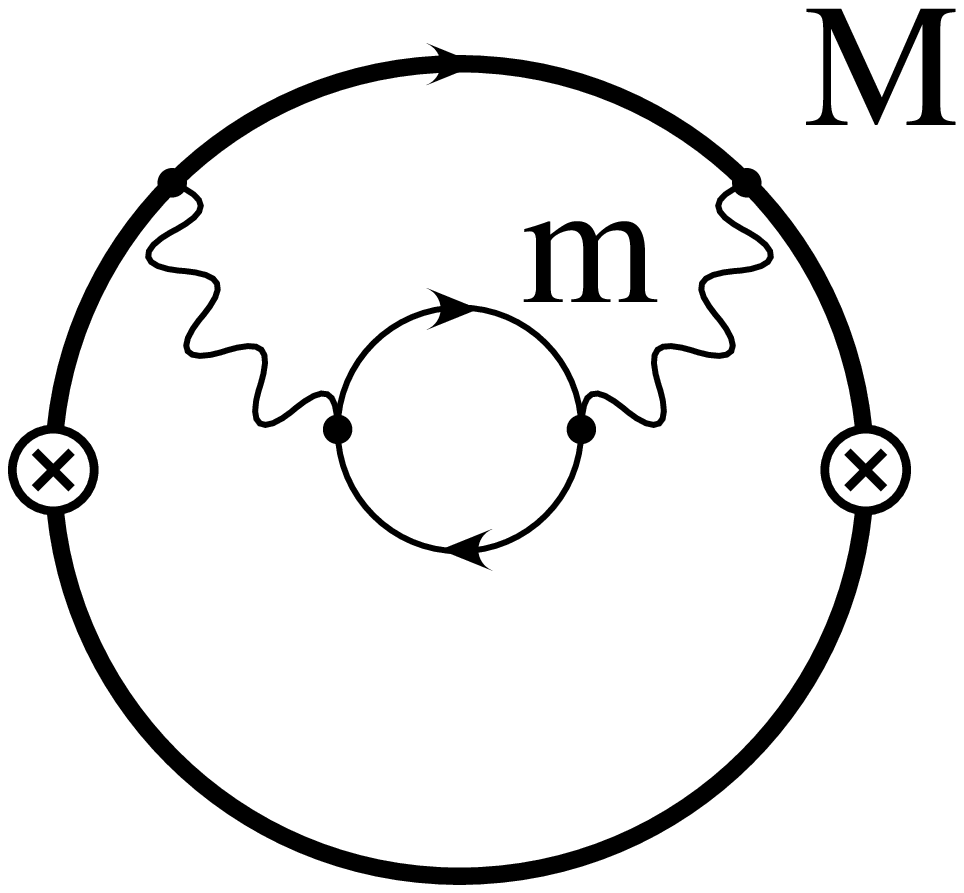}
\vskip -5mm
\caption[]{\label{currentcorrelatordiagrams} 
Double bubble diagrams where a
pair of heavy quarks of mass $M$ is produced through an external
current. The exchanged photon splits into the light quark-antiquark
pair of mass $m$.}
\end{center}
\end{figure}

The virtual light quark ${\cal{O}}(\alpha^2)$ corrections to the
massive quark 
production vertex induced by the vector current
correspond to the sum of the two particle cuts of the current-current
correlator diagrams depicted in Fig.~\ref{currentcorrelatordiagrams}. 
This leads to virtual photon exchange diagrams with an additional 
vacuum polarization insertion of a light quark-antiquark pair. 
\par
The basic idea is to write the light quark-antiquark contributions to
the photon vacuum polarization function 
$\tilde \Pi^{\mu\nu}$ in terms of an one-time subtracted dispersion relation
(which accounts for the correct subtraction of the subdivergence in
the light quark vacuum polarization),
\begin{eqnarray}
\tilde \Pi^{\mu\nu}_{q\bar q}(m^2, q^2) & = &
  -i\,\Big(g^{\mu\nu}\,q^2 - q^\mu\,q^\nu\Big)\,
\tilde \Pi^q(m^2, q^2)\,,
\label{vacp(oldispersion1} \\[2mm]
\tilde \Pi^q(m^2, q^2) & = &
 \frac{1}{3}\,\bigg(\frac{\alpha}{\pi}\bigg)
     \,q^2\int\limits_{4\,m^2}^\infty\,
 \frac{{\rm d}{\lambda}^2}
   {{\lambda}^2\,({\lambda}^2-q^2-i\epsilon)}\,
   \tilde R_{q\bar q}(m^2, \lambda^2)\,,
\label{vacpoldispersion2}
\end{eqnarray}
where 
\begin{equation}
\tilde R_{q\bar q}(m^2, q^2) \, = \, \sqrt{1-4\,\frac{m^2}{q^2}}\,
\left( \, 1+2\,\frac{m^2}{q^2}
 \,\right)
\label{defRff}
\end{equation}
is the absorptive part of the vacuum polarization function of the light
quark-antiquark pair.\footnote{
Via the optical theorem, Eq.~(\ref{defRVA}), $\tilde R_{q\bar q}$ is just
the normalized born cross-section for the production of a $q\bar q$ pair in 
$e^+e^-$ annihilation into a single photon.} 
Thus the effective propagator which has to be
inserted into the virtual photon line can be written as
\begin{eqnarray}
\lefteqn{
\frac{-i}{q^2+i\epsilon}
\left(g^{\mu\sigma}-\xi\frac{q^\mu\,q^\sigma}{q^2}\right)
\,\tilde\Pi^{q\bar q}_{\sigma\rho}(m^2, q^2)\,
  \frac{-i}{q^2+i\epsilon} 
\left(g^{\rho\nu}-\xi\frac{q^\rho\,q^\nu}{q^2}\right)
}\nonumber\\
 & = &
 \frac{1}{3}\left(\frac{\alpha}{\pi}\right) \int\limits_{4\,m^2}^\infty
   \frac{{\rm d}\lambda^2}{\lambda^2}\,
   \bigg[\frac{-\,i\,g^{\mu\nu}}{q^2-\lambda^2+i\epsilon}\bigg]\,
   \tilde R_{q\bar q}(m^2, \lambda^2)\, + \ldots
\,\,.
\label{effectivepropagator}
\end{eqnarray}
The longitudinal terms indicated by the dots are irrelevant 
due to current conservation.
It is evident from Eq.~(\ref{effectivepropagator}) that the effective
propagator is a convolution of a massive vector boson propagator 
in Feynman gauge with
the function $\tilde R_{q\bar q}$. Inserting 
expression~(\ref{effectivepropagator})
into  the vertex diagrams and carrying out the vertex loop integration
first, the intended two-loop corrections can be written as a convolution
integration over the corresponding ${\cal{O}}(\alpha)$ virtual
corrections with exchange of a massive vector boson of mass $\lambda$.
\par
It is customary to parameterize the virtual corrections to the
renormalized vector 
current vertex $\Lambda_\mu^V$ in terms of the Dirac ($F_1$) and the Pauli 
($F_2$) form factors
\begin{equation}
\overline{u}_Q(k_2)\,\Lambda_\mu^V\,v_Q(k_1) \, = \,
\overline{u}_Q(k_2)\,\left[\,
\gamma_\mu\,F_1(q^2) +
\frac{i}{2\,M}\,\sigma_{\mu\nu}\,q^\nu\,F_2(q^2)
\,\right]\,v_Q(k_1)
\,,
\label{defF}
\end{equation}
where $q=k_1+k_2$, $\sigma_{\mu\nu} = \frac{i}{2}[\gamma_\mu,\gamma_\nu]$.
$u_Q$ and $v_Q$ denote the free heavy quark and antiquark Dirac
fields, respectively. 
The perturbative expansions of the form factors
read
\begin{eqnarray}
F_1(q^2) & = & 1 + 
\left(\frac{\alpha}{\pi}\right)\,F_1^{(1)}(q^2) +
\left(\frac{\alpha}{\pi}\right)^2\,F_1^{(2)}(q^2) + \ldots\,,
\label{defF1}
\\[2mm]
F_2(q^2) & = & \hspace{.65cm}
\left(\frac{\alpha}{\pi}\right)\,F_2^{(1)}(q^2) +
\left(\frac{\alpha}{\pi}\right)^2\,F_2^{(2)}(q^2) + \ldots
\,.
\label{defF2}
\end{eqnarray}
The Dirac form factor is normalized to one at zero momentum transfer
$q^2=0$ at all orders in $\alpha$. To ${\cal{O}}(\alpha)$ this is
completely achieved by the on-shell QED quark field wave function
renormalization constant, which is fixed by the requirement that the
residue of the renormalized heavy quark propagator is one. No further
subtraction is required because the vector current as defined in
Eq.~(\ref{defvectorcurrents}) is anomalous dimension free.
Based on the considerations presented above, the light quark
corrections to the ${\cal{O}}(\alpha^2)$ form factors can be written as
an one-dimensional integral,
\begin{equation}
F_{i,q}^{(2)}(M^2, m^2, q^2) = 
 \frac{1}{3} \int\limits_{4\,m^2}^\infty
   \frac{{\rm d}\lambda^2}{\lambda^2}\, 
   F_i^{(1)}(M^2, q^2, \lambda^2)\,
   \tilde R_{q\bar q}(m^2, \lambda^2)\,,\qquad
i=1,2\,,
\label{masterformulavectorvirtual}
\end{equation}
where $F_{1/2}^{(1)}(M^2, q^2, \lambda)$ denotes the first order form factors
with virtual exchange of a vector boson with mass $\lambda$. In
Eq.~(\ref{masterformulavectorvirtual}) we have indicated the mass
dependence of the form factors explicitly in order to illustrate the
mass dependence of the individual factors.
It should be noted that the light quark corrections to the
${\cal{O}}(\alpha^2)$ form factors,
Eq.~(\ref{masterformulavectorvirtual}), are already renormalized
properly and that no additional subtraction is needed. In particular,
the subdivergence from the virtual light quark loop is taken care of
properly by using the subtracted dispersion relation in the
$\lambda^2$~integration, whereas the overall divergences are
subtracted correctly in the ${\cal{O}}(\alpha)$ form factors. The same
remark holds also true for the corresponding virtual light quark
${\cal{O}}(\alpha^2)$ corrections for the axial-vector, scalar and
pseudoscalar cases presented in the subsequent sections.
\par
The contributions of the form factors to the decay rate 
\begin{equation}
\tilde R^V(q^2) \, = \,
\tilde r_V^{(0)}(q^2) 
+\left(\frac{\alpha}{\pi}\right)\,\tilde r_V^{(1)}(q^2)
+\left(\frac{\alpha}{\pi}\right)^2\,\tilde r_V^{(2)}(q^2)
+\ldots
\label{tildeRV}
\end{equation}
read
\begin{eqnarray}
\tilde r_{V}^{(0)}(q^2) & = &
    \frac{1}{2}\,\beta\,(3-\beta^2)
\,,  \\[2mm]
\tilde r_V^{(1),virt}(q^2) & = &
  \beta\,(3-\beta^2)\,\left(\,{\rm Re}F_1^{(1)}(q^2)+{\rm
Re}F_2^{(1)}(q^2)\,\right) +
  \beta^3\,{\rm Re}F_2^{(1)}(q^2)
\,,
\end{eqnarray}
where
\[
\beta \, = \, \sqrt{1-4\,\frac{M^2}{q^2}}
\]
is the velocity of the massive quarks in the c.m.~frame. 
\par
The contribution of $F_{1/2,q}^{(2)}$ to 
$\tilde r_V^{(2)}$ is given by
\begin{equation}
\tilde r_{V,q}^{(2),virt}(q^2) \, = \,
  \beta\,(3-\beta^2)\,\left(\,{\rm Re}F_{1,q}^{(2)}(q^2)+
                     {\rm Re}F_{2,q}^{(2)}(q^2)\,\right) +
  \beta^3\,{\rm Re}F_{2,q}^{(2)}(q^2)
\,.
\end{equation}
The real parts of $F_{1/2}^{(1)}(M^2, q^2, \lambda^2)$ 
have already been presented
in~\cite{HKT1,andrediss}. For the convenience of the reader
the full form of the ${\cal{O}}(\alpha)$ form factors with
massive vector boson exchange above threshold is presented in 
Appendix~\ref{appendixfirstordervector},
Eqs.~(\ref{F1oneloopfinitelambda}) and (\ref{F2oneloopfinitelambda}).
For $\lambda^2\to 0$ the expressions for the ${\cal{O}}(\alpha)$ form
factors relevant for  
$r_V^{(1),virt}$ are obtained. This results in the
well known QED expressions calculated
by~\cite{KallenSabry}, where the IR singularity in $F_1$ is
regularized by the small fictitious photon mass $\lambda$, 
\begin{eqnarray}
\lefteqn{
F_1^{(1)}(M^2, q^2, \lambda^2\rightarrow 0)  \, = \,
}\nonumber \\ & &\mbox{}
-\frac{1}{2}\,\bigg( 1 + \frac{1 + {{\beta}^2}}{2\,\beta}\,\ln p 
  \bigg) \,\ln \frac{\lambda^2}{M^2}   - 
  \frac{1 + {{\beta}^2}}{2\,\beta}\,
   \left( \mbox{Li}_2(1 - p) + \frac{1}{4}\,\ln^2 p - 3\,\zeta(2) 
   \right) \,
\nonumber\,\\ 
 & & \mbox{} - 
  \frac{1 + 2\,{{\beta}^2}}{4\,\beta}\,\ln p - 1 - 
  i\,\pi \,\bigg[\, \frac{1 + {{\beta}^2}}{4\,\beta}\,
   \ln \frac{\lambda^2}{M^2} + 
     \frac{1 + {{\beta}^2}}{4\,\beta}\,
      \ln\Big(\frac{1 - {{\beta}^2}}{4\,{{\beta}^2}}\Big) + 
     \frac{1 + 2\,{{\beta}^2}}{4\,\beta} \,\bigg] 
\,,\label{f1oneloopsl} \\[2mm]
\lefteqn{
F_2^{(1)}(M^2, q^2, \lambda^2\rightarrow 0)  \, = \,
\frac{1 - {\beta^2}}{4\,\beta}\,\Big( \ln p + i\,\pi  \Big)\,, 
}
\label{f2oneloopsl}
\end{eqnarray}
where 
\[ p \, = \, \frac{1-\beta}{1+\beta}\,.
\]
\par
For arbitrary ratios $m/M$ the evaluation of 
Eq.~(\ref{masterformulavectorvirtual}) can be carried out, but is
extremely tedious.\footnote{
The interested reader is referred
to~\cite{andrediss}. In the special case $m = M$ the calculation can
be simplified considerably and leads to the compact results given in
Appendix~\ref{appendixequalmass} (see also \cite{HKT1}).
}
However, in the case of interest, $m/M\to 0$, the task can be
simplified enormously by subtracting and adding $\tilde R_{q\bar q}$ at its
asymptotic value at high energies, $\tilde R_\infty^q\equiv \tilde R_{q\bar
q}(m^2, q^2=\infty)$:\footnote{
To our knowledge this trick has been employed the
first time in~\cite{Kniehl2}.
} 
\begin{eqnarray}
F_{i,f}^{(2)}(M^2, m^2, q^2) & = &
 \frac{1}{3} \int\limits_{4\,m^2}^\infty
   \frac{{\rm d}\lambda^2}{\lambda^2}\, 
   F_i^{(1)}(M^2, q^2, \lambda^2)\,\left[
   \tilde R_{q\bar q}(m^2, \lambda^2)-\tilde R_\infty^q\right] 
\nonumber\\ & &
\,+\, \frac{1}{3}\,\tilde R_\infty^q \int\limits_{4\,m^2}^\infty
   \frac{{\rm d}\lambda^2}{\lambda^2}\, 
   F_i^{(1)}(M^2, q^2, \lambda^2)\,,\qquad
i=1,2\,.
\label{mastermomentvector}
\end{eqnarray}
Because $\tilde R_{q\bar q}$ reaches its asymptotic value already for
values of
$\lambda^2$ much smaller than the heavy scales $M^2$ and $q^2$, which
govern the ${\cal{O}}(\alpha)$ form factors, we can replace
$F_{1/2}^{(1)}(M^2, q^2, \lambda^2)$ in the first integral on the r.h.s. 
of~(\ref{mastermomentvector}) by the QED expressions for $\lambda^2\to
0$, Eqs.~(\ref{f1oneloopsl}) and (\ref{f2oneloopsl}).
The resulting integration can then be trivially expressed in terms of
the moments 
$\tilde R_0^q$ and $\tilde R_1^q$ defined by
\begin{equation}
\tilde R_n^q \, = \,
\frac{1}{n!}\int\limits_0^1 {\rm d}x\,\frac{\ln^n x}{x} \,
\Big[\,
\tilde R_{q\bar q}\Big(m^2, \frac{4\,m^2}{x}\Big)-\tilde
R_\infty^q
\,\Big] 
\,,\qquad n=0,1,2,\ldots\,.
\label{momentsdef}
\end{equation}
The evaluation of the second integral on the r.h.s. 
of~(\ref{mastermomentvector}) constitutes the main effort. The final
results for the light quark ${\cal{O}}(\alpha^2)$ corrections to the
form factors can be cast into the form ($x=q$)
\begin{eqnarray}
F_{1,x}^{(2)} & = &
f_1^{(2)}\,\bigg[\, \tilde R_\infty^x\,\ln^2 \frac{m^2}{q^2} - 
     2\,\tilde R_0^x\,\ln \frac{m^2}{q^2} + 2\,\tilde R_1^x \,\bigg]  + 
  f_1^{(1)}\,\bigg[\, \tilde R_\infty^x\,\ln \frac{m^2}{q^2} - 
                     \tilde R_0^x \,\bigg]  + 
  f_1^{(0)}\,\tilde R_\infty^x
\,,
\label{F12}\\
F_{2,x}^{(2)} & = &
f_2^{(1)}\,\bigg[\, \tilde R_\infty^x\,\ln \frac{m^2}{q^2} - 
                  \tilde R_0^x \,\bigg]  + 
  f_2^{(0)}\,\tilde R_\infty^x
\,,
\label{F22}
\end{eqnarray}
where
\begin{eqnarray}
f_1^{(0)} & = &
A_{virt} + \frac{2}{3}\,( 1 + 2\,{{\beta}^2} ) \,B_{virt} - 
  \frac{1}{3}\,\ln\Big(\frac{1 - {{\beta}^2}}{16}\Big) + 
  \frac{5 + 2\,{{\beta}^2}}{24\,\beta}\,\ln p - 
  \frac{1 - \beta + 2\,{{\beta}^2}}{2\,\beta}\,\zeta(2) + \frac{1}{6}
\\
& & +i\,\pi\,\biggl\{\,
\frac{1}{12\,\beta}\,\biggl[\, 2\,( 1 + {{\beta}^2} ) \,
     \ln^2\Big(\frac{\beta}{2}\Big) - 2\,( 1 + 2\,{{\beta}^2} ) \,
     \ln\Big(\frac{\beta}{2}\Big) + ( 1 + {{\beta}^2} ) \,\zeta(2) + 
    \frac{1}{2}\,( 5 + 2\,{{\beta}^2} )  \,\biggr] 
\,\biggr\} \,,\nonumber\\[2mm]
f_1^{(1)} & = &
\frac{2}{3}\,( 1 + {{\beta}^2} ) \,B_{virt} - 
  \frac{1}{6}\,\ln\Big(\frac{1 - {{\beta}^2}}{16}\Big) + 
  \frac{1 + 2\,{{\beta}^2}}{12\,\beta}\,\ln p - 
  \frac{1 + {{\beta}^2}}{2\,\beta}\,\zeta(2) + \frac{1}{3}
\nonumber\\
& & +i\,\pi\,\biggl\{\,
\frac{1}{12\,\beta}\,\biggl[\, -2\,( 1 + {{\beta}^2} ) \,
     \ln\Big(\frac{\beta}{2}\Big) + 1 + 2\,{{\beta}^2} \,\biggr] 
\,\biggr\} \,,\\[2mm]
f_1^{(2)} & = &
\frac{1 + {{\beta}^2}}{24\,\beta}\,\ln p + \frac{1}{12} 
\, +\,i\,\pi\,\biggl\{\,
\frac{1 + {{\beta}^2}}{24\,\beta}
\,\biggr\} \,, \\[5mm]
f_2^{(0)} & = &
- \frac{2}{3}\,( 1 - {{\beta}^2} ) \,B_{virt}  - 
  \frac{5\,\left( 1 - {{\beta}^2} \right) }{24\,\beta}\,\ln p + 
  \frac{1 - {{\beta}^2}}{2\,\beta}\,\zeta(2)
\, +\,i\,\pi\,\biggl\{\,
\frac{1 - {{\beta}^2}}{6\,\beta}\,
  \biggl[\, \ln\Big(\frac{\beta}{2}\Big) - \frac{5}{4} \,\biggr]
\,\biggr\} \,,\\[2mm]
f_2^{(1)} & = &
-\frac{1 - {{\beta}^2}}{12\,\beta}\,\ln p  
\, +\,i\,\pi\,\biggl\{\,
-\frac{1 - {{\beta}^2}}{12\,\beta}
\,\biggr\} \,,
\end{eqnarray}
and
\begin{eqnarray}
A_{virt} & = &
- \frac{1 + {{\beta}^2}}{6\,\beta}\,
     \bigg[\, 2\,\mbox{Li}_3(1 - p) + \mbox{Li}_3(p) + 
       \mbox{Li}_2(1 - p)\,\ln\Big(\frac{1 - {{\beta}^2}}{16}\Big)\,
\nonumber \\ 
 & & \mbox{}\,\qquad\quad\qquad - \frac{1}{12}\,\ln^3 p + 
       \frac{1}{2}\,\ln^2 p\,\ln\Big(2\,( 1 - {p^2} ) \Big) - 
       \ln p\,\ln^2\Big(2\,( 1 + p ) \Big)\,\nonumber\,\\ 
 & & \mbox{}\,\qquad\quad
        \qquad  - 
       2\,\bigg( 3\,\ln\Big(\frac{\beta}{2}\Big) + \frac{5}{4}\,\ln p \bigg) \,
        \zeta(2) - \zeta(3)\,\bigg]   + 
  \frac{1}{12}\,\ln^2\Big(\frac{1 - {{\beta}^2}}{16}\Big)
\,,\label{defAvirt}\\[2mm]
B_{virt} & = &
\frac{1}{4\,\beta}\,\bigg[\, \mbox{Li}_2(1 - p) + 
    \ln p\,\bigg( \ln(1 + p) - \frac{1}{4}\,\ln p + \ln 2 \bigg) \,\bigg] \,.
\label{defBvirt}
\end{eqnarray}
In~\cite{HKT1} the real parts of the form factors were already presented
in a somewhat more implicit form.
The light quark-antiquark moments read
\begin{eqnarray}
\tilde R_\infty^q &=& 1\,, \nonumber\\[2mm]
\tilde R_0^q & = & \ln 4 - \frac{5}{3}
\,,\nonumber\\[2mm]
\tilde R_1^q &=&  
 \frac{1}{2} \ln^2 4 - \frac{5}{3} \ln 4 + \frac{28}{9}
- \zeta(2) 
\,,\nonumber \\[2mm]
\tilde R_2^q &=&
 \frac{1}{6}\,\ln^3 4 - \frac{5}{6}\,\ln^2 4 + 
  \bigg(\, \frac{28}{9} - \zeta(2) \,\bigg) \,\ln 4 - 
 \frac{164}{27} + \frac{5}{3}\,\zeta(2) + 2\,\zeta(3)
\,. 
\label{momentsfermions}
\end{eqnarray}
The reader should note that
the entire information on the light quark-antiquark vacuum
polarization  
inserted into the photon line is encoded in the three moments
$\tilde R_\infty^q$, $\tilde R_0^q$ and $\tilde R_1^q$. For
completeness we have also given the moment $\tilde R_2^q$.
At this point we would like to mention that the definition of the
moments of the light quark-antiquark vacuum polarization, 
as given in Eq.~(\ref{momentsdef}), strongly relies on the 
dispersion integration approach we used to determine the light quark
${\cal{O}}(\alpha^2)$ corrections to the form factors. 
In this approach the occurrence of
higher moments $\tilde R_n^q$, $n\ge 1$, comes from infrared
singular terms $\propto \ln^n\lambda$ in the ${\cal{O}}(\alpha)$ form
factors for $\lambda\to 0$.\footnote{ 
In fact, for secondary production of massive quarks also the moment
$\tilde R_2^q$ occurs due to an infrared divergence $\propto
\ln^2\lambda$ in the ${\cal{O}}(\alpha)$ form factors describing the
primary 
production of a massless quark-antiquark pair~\cite{HJKT1}.
}
If, on the other hand, dimensional regularization is employed to
regularize ultraviolet as well as infrared divergences, the 
moments~(\ref{momentsfermions}) can be uniquely identified from the
renormalized high energy vacuum polarization function 
in arbitrary $D=4-2\epsilon$ dimension. 
Expanded in terms of small $\epsilon$ the 
vacuum polarization function in the limit $q^2/m^2\to\infty$ ($m$
being the light quark pole mass and $\alpha$ the fine structure
constant) reads
($x=q$)
\begin{eqnarray}
\label{vacpolinddimensions}
\lefteqn{
\tilde\Pi_{light}^x(q^2) \, = \,
 -\frac{\alpha}{3\,\pi}\,\bigg\{\,
   \bigg[\,
    \tilde R_0^x + \tilde R_\infty^x\,\ln\frac{-q^2}{4\,m^2} 
   \,\bigg]
}\nonumber\\[2mm] & & \qquad
-\,\epsilon\,\bigg[\,
\tilde R_1^x 
  + \Big( \tilde R_\infty^x\,\ln\frac{-q^2}{4\,m^2} + \tilde R_0^x \Big)
      \,\ln\frac{-q^2}{4\,\mu^2}
  - \frac{1}{2}\,\tilde R_\infty^x\,\ln^2\frac{-q^2}{4\,m^2}
\,\bigg] \,
\nonumber \\[2mm] & & \qquad
+\,\epsilon^2\,\bigg[\,
\tilde R_2^x 
 + \frac{1}{2}\,\Big( \tilde R_\infty^x\,\ln \frac{-q^2}{4\,m^2} 
 + \tilde R_0^x \Big) \,
 \Big( \ln^2 \frac{-q^2}{4\,\mu^2} + \zeta(2) \Big) 
\nonumber\\ & & \qquad\qquad - 
  \Big( \frac{1}{2}\,\tilde R_\infty^x\,\ln^2 \frac{-q^2}{4\,m^2} 
      - \tilde R_1^x \Big) \,
   \ln \frac{-q^2}{4\,\mu^2} + \frac{1}{6}\,\tilde R_\infty^x\,
   \ln^3 \frac{-q^2}{4\,m^2}
\,\bigg]
\,\bigg\}
 + \, {\cal{O}}(\epsilon^3)
\,,
\label{defmomentsalternative}
\end{eqnarray}
where $\mu$ is the common scale parameter introduced in dimensional 
regularization.
Thus, in the framework of dimensional regularization higher moments
$\tilde R_n^q$, $n\ge 1$, in the expression for the second order
form factors come from $(1/\epsilon)^n$ infrared divergences in the 
${\cal{O}}(\alpha)$ form factors which cancel the corresponding
$\epsilon^n$ terms in $\tilde \Pi_{light}^q$.
\vspace{.5cm}
\subsection{Real Radiation}
\label{subsectionvectorrealcorrections}
The ${\cal{O}}(\alpha^2)$ corrections from real radiation of a light
quark-antiquark pair in primary massive quark pair production
correspond to the 
sum of all four body cuts of the diagrams  shown in 
Fig.~\ref{currentcorrelatordiagrams} and represent the Born rate for
primary massive quark production with additional radiation of a light
quark-antiquark pair. In order to accommodate for the conventions
used in the calculation of the virtual corrections we will for now
remain in the framework of on-shell QED for the calculation of the
real radiation process. 
Again we will determine the contribution to the imaginary part of the  
current-current correlator~(\ref{currentcorrelatorvector}), denoted by 
$\tilde r_{V,q}^{(2),real}$.
This task is accomplished by dividing the four-body phase space into
three different subprocesses: radiation of a photon with virtuality
$\lambda^2$ off one of the massive quarks, propagation of the photon
and decay of the virtual photon into the light quark-antiquark
pair. The first process is described by
$\tilde r_{V}^{(1),real}(M^2, q^2, \lambda^2)$,
the ${\cal{O}}(\alpha)$ contribution to the decay rate
$R^V$, Eq.~(\ref{defR}), from real radiation of a photon with virtuality
(mass squared) $\lambda^2$. The third process is proportional to
$\tilde R_{q\bar q}$, Eq.~(\ref{defRff}). 
Combining both contributions with the photon
propagation function, summing over all photon polarizations and
integrating  over all allowed values of $\lambda^2$, we arrive at the
expression
\begin{equation}
\tilde r_{V,q}^{(2), real}(M^2, m^2, q^2) \, = \,
\frac{1}{3} \int\limits_{4\,m^2}^{(\sqrt{q^2}-2\,M)^2}
   \frac{{\rm d}\lambda^2}{\lambda^2}\,\, 
  \tilde r_{V}^{(1),real} (M^2, q^2, \lambda^2)\,\,
   \tilde R_{q\bar q}(m^2, \lambda^2)
\,.
\label{masterformulavectorreal}
\end{equation}
It is evident that Eq.~(\ref{masterformulavectorreal}) closely
resembles the corresponding formula for the virtual light quark
${\cal{O}}(\alpha^2)$ corrections,
Eq.~(\ref{masterformulavectorvirtual}).  
An integral representation of $\tilde r_V^{(1),real}(M^2, q^2,
\lambda^2)$ can be found in Appendix~\ref{appendixfirstordervector}.
For $\lambda^2\to 0$ the well known QED result of~\cite{KallenSabry} is
recovered:
\begin{eqnarray}
\tilde r_V^{(1), real}(M^2, q^2, \lambda^2) 
& \stackrel{\lambda^2\to 0}{\longrightarrow} &
6\,f_V^{(2), real}\,
    \Big[\, \ln \frac{4\,q^2}{{{\lambda }^2}} \, \Big]  - 
   3\,f_V^{(1), real}
\,,
\label{rV1real}
\end{eqnarray}
where the functions $f_V^{(2), real}$ and $f_V^{(1), real}$ are given 
below (see Eqs.~(\ref{fV2real}) and (\ref{fV1real})). It should be
noted that the leading infrared behavior for $m/M\to 0$ of 
expression~(\ref{masterformulavectorreal}) is determined 
by the logarithmic photon mass singularity in Eq.~(\ref{rV1real}). 
For arbitrary masses $m$ and $M$ the occurrence of three different 
square roots in~(\ref{masterformulavectorreal}) leads to elliptic
integrals (regardless of the order of integration) and makes the
analytic evaluation in terms of polylogarithmic functions (in our
opinion) impossible. Even in the special 
case $m = M \neq 0$ the problem of ellipticity persists and no 
closed result of this four body phase space integration is available. 
This is not surprising as already the massive three particle phase
space of the first order subprocess cannot be expressed in closed
analytic form. However, for any finite ratios $m^2/q^2$ and $M^2/q^2$
the integrations in~(\ref{masterformulavectorreal}) are well behaved
and can be performed in a straightforward way
numerically~\cite{HKT2,ttdiss}.  If (at least) 
one of the quark masses is very light, the problem can be tackled
analytically: the case of primary production of massless quarks and
secondary production of massive ones has
been completely solved in~\cite{HJKT2}, whereas the analytical result
for the opposite mass assignment was first presented
in~\cite{HKT1}. In the following we will describe briefly the
calculation of the latter and give the result in terms of the moments
introduced above.  
\par 
In contrast to the case of virtual radiation even the calculation of
the  three body subprocess leads to elliptic phase space integrals and 
could not be carried out explicitly. In the resulting double 
integral~(\ref{masterformulavectorreal}) singularities (in the 
limit $m \to 0$) arise from both integrations. 
In order to deal with this
complication in a transparent way, we decompose the integration region 
in {\em soft} and {\em hard} radiation by introducing an
infinitesimally small parameter $\Delta$, $m \ll \Delta \ll M$:
\begin{eqnarray}
\tilde r_{V,q}^{(2),real} & = & \tilde r_{V,\,hard}^{(2),real} + 
\tilde r_{V,\,soft}^{(2),real} \nonumber\\[1mm]
 & = & \frac{1}{3} \, 
\Bigg[ 
 \int_{4M^2/q^2}^{(1-\Delta/\sqrt{q^2})^2} {\rm d}y + 
 \int_{(1-\Delta/\sqrt{q^2})^2}^{(1-2m/\sqrt{q^2})^2} {\rm d}y \,
\Bigg] \, \int_{4m^2/q^2}^{(1-\sqrt{y})^2} \frac{{\rm d}z}{z}\,{\cal F}(y,z)\,
  \tilde R_{q\bar q}(m^2, q^2 z)\,,
\label{realphspdecomposition}
\end{eqnarray}
where 
\begin{eqnarray}
{\cal F}(y,z) & \equiv &
\frac{\frac{8M^4}{q^4} + \frac{4M^2}{q^2}(1-y+z) - (1-y+z)^2 - 2(1+z)y}
 {1-y+z} \, 
\ln\frac{1-y+z-\sqrt{1-\frac{4M^2}{q^2 y}}\,\Lambda^{1/2}(y,z)}
        {1-y+z+\sqrt{1-\frac{4M^2}{q^2 y}}\,\Lambda^{1/2}(y,z)}
\nonumber\\[1mm]
&& -\sqrt{1-\frac{4M^2}{q^2 y}}\,\Lambda^{1/2}(y,z)\, 
 \left[ 1 + \frac{ \frac{16M^4}{q^4}+\frac{8M^2}{q^2} + 
        4\left(1+\frac{2M^2}{q^2}\right)z }{
 (1-y+z)^2-\left(1-\frac{4M^2}{q^2 y}\right)\,\Lambda(y,z)} \, \right] 
\end{eqnarray}
with 
\begin{eqnarray}
\Lambda(y,z) & \equiv & 1+y^2+z^2-2(y+z+y z)\,.
\end{eqnarray}
In the {\em soft} region the energy of the secondary quark pair is 
smaller than $\Delta$. The integration area is infinitesimally small and 
hence only the singular part of the integrand contributes. It is therefore
possible to perform the $y$-integration first, taking into account
only the singular terms. 
In the region of {\em hard} radiation, on the other hand, the
singularities arising from $z \to 0$ and $1 - y \to 0$ are well
distinguishable due to our choice $\Delta \gg m$: 
$$
1 - y \ \simeq \ \frac{2\Delta}{\sqrt{s}} \ \gg \
\frac{2m}{\sqrt{s}} 
\ = \ \sqrt{z_{min}}\,.
$$
This enables us to separate the singular part easily. By subtracting 
and adding ${\cal F}(y, z=0)$ we split up the integral in two parts:
the first part containing $[{\cal F}(y, z) - {\cal F}(y, z=0)]$ is 
infrared-save by construction and we can take the limit $m \to 0$ 
immediately. The integrand of the second part has a much simpler $y$
dependence (through ${\cal F}(y, z=0)$ only) and makes the double
integration possible.
\par 
With this decomposition of the phase space in a {\em soft} and a {\em
hard} part the calculation is similar to the case 
of virtual radiation: by subtracting and adding $\tilde R_{\infty}^q$
one splits each of the double integrals in two parts. In the first 
part one benefits from the fast convergence of 
$[\tilde R^q(m^2, q^2)-\tilde R^q_\infty]$ by
taking into account only the leading terms of the integrand. The result of
the first part is proportional to the moments $\tilde R_0^q$ and
$\tilde R_1^q$. The
second part now becomes tractable as there is one square root 
less. This part is proportional to $\tilde R_{\infty}^q$ and is,
similar to  the case of the virtual corrections, the most difficult
part of the calculation.\footnote{
A detailed description of this calculation can be found in
\cite{ttdiss}.}
\par 
The result for the rate describing real radiation of a light
quark-antiquark in terms of the moments is finally given by ($x=q$) 
\begin{eqnarray}
\mbox{}\,\tilde{r}_{V, x}^{(2), real} &=& 
  f_V^{(2), real}\,\bigg[ \ln^2 \frac{{m^2}}{q^2}\,\tilde{R}_{\infty}^x - 
      2\,\ln \frac{{m^2}}{q^2}\,\tilde{R}_0^x + 2\,\tilde{R}_1^x \bigg] + 
\nonumber\\  & &  
 f_V^{(1), real}\,\bigg[ \ln \frac{{m^2}}{q^2}\,\tilde{R}_{\infty}^x - 
      \tilde{R}_0^x \bigg]  + f_V^{(0), real}\,\tilde{R}_{\infty}^x
\,,
\label{finalvectorreal}
\end{eqnarray}
with
\begin{eqnarray}
f_V^{(2), real} &=& \frac{3 - {{\beta}^2}}{2}\,C_{real} \,,
\label{fV2real}\\[2mm]
f_V^{(1), real} &=& 
  \frac{3 - {{\beta}^2}}{2}\,B_{real} + 
   \frac{33 - 10\,{{\beta}^2} + {{\beta}^4}}{48}\,\ln p - 
   \frac{\beta\,\left( 39 - 17\,{{\beta}^2} \right) }{24} \,,
\label{fV1real}\\[2mm]
f_V^{(0), real} &=& 
  \frac{3 - {{\beta}^2}}{2}\,A_{real} + 
   \frac{15 - 6\,{{\beta}^2} - {{\beta}^4}}{24}\,
    \Big( \mbox{Li}_2(p) + \mbox{Li}_2({p^2}) \Big)
   \nonumber\,\\ 
 & & \mbox{}\,
   + \frac{\left( 1 - {{\beta}^2} \right) \,
      \left( 21 - 13\,{{\beta}^2} \right) }{24}\,\mbox{Li}_2(p) + 
   \frac{-51 + 60\,\beta + 46\,{{\beta}^2} - 20\,{{\beta}^3} - 11\,{{\beta}^4}
     }{24}\,\zeta(2) \nonumber\,\\ 
 & & \mbox{}\,
   + \frac{-9 + 24\,\beta + 24\,{{\beta}^2} - 16\,{{\beta}^3} - 
      11\,{{\beta}^4} + 2\,{{\beta}^5}}{24\,\beta}\,\ln^2 p + 
   \frac{\beta\,\left( -3 + {{\beta}^2} \right) }{3}\,
    \ln^2\Big(\frac{2\,{{\beta}^2}}{1 - {{\beta}^2}}\Big) 
   \nonumber\,\\ 
 & & \mbox{}\,
    +\bigg( \,\frac{33 - 10\,{{\beta}^2} + {{\beta}^4}}{24}\,\ln 2 - 
      \frac{\left( 4 - {{\beta}^2} \right) \,{{\beta}^2}}{3}\,\ln \beta - 
       \frac{15 - 22\,{{\beta}^2} + 3\,{{\beta}^4}}{24}\,
       \ln\Big(\frac{1 - {{\beta}^2}}{4}\Big)\, \bigg) \,\ln p 
   \nonumber\,\\ 
 & & \mbox{}\,
    +\frac{237 - 2\,{{\beta}^2} - 51\,{{\beta}^4}}{96}\,\ln p + 
   \frac{\beta\,\left( -39 + 17\,{{\beta}^2} \right) }{12}\,
    \ln\Big(\frac{1 - {{\beta}^2}}{2\,{{\beta}^2}}\Big) + 
   \frac{\beta\,\left( -99 + 37\,{{\beta}^2} \right) }{48}\,,
\end{eqnarray}
and
\begin{eqnarray}
A_{real} &=& \frac{1 + {{\beta}^2}}{6}\,
   \bigg\{ \,8\,\mbox{Li}_3(1 - p) + 6\,\mbox{Li}_3({p^2}) + 
     8\,\mbox{Li}_3\Big(\frac{p}{1 + p}\Big) + 10\,\mbox{Li}_3(1 - {p^2}) - 
     13\,\zeta(3) \nonumber\,\\ 
 & & \mbox{}\,-4\,
      \left( \mbox{Li}_2(p) + \mbox{Li}_2({p^2}) \right) \,
      \ln\Big(\frac{1 - {{\beta}^2}}{2\,{{\beta}^2}}\Big) - 
     \zeta(2)\,\bigg( -8\,\ln 2 + 9\,\ln p + 4\,\ln \beta - 
        6\,\ln\Big(\frac{1 - {{\beta}^2}}{4\,{{\beta}^2}}\Big) \bigg)
     \nonumber\,\\ 
 & & \mbox{}\,+2\,\ln p\,\ln \beta\,
      \bigg( -2\,\ln 2 + 11\,\ln p - 16\,\ln \beta - 
        12\,\ln\Big(\frac{1 - {{\beta}^2}}{4\,{{\beta}^2}}\Big) \bigg)
     \nonumber\,\\ 
 & & \mbox{}\,-\frac{1}{6}\,
      \bigg[ \,12\,\ln^2 2\,\ln p + 30\,\ln 2\,\ln^2 p - 
        46\,\ln^3 p - 36\,\ln 2 \,\ln p\,
        \ln\Big(\frac{1 - {{\beta}^2}}{4}\Big) 
        \nonumber\,\\ 
 & & \mbox{}\,+99\,\ln^2 p\,
         \ln\Big(\frac{1 - {{\beta}^2}}{4}\Big) - 
        51\,\ln p\,\ln^2\Big(\frac{1 - {{\beta}^2}}{4}\Big) - 
        \ln^3\Big(\frac{1 - {{\beta}^2}}{4}\Big)\, \bigg] \, \bigg\} 
\,,\label{defAreal}\\[2mm]
B_{real} &=& - \frac{1 + {{\beta}^2}}{3}\,
      \bigg[ \,\mbox{Li}_2(p) + \mbox{Li}_2({p^2}) - 2\,\zeta(2) 
        \nonumber\,\\ 
 & & \mbox{}\,\qquad\,+\frac{1}{2}\,\ln p\,
         \bigg( \frac{5}{2}\,\ln p + 8\,\ln 2 + 2\,\ln \beta - 
           3\,\ln(1 - {{\beta}^2}) \bigg) \, \bigg]  - 
   \frac{2\,\beta}{3}\,\ln\Big(\frac{1 - {{\beta}^2}}{2\,{{\beta}^2}}\Big)
\,,\label{defBreal}\\[2mm]
C_{real} &=& - \frac{1}{6}\,
     \Big( \beta + \frac{1 + {{\beta}^2}}{2}\,\ln p \Big) 
\,.\label{defCreal}
\end{eqnarray}
\vspace{0.5cm}
\subsection{Contribution to the Total Rate}
\label{subsectionvectorqcd}
Combining the virtual and real corrections determined in the two
previous subsections one arrives at
\begin{eqnarray}
\tilde r_V^{(1)} & = &  \tilde r_V^{(1),virt} + \tilde r_V^{(1),real}
\nonumber\\[2mm] 
 & = & \mbox{} \frac{3 - {{\beta}^2}}{2}\,B - 
   \frac{1 - \beta }{16} \,
      \left( 33 - 39\,\beta - 17\,{{\beta}^2} + 7\,{{\beta}^3} \right) \,
    \ln p + \frac{3}{8}\,\beta\,\left( 5 - 3\,{{\beta}^2} \right)
\end{eqnarray}
with
\begin{eqnarray}
B & = & 
  \Big( 1 + {{\beta}^2} \Big) \,
    \left( 2\,\ln(1 - p)\,\ln p + \ln(1 + p)\,\ln p + 2\,\mbox{Li}_2(p) + 
      \mbox{Li}_2({p^2}) \right) \,\nonumber\,\\ 
 & & \mbox{}\,\quad - 
   2\,\beta\,\Big( 2\,\ln(1 - p) + \ln(1 + p) \Big) 
\label{Bdefinition}
\end{eqnarray}
for the ${\cal{O}}(\alpha)$ corrections to $\tilde R_V$, 
Eq.~(\ref{tildeRV}),  due to virtual and real photon
emission and at 
\begin{eqnarray}
\tilde r_{V,x}^{(2)} & = &  
\tilde r_{V,x}^{(2),virt} + \tilde r_{V,x}^{(2),real}
\nonumber\\[2mm] 
 & = & \mbox{} 
-\frac{1}{3}\Big[\,\tilde R_\infty^x\,\ln\frac{m^2}{q^2} - 
  \tilde R_0^x \,\Big]\, \tilde r_V^{(1)} +
\tilde R_\infty^x\,\delta_V^{(2)}  
\label{realvirtvector}
\end{eqnarray}
with
\begin{eqnarray}
\delta_V^{(2)} & = & \frac{3 - {{\beta}^2}}{2}\,A + 
   \frac{1}{24}\,\bigg\{\,
       \left( 15 - 6\,{{\beta}^2} - {{\beta}^4} \right) \,
       \left( \mbox{Li}_2(p) + \mbox{Li}_2({p^2}) \right)  + 
      3\,\left( 7 - 22\,{{\beta}^2} + 7\,{{\beta}^4} \right) \,
       \mbox{Li}_2(p)\,\nonumber\,\\ 
 & & \mbox{}\,\qquad - 
      \left( 1 - \beta \right) \,
       \left( 51 - 45\,\beta - 27\,{{\beta}^2} + 5\,{{\beta}^3} \right) \,
       \zeta(2)\,\nonumber\,\\ 
 & & \mbox{}\,\qquad + 
      \frac{\left( 1 + \beta \right)}{\beta}\,
         \left( -9 + 33\,\beta - 9\,{{\beta}^2} - 15\,{{\beta}^3} + 
           4\,{{\beta}^4} \right)\,\ln^2 p\,\nonumber\,
       \\ 
 & & \mbox{}\,\qquad + 
      \bigg[\,\left( 33 + 22\,{{\beta}^2} - 7\,{{\beta}^4} \right) \,
          \ln 2 - 10\,\left( 3 - {{\beta}^2} \right) \,
          \left( 1 + {{\beta}^2} \right) \,\ln\beta\,\nonumber\,
          \\ 
 & & \mbox{}\,\qquad\,\qquad\,\qquad - 
         \left( 15 - 22\,{{\beta}^2} + 3\,{{\beta}^4} \right) \,
          \ln\Big(\frac{1 - {{\beta}^2}}{4\,{{\beta}^2}}\Big)\,\bigg] \,
       \ln p\,\nonumber\,\\ 
 & & \mbox{}\,\qquad + 
      2\,\beta\,\left( 3 - {{\beta}^2} \right) \,
       \bigg( 2\,\ln\Big(\frac{{{\beta}^2}}{2}\Big) - 
      \ln(1 - {{\beta}^2}) \bigg) \,
       \bigg( 3\,\ln\Big(\frac{1 - {{\beta}^2}}{4}\Big) - 
      4\,\ln\beta \bigg) \, \nonumber\,\\ 
 & & \mbox{}\,\qquad + 
      \frac{237 + 62\,{{\beta}^2} - 59\,{{\beta}^4}}{4}\,\ln p + 
      8\,\beta\,\left( 3 - {{\beta}^2} \right) \,
       \Big( 3\,\ln 2 - 2\,\ln\beta \Big) \,\nonumber\,
       \\ 
 & & \mbox{}\,\qquad - 
      6\,\beta\,\left( 17 - 7\,{{\beta}^2} \right) \,
       \ln\Big(\frac{1 - {{\beta}^2}}{2\,{{\beta}^2}}\Big) - 
      \frac{\beta\,\left( 75 - 29\,{{\beta}^2} \right) }{2}\,\bigg\} 
\,,
\end{eqnarray}
\begin{eqnarray}
A &=& - \frac{1 + {{\beta}^2}}{3}\, 
     \bigg\{\,\mbox{Li}_3(p) - 2\,\mbox{Li}_3(1 - p) - 
       3\,\mbox{Li}_3({p^2}) - 4\,\mbox{Li}_3\Big({p\over {1 + p}}\Big) - 
       5\,\mbox{Li}_3(1 - {p^2}) + 
       \frac{11}{2}\,\zeta(3)\,\nonumber\,\\ 
 & & \mbox{}\,\qquad + 
       \mbox{Li}_2(p)\,\ln\Big(\frac{4\,\left( 1 - {{\beta}^2} \right) }{
          {{\beta}^4}}\Big) + 2\,\mbox{Li}_2({p^2})\,
        \ln\Big(\frac{1 - {{\beta}^2}}{2\,{{\beta}^2}}\Big) + 
       2\,\zeta(2)\,\bigg( \ln p - 
    \ln\Big(\frac{1 - {{\beta}^2}}{4\,\beta}\Big)
           \bigg) \,\nonumber\,\\ 
 & & \mbox{}\,\qquad - 
       \frac{1}{6}\,\ln\Big(\frac{1 + \beta}{2}\Big)\,
        \bigg( 36\,\ln 2\,\ln p - 44\,\ln^2 p + 
          49\,\ln p\,\ln\Big(\frac{1 - {{\beta}^2}}{4}\Big) + 
          \ln^2\Big(\frac{1 - {{\beta}^2}}{4}\Big) \bigg) \,\nonumber\,
        \\ 
 & & \mbox{}\,\qquad - 
       \frac{1}{2}\,\ln p\,\ln\beta\,
        \left( 36\,\ln 2 + 21\,\ln p + 16\,\ln\beta - 
          22\,\ln(1 - {{\beta}^2}) \right) \,\bigg\}  
\label{Adefinition}
\end{eqnarray}
for the ${\cal{O}}(\alpha^2)$ corrections from virtual and real
emission of a light quark-antiquark pair ($x=q$). It is evident that
the quadratic logarithms of the ratio $m^2/q^2$ in Eqs.~(\ref{F12})
and (\ref{finalvectorreal}) cancel in the sum~(\ref{realvirtvector}). 
Further, all terms proportional to $\tilde R^q_1$ add up to zero. As
already indicated at the end of 
Subsection~\ref{subsectionvectorvirtualcorrections} this corresponds to
the cancellation of the infrared singularity $\propto \ln(\lambda^2)$ in
the ${\cal{O}}(\alpha)$ result $\tilde r^{(1)}_V$, when the real and
virtual contributions are combined. 
However, one single logarithm of $m^2/q^2$ (multiplying $\tilde
r^{(1)}_V$) remains, rendering the 
perturbative expansion unreliable if the ratio $m^2/q^2$ is very small.
This is a consequence of the on-shell
renormalization scheme, where $\alpha$ is defined as
the fine structure constant, i.e.\ at momentum transfer zero. Thus, to
determine the correction due to real and virtual radiation of massless
quarks a mass independent
definition of the coupling like the ${\overline{\mbox{MS}}}$ scheme
is more appropriate. 
The transition to the ${\overline{\mbox{MS}}}$ scheme and to QCD will
be carried out in Section~\ref{sectiontransitiontoqcd}.
\par
\vspace{1cm}
\section{Heavy Quark Production through the Axial-Vector Current}
\label{sectionaxialvectorcurrent}
This section is devoted to the presentation of the light quark
${\cal{O}}(\alpha^2)$ corrections to the massive
quark-antiquark axial-vector vertex. As indicated in the
introduction only non-singlet contributions are considered. 
We follow the lines of Section~\ref{sectionvectorcurrent} and present
the results for the virtual and real radiation corrections in the framework
of on-shell renormalized QED. The transition to the 
$\overline{\mbox{MS}}$ scheme and to QCD is indicated in
Section~\ref{sectiontransitiontoqcd}.
\par
Let us start with the presentation of the virtual corrections to the
axial-vector vertex, $\Lambda_\mu^A$, parametrized in terms of the form
factors $F_3$ and $F_4$,
\begin{equation}
\overline{u}_Q(k_2)\,\Lambda_\mu^A\,v_Q(k_1) \, = \,
\overline{u}_Q(k_2)\,\left[\,
\gamma_\mu\,\gamma_5\,F_3(q^2) - 
\frac{1}{2\,M}\,q^\mu\,\gamma_5\,F_4(q^2)
\,\right]\,v_Q(k_1)
\,,
\end{equation} 
where $q=k_1+k_2$.
The perturbative expansions of the form factors read 
\begin{eqnarray}
F_3(q^2) & = & 1 + 
\left(\frac{\alpha}{\pi}\right)\,F_3^{(1)}(q^2) +
\left(\frac{\alpha}{\pi}\right)^2\,F_3^{(2)}(q^2) + \ldots\,,
\\[2mm]
F_4(q^2) & = & \hspace{.65cm}
\left(\frac{\alpha}{\pi}\right)\,F_4^{(1)}(q^2) +
\left(\frac{\alpha}{\pi}\right)^2\,F_4^{(2)}(q^2) + \ldots
\,.
\end{eqnarray}
To ${\cal{O}}(\alpha)$ $F_3$ is completely renormalized 
by the QED quark field wave function
renormalization constant. In contrast to the Dirac form factor this 
leads to a non-vanishing value of $F_3$
for zero momentum transfer. Only $F_3$ leads to contributions to the
rate $\tilde R^A$, 
\begin{eqnarray}
\tilde R^A(q^2) & = &
\tilde r_A^{(0)}(q^2) 
+\left(\frac{\alpha}{\pi}\right)\,\tilde r_A^{(1)}(q^2)
+\left(\frac{\alpha}{\pi}\right)^2\,\tilde r_A^{(2)}(q^2)
+\ldots\,,
\label{tildeRA}
\end{eqnarray}
where
\begin{eqnarray}
\tilde r_{A}^{(0)} & = &
   \beta^3
\,,\\[2mm]
\tilde r_A^{(1),virt} & = &
  2\,\beta^3\,{\rm Re}F_3^{(1)}
\,.
\end{eqnarray}
The contribution of the virtual light quark 
contributions to $\tilde r_{A}$ is given by
\begin{equation}
\tilde r_{A,q}^{(2),virt} \, = \,
  2\,\beta^3\,{\rm Re}F_{3,q}^{(2)}
\,.
\end{equation}
In analogy to the calculations for the vector current vertex,
$F_{3/4,q}^{(2)}$ can be written as an integral over the
${\cal{O}}(\alpha)$ form factors with exchange of a vector boson with
mass $\lambda$,
\begin{equation}
F_{i,q}^{(2)}(M^2, m^2, q^2) = 
 \frac{1}{3} \int\limits_{4\,m^2}^\infty
   \frac{{\rm d}\lambda^2}{\lambda^2}\, 
   F_i^{(1)}(M^2, q^2, \lambda^2)\,
   \tilde R_{q\bar q}(m^2, \lambda^2)\,,\qquad
i=3,4\,.
\label{masterformulaaxialvectorvirtual}
\end{equation}
As in the vector current case
Eq.~(\ref{masterformulaaxialvectorvirtual}) 
accounts for all necessary subtractions.
Explicit formulae for
$F_{3/4}^{(1)}(M^2, q^2, \lambda^2)$ are presented  
in Appendix~\ref{appendixfirstorderaxialvector}
for energies above threshold, 
Eqs.~(\ref{F3oneloopfinitelambda}) and (\ref{F4oneloopfinitelambda}).
$\tilde R_{q\bar q}$ is given in Eq.~(\ref{defRff}).
The well known ${\cal{O}}(\alpha)$ QED expressions~\cite{Jersak1} are
obtained for $\lambda^2\to 0$, 
\begin{eqnarray}
\lefteqn{
F_{3}^{(1)}(M^2, q^2, \lambda^2\to 0) \, = \,}
\nonumber\\ & &
-\frac{1}{2}\,\bigg( 1 + \frac{1 + {{\beta}^2}}{2\,\beta}\,\ln p 
  \bigg) \,\ln \frac{\lambda^2}{M^2}   - 
  \frac{1 + {{\beta}^2}}{2\,\beta}\,
   \bigg( \mbox{Li}_2(1 - p) + \frac{1}{4}\,\ln^2 p - 3\,\zeta(2) \bigg) \,
   \nonumber\,\\ 
 & & \mbox{} - \frac{2 + {{\beta}^2}}{4\,\beta}\,\ln p - 
  1 - i\,\pi \,\bigg[\, \frac{1 + {{\beta}^2}}{4\,\beta}\,
     \ln \frac{\lambda^2}{M^2} + 
     \frac{1 + {{\beta}^2}}{4\,\beta}\,
      \ln\Big(\frac{1 - {{\beta}^2}}{4\,{{\beta}^2}}\Big) + 
     \frac{2 + {{\beta}^2}}{4\,\beta} \,\bigg]  
\,,\\[2mm]
\lefteqn{
F_{4}^{(1)}(M^2, q^2, \lambda^2\to 0) \, = \,
-\,\frac{\left( 1 - {{\beta}^2} \right) \,
      \left( 2 + {{\beta}^2} \right) }{4\,\beta}\,
    \bigg( \ln p + \frac{2\,\beta}{2 + {{\beta}^2}} + i\,\pi  \bigg)  
\,.}
\end{eqnarray}
Expressed in terms of the moments introduced in
Section~\ref{sectionvectorcurrent}, Eqs.~(\ref{momentsfermions}), 
the virtual light quark ${\cal{O}}(\alpha^2)$  corrections read
($x=q$) 
\begin{eqnarray}
F_{3,x}^{(2)} & = &
f_3^{(2)}\,\bigg[\, \tilde R_\infty^x\,\ln^2 \frac{m^2}{q^2} - 
     2\,\tilde R_0^x\,\ln \frac{m^2}{q^2} + 2\,\tilde R_1^x \,\bigg]  + 
  f_3^{(1)}\,\bigg[\, \tilde R_\infty^x\,\ln \frac{m^2}{q^2} - 
                     \tilde R_0^x \,\bigg]  + 
  f_3^{(0)}\,\tilde R_\infty^x
\,,\\
F_{4,x}^{(2)} & = &
f_4^{(1)}\,\bigg[\, \tilde R_\infty^x\,\ln \frac{m^2}{q^2} - 
                  \tilde R_0^x \,\bigg]  + 
   f_4^{(0)}\,\tilde R_\infty^x
\,,
\end{eqnarray}
where
\begin{eqnarray}
f_3^{(0)} & = &
A_{virt} + \frac{2}{3}\,( 2 + {{\beta}^2}) \,B_{virt} - 
  \frac{1}{3}\,\ln\Big(\frac{1 - {{\beta}^2}}{16}\Big) + 
  \frac{4 + 3\,{{\beta}^2}}{24\,\beta}\,\ln p - 
  \frac{2 - \beta + {{\beta}^2}}{2\,\beta}\,\zeta(2) + \frac{1}{6}
\\ & &
+i\,\pi\,\bigg\{\,
\frac{1}{12\,\beta}\,\bigg[\, 2\,( 1 + {{\beta}^2}) \,
     \ln^2\Big(\frac{\beta}{2}\Big) - 2\,(2 + {{\beta}^2} ) \,
     \ln\Big(\frac{\beta}{2}\Big) + ( 1 + {{\beta}^2} ) \,\zeta(2) + 
    \frac{1}{2}\,( 4 + 3\,{{\beta}^2} )  \,\bigg]
\,\bigg\}
\,,\nonumber\\[2mm]
f_3^{(1)} & = &
\frac{2}{3}\,( 1 + {{\beta}^2} ) \,B_{virt} - 
  \frac{1}{6}\,\ln\Big(\frac{1 - {{\beta}^2}}{16}\Big) + 
  \frac{2 + {{\beta}^2}}{12\,\beta}\,\ln p - 
  \frac{1 + {{\beta}^2}}{2\,\beta}\,\zeta(2) + \frac{1}{3}
\nonumber\\ & &
+i\,\pi\,\bigg\{\,
\frac{1}{12\,\beta}\,\bigg[\, -2\,( 1 + {{\beta}^2} ) \,
     \ln\Big(\frac{\beta}{2}\Big) + 2 + {{\beta}^2} \,\bigg]
\,\bigg\}
\,,\\[2mm]
f_3^{(2)} & = &
\frac{1 + {{\beta}^2}}{24\,\beta}\,\ln p + \frac{1}{12}
\, + \, i\,\pi\,\bigg\{\,
\frac{1 + {{\beta}^2}}{24\,\beta}
\,\bigg\}
\,,\\[5mm]
f_4^{(0)} & = &
\frac{2}{3}\,( 1 - {{\beta}^2} ) \,
  \bigg[\, ( 2 + {{\beta}^2} ) \,B_{virt} - 
    \frac{1}{4}\,\ln\Big(\frac{1 - {{\beta}^2}}{16}\Big) + 
    \frac{4 + 3\,{{\beta}^2}}{16\,\beta}\,\ln p - 
    \frac{3\,\left( 2 + {{\beta}^2} \right) }{4\,\beta}\,\zeta(2) + 
    \frac{3}{8} \,\bigg]
\nonumber\\ & &
+i\,\pi\,\bigg\{\,
\frac{1 - {{\beta}^2}}{6\,\beta}\,
  \bigg[\, -( 2 + {{\beta}^2}) \,\ln\Big(\frac{\beta}{2}\Big) 
    + \frac{1}{4}\,( 4 + 3\,{{\beta}^2})  \,\bigg] 
\,\bigg\}
\,,\\[2mm]
f_4^{(1)} & = &
\frac{1 - {{\beta}^2}}{6}\,\bigg[\, 
  \frac{2 + {{\beta}^2}}{2\,\beta}\,\ln p + 1 \,\bigg] 
\, + \, i\,\pi\,\bigg\{\,
\frac{\left( 1 - {{\beta}^2} \right) \,\left( 2 + {{\beta}^2} \right) }{
  12\,\beta}
\,\bigg\}
\,.
\end{eqnarray}
The functions $A_{virt}$ and $B_{virt}$ are defined
in Eqs.~(\ref{defAvirt}) and (\ref{defBvirt}), respectively.
\par
The corresponding real radiation corrections can be written as an
integral similar to~(\ref{masterformulavectorreal})
\begin{equation}
\tilde r_{A,q}^{(2),real}(M^2, m^2, q^2) \, = \,
\frac{1}{3} \int\limits_{4\,m^2}^{(\sqrt{q^2}-2\,M)^2}
   \frac{{\rm d}\lambda^2}{\lambda^2}\,\, 
  \tilde r_{A}^{(1),real} (M^2, q^2, \lambda^2)\,\,
   \tilde R_{q\bar q}(m^2, \lambda^2)
\,,
\label{masterformulaaxialvectorreal}
\end{equation}
where $\tilde r_{A}^{(1),real}(M^2, q^2, \lambda^2)$ describes the
real radiation of a vector boson with mass $\lambda$ off one of the
primary heavy quarks. An integral representation of 
$\tilde r_{A}^{(1),real}(M^2, q^2, \lambda^2)$ is given in the
Appendix~\ref{appendixfirstorderaxialvector}. For  
$\lambda^2\to 0$ the result of~\cite{Jersak1} is recovered,
\begin{eqnarray}
\tilde r_{A}^{(1),real} & \stackrel{\lambda^2\to 0}{\longrightarrow} &
6\,f_A^{(2), real}\,
    \Big[ \ln \frac{4\,q^2}{{{\lambda }^2}}  \Big]  - 
   3\,f_A^{(1), real}
\,,
\end{eqnarray}
with $f_A^{(2), real}$ and $f_A^{(1), real}$ given below in 
Eqs.~(\ref{f2areal}) and (\ref{f1areal}), respectively. 
The complete form of the light quark second order real corrections
expressed in terms of the moments reads ($x = q$)
\begin{eqnarray}
\tilde r_{A, x}^{(2), real} & = &
  f_A^{(2), real}\,\bigg[ \ln^2 \frac{{m^2}}{q^2}\,\tilde{R}_{\infty}^x - 
      2\,\ln \frac{{m^2}}{q^2}\,\tilde{R}_0^x + 2\,\tilde{R}_1^x \bigg]  + 
\nonumber\\& &  
 f_A^{(1), real}\,\bigg[ \ln \frac{{m^2}}{q^2}\,\tilde{R}_{\infty}^x - 
      \tilde{R}_0^x \bigg]  + f_A^{(0), real}\,\tilde{R}_{\infty}^x
\,,
\end{eqnarray}
where
\begin{eqnarray}
f_A^{(2), real} &=& {{\beta}^2}\,C_{real}
\label{f2areal}\,,\\[2mm]
f_A^{(1), real} &=& 
  {{\beta}^2}\,B_{real} + \frac{
     1 + {{\beta}^2}}{96} \,
      \left( 21 + 6\,{{\beta}^2} - 3\,{{\beta}^4} \right)\,\ln p  + 
   \frac{\beta}{48}\,\left( 21 + {{\beta}^2} \right) \,
     \left( 1 - 3\,{{\beta}^2} \right)
\,,\label{f1areal}\\[2mm]
f_A^{(0), real} &=& 
  {{\beta}^2}\,A_{real} + \frac{
     5 - {{\beta}^2}}{96} \,
      \left( 3 + 2\,{{\beta}^2} + 3\,{{\beta}^4} \right)\,
    \left( \mbox{Li}_2(p) + \mbox{Li}_2({p^2}) \right)
   \nonumber\,\\ 
 & & \mbox{}\,
    +\frac{1 - {{\beta}^2}}{96} \,
      \left( 39 + 62\,{{\beta}^2} + 3\,{{\beta}^4} \right)\,
    \mbox{Li}_2(p) + \frac{1}{96}\,\left(-69 - 
 37\,{{\beta}^2} + 160\,{{\beta}^3} + 
      33\,{{\beta}^4} + 9\,{{\beta}^6}\right)\,\zeta(2)
   \nonumber\,\\ 
 & & \mbox{}\,
    +\frac{1}{192}\,\left(57 - 144\,\beta + 
29\,{{\beta}^2} + 176\,{{\beta}^3} + 
      3\,{{\beta}^4} - 9\,{{\beta}^6}\right)\,\ln^2 p - 
   \frac{2\,{{\beta}^3}}{3}\,
 \ln^2\Big(\frac{2\,{{\beta}^2}}{1 - {{\beta}^2}}\Big)
   \nonumber\,\\ 
 & & \mbox{}\,
    +\bigg(\,\frac{1 + {{\beta}^2}}{16}\,
         \left( 7 + 2\,{{\beta}^2} - {{\beta}^4} \right)\,\ln 2  - 
      \frac{{{\beta}^2}\,\left( 2 + {{\beta}^2} \right) }{3}\,\ln \beta
      \nonumber\,\\ 
 & & \mbox{}\,\quad\,
       +\frac{1}{96}\,\left(-15 + 25\,{{\beta}^2} + 3\,{{\beta}^4} 
+ 3\,{{\beta}^6}\right)\,
       \ln\Big(\frac{1 - {{\beta}^2}}{4}\Big)\, \bigg) \,\ln p 
   \nonumber\,\\ 
 & & \mbox{}\,
    +\frac{1}{96}\,\left(-58 + 151\,{{\beta}^2} + 
 92\,{{\beta}^4} - {{\beta}^6}\right)\,\ln p  
    + \frac{\beta}{24}\,\left( 1 - 3\,{{\beta}^2} \right) \,
      \left( 21 + {{\beta}^2} \right)\,
    \ln\Big(\frac{1 - {{\beta}^2}}{2\,{{\beta}^2}}\Big)
   \nonumber\,\\ 
 & & \mbox{}\,
    +\frac{\beta}{48}\,\left( 26 - 81\,{{\beta}^2} - 7\,{{\beta}^4} \right)
\,.
\end{eqnarray}
The functions $A_{real}$, $B_{real}$ and $C_{real}$ are defined in 
Eqs.~(\ref{defAreal}), (\ref{defBreal}) and (\ref{defCreal}),
respectively.
\par
By adding the corresponding virtual and real contributions
one now obtains the ${\cal{O}}(\alpha)$ corrections to $\tilde R_A$
\begin{eqnarray}
\tilde r_A^{(1)} & = & \tilde r_A^{(1),virt} + \tilde r_A^{(1),real}
\\ & = &
 {{\beta}^2}\,B - \frac{
     1 - \beta}{32} \,
      \left( 21 + 21\,\beta + 80\,{{\beta}^2} - 16\,{{\beta}^3} + 
        3\,{{\beta}^4} + 3\,{{\beta}^5} \right)\,\ln p + 
   \frac{3\,\beta}{16}\,\left( -7 + 10\,{{\beta}^2} + {{\beta}^4} \right)
\nonumber
\end{eqnarray}
and the  ${\cal{O}}(\alpha^2)$ corrections due to virtual and real
emission of a light quark-antiquark pair ($x = q$):
\begin{eqnarray}
\tilde r_{A,q}^{(2)} & = &  
\tilde r_{A,q}^{(2),virt} + \tilde r_{A,q}^{(2),real}
\nonumber\\[2mm] 
 & = & \mbox{} 
-\frac{1}{3}\Big[\,\tilde R_\infty^x\,\ln\frac{m^2}{q^2} - 
  \tilde R_0^x \,\Big]\, \tilde r_A^{(1)} +
\tilde R_\infty^x\,\delta_A^{(2)}  
\label{realvirtaxialvector}
\end{eqnarray}
with
\begin{eqnarray}
\delta_A^{(2)} & = & 
 {{\beta}^2}\,A + \frac{1}{96}\,
    \bigg\{\,\left( 5 - {{\beta}^2} \right) \,
       \left( 3 + 2\,{{\beta}^2} + 3\,{{\beta}^4} \right) \,
       \left( \mbox{Li}_2(p) + \mbox{Li}_2({p^2}) \right) \,\nonumber\,
       \\ 
 & & \mbox{}\,\quad + 
      \left( 39 - 41\,{{\beta}^2} - 91\,{{\beta}^4} - 3\,{{\beta}^6} \right)\,
       \mbox{Li}_2(p) \,\nonumber\,
       \\ 
 & & \mbox{}\,\quad - 
      \left( 1 - \beta \right) \,
       \left( 69 + 69\,\beta + 234\,{{\beta}^2} - 22\,{{\beta}^3} + 
         9\,{{\beta}^4} + 9\,{{\beta}^5} \right) \,\zeta(2)\,\nonumber\,
       \\ 
 & & \mbox{}\,\quad + 
      \frac{1 + \beta}{2} \,
         \left( 57 - 201\,\beta + 198\,{{\beta}^2} - 22\,{{\beta}^3} + 
           9\,{{\beta}^4} - 9\,{{\beta}^5} \right)\,\ln^2 p\,\nonumber\,
       \\ 
 & & \mbox{}\,\quad + 
      \bigg[\,2\,\left( 21 + 59\,{{\beta}^2} + 19\,{{\beta}^4} - 
            3\,{{\beta}^6} \right) \,\ln 2 + 
         2\,\left( -15 - 39\,{{\beta}^2} - 29\,{{\beta}^4} + 
            3\,{{\beta}^6} \right) \,\ln\beta\,\nonumber\,\\ 
 & & \mbox{}\,
          \qquad + \left( -15 + 25\,{{\beta}^2} + 3\,{{\beta}^4} + 
            3\,{{\beta}^6} \right) \,
          \ln\Big(\frac{1 - {{\beta}^2}}{4\,{{\beta}^2}}\Big)\,\bigg] \,
       \ln p\,\nonumber\,\\ 
 & & \mbox{}\,\quad + 
      16\,{{\beta}^3}\,\bigg( 2\,\ln\Big(\frac{{{\beta}^2}}{2}\Big) - 
         \ln(1 - {{\beta}^2}) \bigg) \,
       \bigg( 3\,\ln\Big(\frac{1 - {{\beta}^2}}{4}\Big) - 4\,\ln\beta \bigg) \,
       \nonumber\,\\ 
 & & \mbox{}\,\quad + 
      \left( -58 + 183\,{{\beta}^2} + 116\,{{\beta}^4} - {{\beta}^6} \right)\,
       \ln p + 12\,\beta\,\left( 7 - 26\,{{\beta}^2} - {{\beta}^4} \right) \,
       \ln\Big(\frac{1 - {{\beta}^2}}{2\,{{\beta}^2}}\Big)\,\nonumber\,
       \\ 
 & & \mbox{}\,\quad + 
      64\,{{\beta}^3}\,\Big( 3\,\ln 2 - 2\,\ln\beta \Big)  + 
      2\,\beta\,\left( 26 - 65\,{{\beta}^2} - 7\,{{\beta}^4} \right)\, 
 \bigg\}\,.
\end{eqnarray}
The functions $A$ and $B$ are defined in Eqs.~(\ref{Adefinition}) and
(\ref{Bdefinition}), respectively.
\par
\vspace{1cm}
\section{Heavy Quark Production through the Scalar and Pseudoscalar
Currents}
\label{sectionscalarcurrent}
This section is devoted to the presentation of the light quark
${\cal{O}}(\alpha^2)$ corrections to the massive
quark-antiquark production rate through the scalar and pseudoscalar
currents, Eqs.~(\ref{defscalarcurrents}) and (\ref{defRSP}). As
indicated in the introduction only non-singlet contributions are considered. 
We follow the lines of Section~\ref{sectionvectorcurrent} and present
the results for the virtual and real radiation corrections in the framework
of on-shell renormalized QED. The transition to the 
$\overline{\mbox{MS}}$ scheme and to QCD is carried out in 
Section~\ref{sectiontransitiontoqcd}. 
\par
The virtual corrections to the scalar and pseudoscalar vertices are
parametrized in terms of the form factors $S$ and $P$, respectively.
$S$ and $P$ are normalized to one at the Born level and thus have the
following perturbative expansions
\begin{eqnarray}
S(q^2) & = & 1 + 
\left(\frac{\alpha}{\pi}\right)\,S^{(1)}(q^2) +
\left(\frac{\alpha}{\pi}\right)^2\,S^{(2)}(q^2) + \ldots
\,,\\[2mm]
P(q^2) & = & 1 + 
\left(\frac{\alpha}{\pi}\right)\,P^{(1)}(q^2) +
\left(\frac{\alpha}{\pi}\right)^2\,P^{(2)}(q^2) + \ldots
\,.
\end{eqnarray}
We would like to remind the reader that, as defined in
Eq.~(\ref{defscalarcurrents}), the overall divergences for the scalar
and pseudoscalar form factors are subtracted by the massive quark
wave function renormalization constant and the massive quark (pole)
mass counterterm.
The contributions of the form factors to the rates
\begin{eqnarray}
\tilde R^{S/P}(q^2) & = &
\tilde r_{S/P}^{(0)}(q^2) 
+\left(\frac{\alpha}{\pi}\right)\,\tilde r_{S/P}^{(1)}(q^2)
+\left(\frac{\alpha}{\pi}\right)^2\,\tilde r_{S/P}^{(2)}(q^2)
+\ldots
\label{tildeRSP}
\end{eqnarray}
read   
\begin{eqnarray}
\tilde r_{S}^{(0)} & = & \beta^3
\,,  \hspace{4cm}
\tilde r_{P}^{(0)} \,\,\, = \,\,\, \beta  
\,,\\[2mm]
\tilde r_S^{(1),virt} & = &
  2\,\beta^3\,{\rm Re}S^{(1)}
\,,  \hspace{2cm}
\tilde r_P^{(1),virt} \,\,\, = \,\,\,
  2\,\beta\,{\rm Re}P^{(1)}
\end{eqnarray}
for the Born and ${\cal{O}}(\alpha)$ coefficients and
\begin{eqnarray}
\tilde r_{S,q}^{(2),virt} & = &
  2\,\beta^3\,{\rm Re}S_q^{(2)}
\,,  \hspace{2cm}
\tilde r_{P,q}^{(2),virt} \,\,\, = \,\,\,
  2\,\beta\,{\rm Re}P_q^{(2)}
\end{eqnarray}
for the virtual light quark ${\cal{O}}(\alpha^2)$ 
corrections. The integral representation for $S_q^{(2)}$ and
$P_q^{(2)}$ is given by 
\begin{equation}
X_{q}^{(2)}(M^2, m^2, q^2) = 
 \frac{1}{3} \int\limits_{4\,m^2}^\infty
   \frac{{\rm d}\lambda^2}{\lambda^2}\, 
   X^{(1)}(M^2, q^2, \lambda^2)\,
   \tilde R_{q\bar q}(m^2, \lambda^2)\,,\qquad
X=S,P
\,.
\label{masterformulascalarvirtual}
\end{equation}
$S^{(1)}(M^2, q^2, \lambda^2)$ and $P^{(1)}(M^2, q^2, \lambda^2)$, the
${\cal{O}}(\alpha)$ 
form factors with exchange of a vector boson with mass $\lambda$, are
presented in the Appendix~\ref{appendixfirstorderscalar} and
\ref{appendixfirstorderpseudoscalar} , 
Eqs.~(\ref{Soneloopfinitelambda}) and (\ref{Poneloopfinitelambda}),
respectively. For $\lambda^2\to 0$ the corresponding QED expressions
are recovered, 
\begin{eqnarray}
\lefteqn{
S^{(1)}(M^2, q^2, \lambda^2\to 0) \, = \,}
\nonumber\\ & &
-\frac{1}{2}\,\bigg( 1 + \frac{1 + {{\beta}^2}}{2\,\beta}\,\ln p 
  \bigg) \,\ln \frac{\lambda^2}{M^2}   - 
  \frac{1 + {{\beta}^2}}{2\,\beta}\,
   \bigg( \mbox{Li}_2(1 - p) + \frac{1}{4}\,\ln^2 p - 3\,\zeta(2) \bigg) \,
   \nonumber\,\\ 
 & & \mbox{} - \frac{1 - {{\beta}^2}}{2\,\beta}\,\ln p - 
  \frac{1}{2} - i\,\pi \,\bigg[\,
      \frac{1 + {{\beta}^2}}{4\,\beta}\,\ln \frac{\lambda^2}{M^2} + 
     \frac{1 + {{\beta}^2}}{4\,\beta}\,
      \ln\Big(\frac{1 - {{\beta}^2}}{4\,{{\beta}^2}}\Big) + 
     \frac{1 - {{\beta}^2}}{2\,\beta} \,\bigg] 
\,,\\[3mm]
\lefteqn{
P^{(1)}(M^2, q^2, \lambda^2\to 0) \, = \,}
\nonumber\\ & &
-\frac{1}{2}\,\bigg( 1 + \frac{1 + {{\beta}^2}}{2\,\beta}\,\ln p 
  \bigg) \,\ln \frac{\lambda^2}{M^2}   - 
  \frac{1 + {{\beta}^2}}{2\,\beta}\,
   \bigg( \mbox{Li}_2(1 - p) + \frac{1}{4}\,\ln^2 p - 3\,\zeta(2) \bigg)  - 
  \frac{1}{2}\,\nonumber\,\\ 
 & & \mbox{} - 
  i\,\pi \,\bigg[\, \frac{1 + {{\beta}^2}}{4\,\beta}\,\ln \frac{\lambda^2}{M^2} + 
     \frac{1 + {{\beta}^2}}{4\,\beta}\,
      \ln\Big(\frac{1 - {{\beta}^2}}{4\,{{\beta}^2}}\Big) \,\bigg]
\,,
\end{eqnarray}
in agreement with~\cite{Braaten1,Drees1}.
The final expressions for the virtual light quark
${\cal{O}}(\alpha^2)$ corrections
expressed in terms of the moments $\tilde R_\infty^q$, $\tilde R_0^q$ and
$\tilde R_1^q$, Eqs.~(\ref{momentsfermions}), 
read ($x=q$)
\begin{eqnarray}
S_{x}^{(2)} & = &
s^{(2)}\,\bigg[\, \tilde R_\infty^x\,\ln^2 \frac{m^2}{q^2} - 
     2\,\tilde R_0^x\,\ln \frac{m^2}{q^2} + 2\,\tilde R_1^x \,\bigg]  + 
  s^{(1)}\,\bigg[\, \tilde R_\infty^x\,\ln \frac{m^2}{q^2} - 
                     \tilde R_0^x \,\bigg]  + 
  s^{(0)}\,\tilde R_\infty^x
\,,\\
P_{x}^{(2)} & = &
p^{(2)}\,\bigg[\, \tilde R_\infty^x\,\ln^2 \frac{m^2}{q^2} - 
     2\,\tilde R_0^x\,\ln \frac{m^2}{q^2} + 2\,\tilde R_1^x \,\bigg]  + 
  p^{(1)}\,\bigg[\, \tilde R_\infty^x\,\ln \frac{m^2}{q^2} - 
                     \tilde R_0^x \,\bigg]  + 
  p^{(0)}\,\tilde R_\infty^x
\,,
\end{eqnarray}
where 
\begin{eqnarray}
s^{(0)} & = &
A_{virt} + \frac{4}{3}\,( 1 - {{\beta}^2} ) \,B_{virt} - 
  \frac{1}{6}\,\ln\Big(\frac{1 - {{\beta}^2}}{16}\Big) + 
  \frac{1 - {{\beta}^2}}{6\,\beta}\,\ln p - 
  \frac{2 - \beta - 2\,{{\beta}^2}}{2\,\beta}\,\zeta(2) - \frac{1}{12}
\nonumber\\ & &
+ i\,\pi\,\bigg\{\,
\frac{1}{12\,\beta}\,\bigg[\, 2\,( 1 + {{\beta}^2} ) \,
     \ln^2\Big(\frac{\beta}{2}\Big) - 4\,( 1 - {{\beta}^2} ) \,
     \ln\Big(\frac{\beta}{2}\Big) + ( 1 + {{\beta}^2} ) \,\zeta(2) + 
    2\,( 1 - {{\beta}^2} ) \,\bigg] 
\,\bigg\}
\,,\\[2mm]
s^{(1)} & = &
\frac{2}{3}\,( 1 + {{\beta}^2} ) \,B_{virt} - 
  \frac{1}{6}\,\ln\Big(\frac{1 - {{\beta}^2}}{16}\Big) + 
  \frac{1 - {{\beta}^2}}{6\,\beta}\,\ln p - 
  \frac{1 + {{\beta}^2}}{2\,\beta}\,\zeta(2) + \frac{1}{6}
\nonumber\\ & &
+ i\,\pi\,\bigg\{\,
\frac{1}{12\,\beta}\,\bigg[\, -2\,( 1 + {{\beta}^2} ) \,
     \ln\Big(\frac{\beta}{2}\Big) + 
    2\,( 1 - {{\beta}^2} )  \,\bigg] 
\,\bigg\}
\,,\\[2mm]
s^{(2)} & = &
\frac{1 + {{\beta}^2}}{24\,\beta}\,\ln p + \frac{1}{12}
\, + \, i\,\pi\,\bigg\{\,
\frac{1 + {{\beta}^2}}{24\,\beta}
\,\bigg\}
\,,\\[5mm]
p^{(0)} & = &
A_{virt} - \frac{1}{6}\,\ln\Big(\frac{1 - {{\beta}^2}}{16}\Big) + 
  \frac{1}{2}\,\zeta(2) - \frac{1}{12}
\nonumber\\ & &
+ i\,\pi\,\bigg\{\,
\frac{1}{12\,\beta}\,\bigg[\, 2\,( 1 + {{\beta}^2} ) \,
     \ln^2\Big(\frac{\beta}{2}\Big) + ( 1 + {{\beta}^2} ) \,\zeta(2) 
 \,\bigg]
\,\bigg\}
\,,\\[2mm]
p^{(1)} & = &\
\frac{2}{3}\,( 1 + {{\beta}^2} ) \,B_{virt} - 
  \frac{1}{6}\,\ln\Big(\frac{1 - {{\beta}^2}}{16}\Big) - 
  \frac{1 + {{\beta}^2}}{2\,\beta}\,\zeta(2) + \frac{1}{6}
\nonumber\\ & &
+ i\,\pi\,\bigg\{\,
-\frac{2\,\left( 1 + {{\beta}^2} \right) }{12\,\beta}\,
    \ln\Big(\frac{\beta}{2}\Big) 
\,\bigg\}
\,,\\[2mm]
p^{(2)} & = &
\frac{1 + {{\beta}^2}}{24\,\beta}\,\ln p + \frac{1}{12}
+ i\,\pi\,\bigg\{\,
\frac{1 + {{\beta}^2}}{24\,\beta}
\,\bigg\}
\,.
\end{eqnarray}
The functions $A_{virt}$ and $B_{virt}$ are defined in 
Eqs.~(\ref{defAvirt}) and (\ref{defBvirt}), respectively.
\par
The integral representation of the ${\cal{O}}(\alpha^2)$ corrections
to $\tilde R^{S/P}$ from the real radiation of a light quark pair via
photon emission reads
\begin{equation}
\tilde r_{S/P,q}^{(2),real}(M^2, m^2, q^2) \, = \,
\frac{1}{3} \int\limits_{4\,m^2}^{(\sqrt{q^2}-2\,M)^2}
   \frac{{\rm d}\lambda^2}{\lambda^2}\,\, 
  \tilde r_{S/P}^{(1),real} (M^2, q^2, \lambda^2)\,\,
   \tilde R_{q\bar q}(m^2, \lambda^2)
\,,
\label{masterformulascalarreal}
\end{equation}
where $\tilde r_{S/P}^{(1),real}(M^2,q^2,\lambda^2)$, the first order 
corrections due to radiation of a vector boson with mass $\lambda$, 
are given in the Appendices~\ref{appendixfirstorderscalar} and
\ref{appendixfirstorderpseudoscalar} in terms of a
one-dimensional integral. For $\lambda^2\to 0$ the ${\cal{O}}(\alpha)$ 
corrections from real photon emission are obtained 
\begin{eqnarray}
\tilde r_{S/P}^{(1),real} & \stackrel{\lambda^2\to 0}{\longrightarrow} &
 6\,f_{S/P}^{(2), real}\,
    \Big[\, \ln \frac{4\,q^2}{{{\lambda }^2}} \, \Big]  - 
   3\,f_{S/P}^{(1), real}
\end{eqnarray}
in agreement with~\cite{Braaten1,Drees1}.
The final results for the real light quark ${\cal{O}}(\alpha^2)$
corrections are ($x = q$)
\begin{eqnarray}
\tilde r_{S/P, x}^{(2), real}(q^2) & = &
  f_{S/P}^{(2), real}\,\bigg[ \ln^2 \frac{{m^2}}{q^2}\,\tilde{R}_{\infty}^x - 
      2\,\ln \frac{{m^2}}{q^2}\,\tilde{R}_0^x + 2\,\tilde{R}_1^x \bigg]  + 
\nonumber\\
& &   f_{S/P}^{(1), real}\,\bigg[ \ln \frac{{m^2}}{q^2}\,\tilde{R}_{\infty}^x - 
      \tilde{R}_0^x \bigg]  + f_{S/P}^{(0), real}\,\tilde{R}_{\infty}^x
\,,
\end{eqnarray}
where
\begin{eqnarray}
f_S^{(2), real} &=& {{\beta}^2}\,C_{real}
\,,\\[2mm]
f_S^{(1), real} &=& 
  {{\beta}^2}\,B_{real} + \frac{1 + 6\,{{\beta}^2} + {{\beta}^4}}{16}\,
    \ln p  + \frac{\beta\,\left( 3 - 29\,{{\beta}^2} \right) }{24}
\,,\\[2mm]
f_S^{(0), real} &=& 
  {{\beta}^2}\,A_{real} + \frac{-9 + 26\,{{\beta}^2} + 23\,{{\beta}^4}}{48}\,
    \left( \mbox{Li}_2(p) + \mbox{Li}_2({p^2}) \right)
   \nonumber\,\\ 
 & & \mbox{}\,
    +\frac{39 - 38\,{{\beta}^2} - 25\,{{\beta}^4}}{48}\,\mbox{Li}_2(p) + 
   \frac{-21 - 14\,{{\beta}^2} + 80\,{{\beta}^3} - 21\,{{\beta}^4}}{48}\,
    \zeta(2) \nonumber\,\\ 
 & & \mbox{}\,
    +\frac{-3 - 72\,\beta + 46\,{{\beta}^2} + 136\,{{\beta}^3} + 
      45\,{{\beta}^4}}{96}\,\ln^2 p - 
   \frac{2\,{{\beta}^3}}{3}\,
 \ln^2\Big(\frac{2\,{{\beta}^2}}{1 - {{\beta}^2}}\Big) 
   \nonumber\,\\ 
 & & \mbox{}\,
    +\bigg( \,\frac{1 + 6\,{{\beta}^2} + {{\beta}^4}}{8}\,\ln 2  - 
      \frac{2\,{{\beta}^2}\,\left( 1 - {{\beta}^2} \right) }{3}\,\ln \beta 
      \nonumber\,\\ 
 & & \mbox{}\,\quad\,
       +\frac{9 - 10\,{{\beta}^2} - 39\,{{\beta}^4}}{48}\,
       \ln\Big(\frac{1 - {{\beta}^2}}{4}\Big)\,\bigg) \,\ln p 
   \nonumber\,\\ 
 & & \mbox{}\,
    +\frac{-19 + 98\,{{\beta}^2} + 29\,{{\beta}^4}}{48}\,\ln p 
  + \frac{\beta\,\left( 3 - 29\,{{\beta}^2} \right) }{12}\,
    \ln\Big(\frac{1 - {{\beta}^2}}{2\,{{\beta}^2}}\Big) 
   \nonumber\,\\ 
 & & \mbox{}\,
    +\frac{-7\,\beta\,\left( 1 + 3\,{{\beta}^2} \right) }{24}
\,,\\[5mm]
f_P^{(2), real} &=& C_{real}
\,,\\[2mm]
f_P^{(1), real} &=& 
  B_{real} + \frac{19 + 2\,{{\beta}^2} + 3\,{{\beta}^4}}{48}\,\ln p  - 
   \frac{\beta\,\left( 29 - 3\,{{\beta}^2} \right) }{24}
\,,\\[2mm]
f_P^{(0), real} &=& 
  A_{real} + \frac{19 + 18\,{{\beta}^2} + 3\,{{\beta}^4}}{48}\,
    \left( \mbox{Li}_2(p) + \mbox{Li}_2({p^2}) \right) 
   \nonumber\,\\ 
 & & \mbox{}\,
    +\frac{19 - 46\,{{\beta}^2} + 3\,{{\beta}^4}}{48}\,\mbox{Li}_2(p) + 
   \frac{-57 + 80\,\beta + 10\,{{\beta}^2} - 9\,{{\beta}^4}}{48}\,\zeta(2)
   \nonumber\,\\ 
 & & \mbox{}\,
    +\frac{-24 + 57\,\beta + 88\,{{\beta}^2} + 22\,{{\beta}^3} + 9\,{{\beta}^5}
     }{96\,\beta}\,\ln^2 p - \frac{2\,\beta}{3}\,
    \ln^2\Big(\frac{2\,{{\beta}^2}}{1 - {{\beta}^2}}\Big)
   \nonumber\,\\ 
 & & \mbox{}\,
    +\bigg( \frac{19 + 2\,{{\beta}^2} + 3\,{{\beta}^4}}{24}\,\ln 2  - 
      \frac{19 + 18\,{{\beta}^2} + 3\,{{\beta}^4}}{48}\,
       \ln\Big(\frac{1 - {{\beta}^2}}{4}\Big) \bigg) \,\ln p 
   \nonumber\,\\ 
 & & \mbox{}\,
    +\frac{89 + 18\,{{\beta}^2} + {{\beta}^4}}{48}\,\ln p  + 
   \frac{\beta\,\left( -29 + 3\,{{\beta}^2} \right) }{12}\,
    \ln\Big(\frac{1 - {{\beta}^2}}{2\,{{\beta}^2}}\Big) 
   \nonumber\,\\ 
 & & \mbox{}\,
    +\frac{7\,\beta\,\left( -5 + {{\beta}^2} \right) }{24}
\,.
\end{eqnarray}
The functions $A_{real}$, $B_{real}$ and $C_{real}$ are defined in 
Eqs.~(\ref{defAreal}), (\ref{defBreal}) and (\ref{defCreal}), respectively.
\par
To ${\cal{O}}(\alpha)$ the sums of virtual and real corrections read
\begin{eqnarray}
\tilde r_S^{(1)} & = & \tilde r_S^{(1),virt} + \tilde r_S^{(1),real}
\nonumber\\ & = &
{{\beta}^2}\,B - \frac{
     3 + 34\,{{\beta}^2} - 48\,{{\beta}^3} - 13\,{{\beta}^4}}{16}\,\ln p + 
   \frac{3\,\beta\,\left( -1 + 7\,{{\beta}^2} \right) }{8}
\,,\\[2mm]
\tilde r_P^{(1)} & = & \tilde r_P^{(1),virt} + \tilde r_P^{(1),real}
\nonumber\\ & = &
B - \frac{19 - 48\,\beta + 2\,{{\beta}^2} + 3\,{{\beta}^4}}{16}\,
    \ln p + \frac{3\,\beta\,\left( 7 - {{\beta}^2} \right) }{8}
\,,
\end{eqnarray}
whereas the light quark ${\cal{O}}(\alpha^2)$ contributions to
$r_{S/P}^{(2)}$ are ($x = q$)
\begin{eqnarray}
\tilde r_{S/P,x}^{(2)} & = &  
\tilde r_{S/P,x}^{(2),virt} + \tilde r_{S/P,x}^{(2),real}
\nonumber\\[2mm] 
 & = & \mbox{} 
-\frac{1}{3}\Big[\,\tilde R_\infty^x\,\ln\frac{m^2}{q^2} - 
  \tilde R_0^x \,\Big]\, \tilde r_{S/P}^{(1)} +
\tilde R_\infty^x\,\delta_{S/P}^{(2)}  
\label{realvirtscalar}
\end{eqnarray}
with
\begin{eqnarray}
\delta_S^{(2)} & = & 
{{\beta}^2}\,A + \frac{1}{48}\,
    \bigg\{\,\left( -9 + 26\,{{\beta}^2} + 23\,{{\beta}^4} \right) \,
       \left( \mbox{Li}_2(p) + \mbox{Li}_2({p^2}) \right) \,\nonumber\,
       \\ 
 & & \mbox{}\,\quad + 
      \left( 39 - 70\,{{\beta}^2} + 7\,{{\beta}^4} \right) \,\mbox{Li}_2(p) + 
      \left( -21 - 78\,{{\beta}^2} + 128\,{{\beta}^3} + 43\,{{\beta}^4}
          \right) \,\zeta(2)\,\nonumber\,\\ 
 & & \mbox{}\,\quad + 
      \frac{1}{2}\,\left(-3 - 72\,\beta + 30\,{{\beta}^2} + 136\,{{\beta}^3} + 
         61\,{{\beta}^4}\right)\,\ln^2 p\,\nonumber\,\\ 
 & & \mbox{}\,\quad + 
      \bigg[\,2\,\left( 3 + 34\,{{\beta}^2} - 13\,{{\beta}^4} \right)
           \,\ln 2 + 2\,\left( 9 - 42\,{{\beta}^2} - 7\,{{\beta}^4} \right)\,
          \ln\beta\,\nonumber\,\\ 
 & & \mbox{}\,\qquad + 
         \left( 9 - 10\,{{\beta}^2} - 39\,{{\beta}^4} \right) \,
          \ln\Big(\frac{1 - {{\beta}^2}}{4\,{{\beta}^2}}\Big)\,\bigg] \,
       \ln p\,\nonumber\,\\ 
 & & \mbox{}\,\quad + 
      8\,{{\beta}^3}\,\bigg( 2\,\ln\Big(\frac{{{\beta}^2}}{2}\Big) - 
         \ln(1 - {{\beta}^2}) \bigg) \,
       \bigg( 3\,\ln\Big(\frac{1 - {{\beta}^2}}{4}\Big) - 4\,\ln\beta \bigg) \,
       \nonumber\,\\ 
 & & \mbox{}\,\quad + 
      \left( -19 + 114\,{{\beta}^2} + 13\,{{\beta}^4} \right) \,\ln p + 
      12\,\beta\,\left( 1 - 11\,{{\beta}^2} \right) \,
       \ln\Big(\frac{1 - {{\beta}^2}}{2\,{{\beta}^2}}\Big)\,\nonumber\,
       \\ 
 & & \mbox{}\,\quad + 
      16\,{{\beta}^3}\,\Big( 3\,\ln 2 - 2\,\ln\beta \Big)  - 
      2\,\beta\,\left( 7 + 25\,{{\beta}^2} \right) \,\bigg\} 
\,,\\[2mm]
\delta_P^{(2)} & = & 
 A + \frac{1}{48}\,\bigg\{\,
       \left( 19 + 18\,{{\beta}^2} + 3\,{{\beta}^4} \right) \,
       \left( \mbox{Li}_2(p) + \mbox{Li}_2({p^2}) \right) \,\nonumber\,
       \\ 
 & & \mbox{}\,\quad + 
      \left( 19 - 46\,{{\beta}^2} + 3\,{{\beta}^4} \right) \,\mbox{Li}_2(p) + 
      \left( -57 + 128\,\beta + 10\,{{\beta}^2} - 9\,{{\beta}^4} \right) \,
       \zeta(2)\,\nonumber\,\\ 
 & & \mbox{}\,\quad + 
      \frac{1}{2\,\beta}\,\left(-24 + 57\,\beta + 88\,{{\beta}^2} + 22\,
{{\beta}^3} + 
         9\,{{\beta}^5}\right)\,\ln^2 p\,\nonumber\,\\ 
 & & \mbox{}\,
       \quad + \bigg[ 2\,\left( 19 + 2\,{{\beta}^2} + 3\,{{\beta}^4} \right)\,
          \ln 2 - \left( 19 + 18\,{{\beta}^2} + 3\,{{\beta}^4} \right) \,
          \ln\Big(\frac{1 - {{\beta}^2}}{4}\Big) \bigg] \,\ln p\,\nonumber\,
       \\ 
 & & \mbox{}\,\quad + 
      8\,\beta\,\bigg( 2\,\ln\Big(\frac{{{\beta}^2}}{2}\Big) - 
         \ln(1 - {{\beta}^2}) \bigg) \,
       \bigg( 3\,\ln\Big(\frac{1 - {{\beta}^2}}{4}\Big) - 4\,\ln\beta \bigg) \,
       \nonumber\,\\ 
 & & \mbox{}\,\quad + 
      \left( 89 + 18\,{{\beta}^2} + {{\beta}^4} \right) \,\ln p + 
      16\,\beta\,\Big( 3\,\ln 2 - 2\,\ln\beta \Big) \,\nonumber\,
       \\ 
 & & \mbox{}\,\quad + 
      12\,\beta\,\left( -11 + {{\beta}^2} \right) \,
       \ln\Big(\frac{1 - {{\beta}^2}}{2\,{{\beta}^2}}\Big) + 
      2\,\beta\,\left( -39 + 7\,{{\beta}^2} \right) \,\bigg\} 
\,.
\end{eqnarray}
The functions $A$ and $B$ are defined in Eqs.~(\ref{Adefinition}) and
(\ref{Bdefinition}), respectively.
\par
\vspace{1cm}
\section{Transition to ${\overline{\mbox{MS}}}$ and QCD}
\label{sectiontransitiontoqcd}
In Sections~\ref{sectionvectorcurrent} to \ref{sectionscalarcurrent} we
have presented results for the light quark second order
corrections to the absorptive part of the current-current
correlators~(\ref{currentcorrelatorvector}) and 
(\ref{currentcorrelatorscalar}) in the framework of on-shell
renormalized QED. We have seen that in the sum of the virtual and real
contributions the quadratic logarithms of the ratio $m^2/q^2$ cancel as
a consequence of the cancellation of the soft photon singularities in
the sum of virtual and real ${\cal{O}}(\alpha)$ corrections. However, a single 
logarithm remains which renders the perturbative expansion unreliable if
the ratio $m^2/q^2$ is very small, see Eqs.~(\ref{realvirtvector}),
(\ref{realvirtaxialvector}) and (\ref{realvirtscalar}). This mass
singularity is a consequence of the on-shell
renormalization scheme we used so far, where the coupling $\alpha$ is 
the fine structure constant, i.e. defined at zero momentum transfer.
However, for very small ratios $m^2/q^2$ a mass independent definition of
the coupling like the ${\overline{\mbox{MS}}}$ prescription is more
appropriate. 
Because the rates $\tilde R^\Theta$, $\Theta=V,A,S,P$, are
renormalization group invariant and have vanishing anomalous dimensions,
the fine structure constant $\alpha$ just has to be expressed in terms
of the ${\overline{\mbox{MS}}}$ coupling in order to perform the
transition to the ${\overline{\mbox{MS}}}$ scheme.\footnote{
We would like to remind the reader that we keep $M$ to be the massive
quark pole mass.
}
\par
The ${\cal{O}}(\alpha^2)$ relation between the fine structure constant
and the running ${\overline{\mbox{MS}}}$ QED coupling\footnote{
It should be noted that $\alpha_{\overline{\mbox{\tiny MS}}}$
is not equivalent to the effective electric charge sometimes used in
QED calculations.
}
at the scale $\mu$
reads ($x=q$)
\begin{equation}
\alpha \, = \, \alpha_{\overline{\mbox{\tiny MS}}}(\mu^2) \,
           \left(\,1+\frac{\alpha_{\overline{\mbox{\tiny MS}}}(\mu^2)}{\pi}\,
            \frac{1}{3}\,\tilde R_\infty^x\,\ln\frac{m^2}{\mu^2}\,\right)
          + {\cal O}(\alpha_{\overline{\mbox{\tiny MS}}}^3)
\label{alphaqedtorun}
\end{equation}
and leads to 
\begin{equation}
\tilde R^{\Theta}\, = \, \tilde r_{\Theta}^{(0)} + 
\,\Big(\frac{\alpha_{\overline{\mbox{\tiny MS}}}(\mu^2)}
           {\pi}\Big)\,\tilde r_{\Theta}^{(1)} +
\,\Big(\frac{\alpha_{\overline{\mbox{\tiny MS}}}(\mu^2)}
     {\pi}\Big)^2\,\tilde r^{(2),\overline{\mbox{\tiny MS}}}_{\Theta} +
   \, ...
\end{equation}
as the perturbative expansion for the rates $\tilde R^\Theta$, 
$\Theta=V,A,S,P$, in the ${\overline{\mbox{MS}}}$ scheme, where
\begin{equation}
\tilde r^{(2),\overline{\mbox{\tiny MS}}}_{\Theta,x} \, = \,
-\frac{1}{3}\Big[\,\tilde R_\infty^{x,\overline{\mbox{\tiny MS}}}\,
   \ln\frac{\mu^2}{q^2} - 
                  \tilde R_0^{x,\overline{\mbox{\tiny MS}}} \,\Big]\,
\tilde r_\Theta^{(1)} +
\tilde R_\infty^{x,\overline{\mbox{\tiny MS}}}\,\delta_{\Theta}^{(2)}  
  \,
\,.
\label{r2msbar}
\end{equation}
It is evident that the logarithmic divergence for $m\to 0$ has been removed. 
The moments $\tilde R_\infty^{q,\overline{\mbox{\tiny MS}}}$ and
$\tilde R_0^{q,\overline{\mbox{\tiny MS}}}$ are equal to the
corresponding on-shell ones given in Eqs.~(\ref{momentsfermions}). 
In particular, the equality between the zero-moments 
$\tilde R_0^{q,\overline{\mbox{\tiny MS}}}$ and $\tilde R_0^{q}$
corresponds to the fact that there is no non-logarithmic contribution
in the relation between the fine structure constant
and the ${\overline{\mbox{MS}}}$ coupling,
Eq.~(\ref{alphaqedtorun}). The moments  
$\tilde R_\infty^{q,\overline{\mbox{\tiny MS}}}$ and $\tilde
R_\infty^{q}$,
on the other hand, have to be equal because of renormalization group
invariance. In fact, the combination
$\alpha^2\,\tilde R_\infty^{q}/(3\,\pi)$ is equal to the contribution
of a light quark to the ${\cal{O}}(\alpha^2)$ $\beta$-function of the
$\overline{\mbox{MS}}$ coupling.
However, we used different names for the on-shell and 
the $\overline{\mbox{MS}}$ moments in order to indicate that the 
$\overline{\mbox{MS}}$ moments could be directly extracted from the 
${\overline{\mbox{MS}}}$ renormalized
vacuum polarization function from a massless quark ($x=q$)
\begin{eqnarray}
\tilde \Pi_{massless}^{x,\overline{\mbox{\tiny MS}}}(q^2)  
         & = &
 -\frac{\alpha_{\overline{\mbox{\tiny MS}}}(\mu^2)}{3\,\pi}\,\Big[\,
   \tilde R_\infty^{x,\overline{\mbox{\tiny MS}}}\,
   \ln\frac{-q^2}{4\,\mu^2} + \tilde R_0^{x,\overline{\mbox{\tiny MS}}}
   \,\Big]\,,
\label{vacuumpolx}
\end{eqnarray}
without referring back to the on-shell and dispersion integration 
definition, Eq.~(\ref{momentsdef}).\footnote{For moments of vacuum
polarizations of higher order as well as for higher moments ($n > 1$)
of the one loop vacuum polarizations discussed here the situation is
more complicated and the on-shell moments are in general different
from the $\overline{\rm MS}$ ones.}
\par
To finally arrive at the corrections to $R^{\Theta}$, Eq.~(\ref{defR}), 
in the framework of QCD, where the exchanged vector bosons in the 
double bubble diagrams of Fig.~\ref{currentcorrelatordiagrams} are 
gluons, we have to multiply all previous results with the proper SU(3)
group theoretical factors, $N_c=3$, $T=1/2$ and $C_F=4/3$. We arrive
at ($x=q$)
\begin{eqnarray}
r_{\Theta}^{(0)} & = & N_c\,\tilde r_{\Theta}^{(0)}
\,,\qquad\qquad
r_{\Theta}^{(1)} \, = \, N_c\,C_F\,\tilde r_{\Theta}^{(1)}
\,,\\[2mm]
r_{{\Theta},x}^{(2)} & = & 
-\frac{1}{3}\Big[\,
     R_\infty^{x,\overline{\mbox{\tiny MS}}}\,\ln\frac{\mu^2}{q^2} - 
     R_0^{x,\overline{\mbox{\tiny MS}}} \,\Big]\,
   r_{\Theta}^{(1)} +
R_\infty^{x,\overline{\mbox{\tiny MS}}}\,N_c\,C_F\,\delta_{\Theta}^{(2)}  
\,,\qquad \Theta=V,A,S,P\,,
\label{rvectorqcd}
\end{eqnarray}
with the moments
\begin{eqnarray}
R_n^{x,\overline{\mbox{\tiny MS}}} & = & 
    T\,\tilde R_n^{x,\overline{\mbox{\tiny MS}}}\,,\qquad n=0,1,\ldots 
\,,\nonumber\\[2mm]
R_\infty^{x,\overline{\mbox{\tiny MS}}} & = & 
    T\,\tilde R_\infty^{x,\overline{\mbox{\tiny MS}}} 
\,.
\label{defmomentsqcd}
\end{eqnarray}
We would like to remind the reader that all quantities with a tilde
refer to QED, whereas those without a tilde are QCD quantities.
\par
\vspace{1cm}
\section{The Method of the Moments}
\label{sectionmoments}
Eqs.~(\ref{realvirtvector}), (\ref{realvirtaxialvector}), 
(\ref{realvirtscalar}) and (\ref{rvectorqcd}) 
provide a simple and unambiguous method to determine the non-singlet
light quark second order corrections to the rates $R^\Theta$ and
$\tilde R^\Theta$, $\Theta=V,A,S,P$: one computes the moments of the
vacuum polarization function from the light quarks, either through the
dispersion definition, Eq.~(\ref{momentsdef}), or directly by the
calculation of the vacuum polarization function in the high energy
limit or for vanishing quark mass,
Eqs.~(\ref{vacpolinddimensions}) or 
(\ref{vacuumpolx}), and then inserts the moments into
Eqs.~(\ref{realvirtvector}), (\ref{realvirtaxialvector}),
(\ref{realvirtscalar}) and (\ref{rvectorqcd}) .
It is obvious that this ``method of the moments'' can also be applied
to determine second order non-singlet corrections from other light
particle vacuum polarization insertions into the photon or gluon
line. The only condition is that the absorptive part of the inserted
vacuum polarization function approaches a constant for large momentum
transfer. This is equivalent to the occurrence of at most one
single power of the logarithm $\ln(q^2)$ in the renormalized vacuum 
polarization function in $D=4$ dimensions.
We would like to emphasize that the method of the moments and some
applications have already been presented before in~\cite{CHKST1} for the
vector current induced rate, $R^V$, using exactly the same convention
as employed here. However, in view of the new
results in this work and for the convenience of the reader we will
illustrate the method of the moments in the following section by
applying it to determine the
second order non-singlet corrections to the rates from the light
scalar and the gluonic self energy insertions by identifying the
corresponding moments. In~\cite{CHKST1} the method of the moments 
has even been applied to calculate some third and even fourth order
corrections to $R^V$. The reader interested in those higher order
applications is referred to~\cite{CHKST1}. The higher order moments
presented there can be directly transferred to the formulae presented
in this work. We would also like to mention that in the  
framework of QED the method of the moments can be used to calculate 
hadronic vacuum polarization corrections by determining  the moments
of the experimental data to the total hadronic cross-section 
$R_{had} = \sigma(e^+e^-\to hadrons)/\sigma_{point}$
via Eq.~(\ref{momentsdef}). This has been exploited
in~\cite{Kniehl2,HKT2,HJKT1} and shall not be discussed in this work. 
\vspace{.5cm}
\subsection{Light Scalar Second Order Corrections}
\label{subsectionscalar}
In the framework of on-shell QED the moments of the vacuum
polarization of a unit charged scalar-antiscalar pair can be easily
determined via the absorptive part of the vacuum polarization function 
defined in analogy to Eq.~(\ref{vacpoldispersion2}),
\begin{equation}
\tilde R_{s s^*}(m^2, q^2) \, = \, \frac{1}{4}\, 
\sqrt{1-4\,\frac{m^2}{q^2}}\,
\left( \, 1-4\,\frac{m^2}{q^2}
 \,\right)
\,.
\label{defRss}
\end{equation}
The moments needed for the ${\cal{O}}(\alpha^2)$ corrections to the
rates can then be easily determined via relation~(\ref{momentsdef}),
\begin{eqnarray}
\tilde R_\infty^s &=& \frac{1}{4}\,, \nonumber\\[2mm]
\tilde R_0^s & = & \frac{1}{4}\,\ln 4 - \frac{2}{3}
\,,\nonumber\\[2mm]
\tilde R_1^s &=&  
 \frac{1}{8}\,\ln^2 4 
    - \frac{2}{3} \ln 4 + \frac{13}{9}
- \frac{1}{4}\,\zeta(2)
\,,\nonumber\\[2mm]
\tilde R_2^s &=& 
 \frac{1}{24}\,\ln^3 4 - \frac{1}{3}\,\ln^2 4 + 
  \bigg(\, \frac{13}{9} - \frac{1}{4}\,\zeta(2) \,\bigg) \,\ln 4 - 
  \frac{80}{27} + \frac{2}{3}\,\zeta(2) 
    + \frac{1}{2}\,\zeta(3) 
\,.  
\label{momentsscalars}
\end{eqnarray}
For completeness we have also given the moment $\tilde R_1^s$  and 
$\tilde R_2^s$ although
they are not necessary for the light scalar ${\cal{O}}(\alpha^2)$
corrections to the total rates.
As far as the transition to the ${\overline{\mbox{MS}}}$-scheme and to
QCD is concerned the situation for light scalars is in complete
analogy to the one for light quarks. Therefore, all statements and  
formulae of Section~\ref{sectiontransitiontoqcd} are also valid for 
the scalar case ($x=s$). 
\vspace{.5cm}
\subsection{Gluonic Second Order Corrections}
\label{subsectiongluon}
\begin{figure}[htb]
\begin{center}
\epsfxsize=3.0cm
\leavevmode
\epsffile[170 270 420 520]{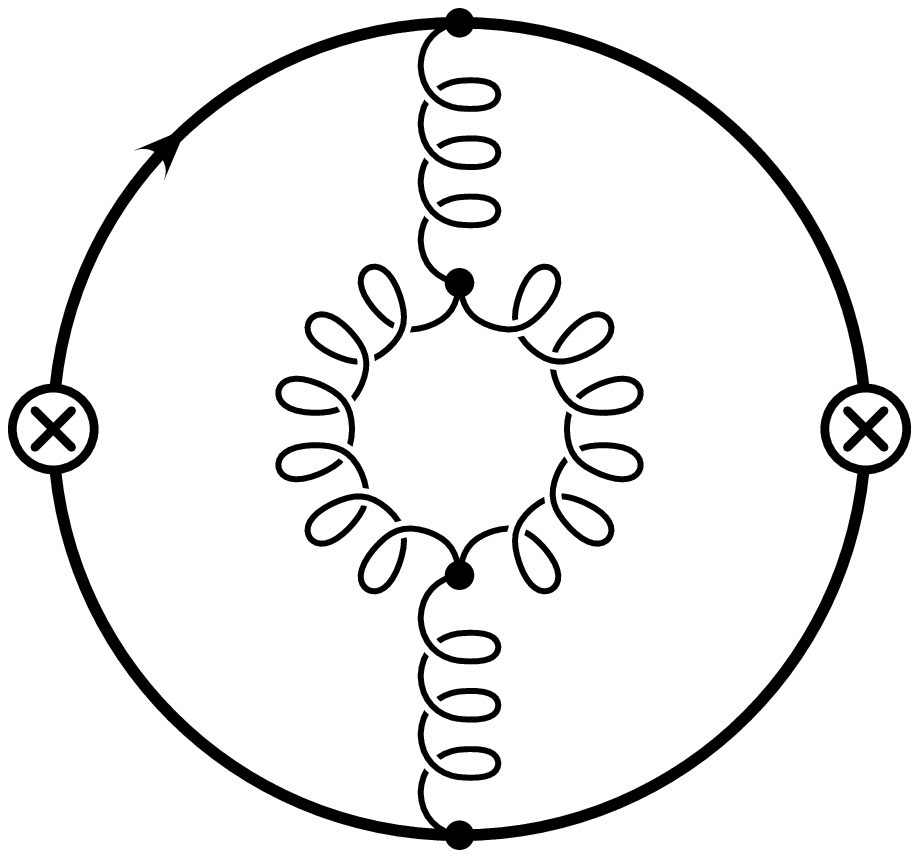}
\hspace{10ex}
\epsfxsize=3.0cm
\leavevmode
\epsffile[170 270 420 520]{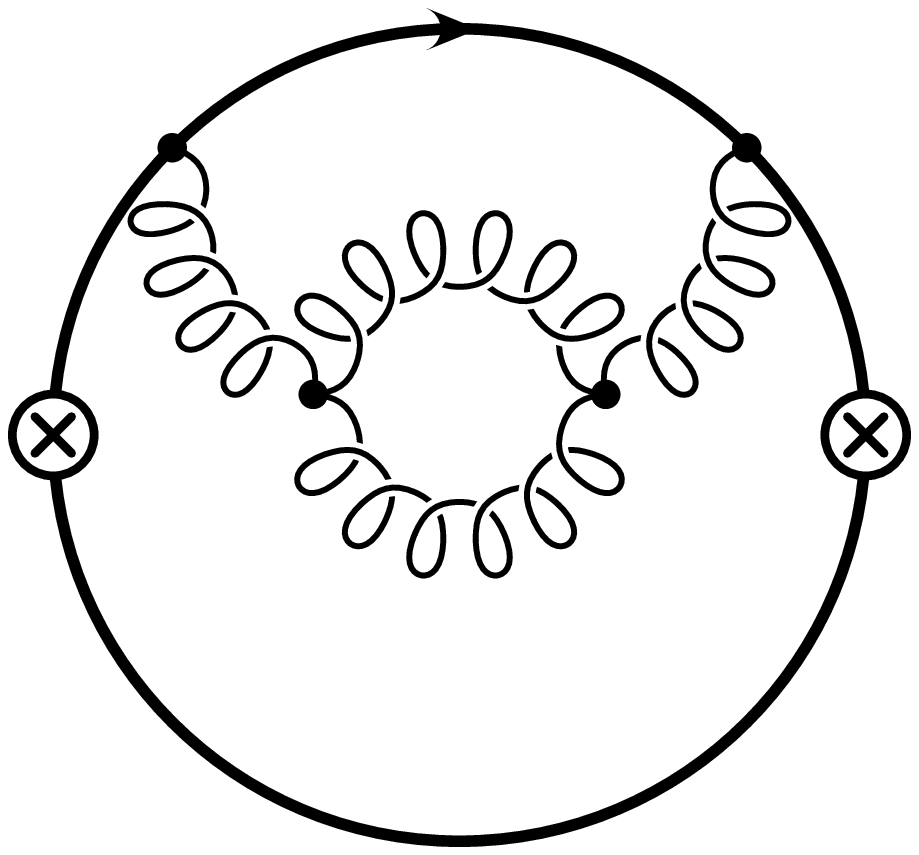}
\vskip -5mm
\caption{\label{curcorglu} Double bubble diagrams, where the massive
quark-antiquark pair is produced by the external current and the
internal bubble is purely gluonic. The ghost diagrams are not depicted
explicitly.}
\end{center}
\end{figure}

The moments of the gluonic (and ghost) contributions to the
${\cal{O}}(\alpha_s)$ gluon propagator corrections, which are needed
to calculate the ${\cal{O}}(\alpha_s^2)$ gluonic (and ghost) selfenergy
contributions to the rates, as illustrated in Fig.~\ref{curcorglu},
can be determined by identifying the corresponding moments in the
gluonic (and ghost) vacuum polarization function as defined in
Eq.~(\ref{vacuumpolx}),
\begin{eqnarray}
R_\infty^{g,\overline{\mbox{\tiny MS}}} 
& = & C_A\left(-\frac{5}{4} - \frac{3}{8}\xi\right),
\nonumber \\
R_0^{g,\overline{\mbox{\tiny MS}}} 
& = & C_A\left(\frac{31}{12}-\frac{3}{4}\xi+\frac{3}{16}\xi^2
                    +\left(-\frac{5}{4}-\frac{3}{8}\xi\right)\ln 4\right)
\,,
\end{eqnarray}
where the gauge parameter $\xi$ is defined via the gluon propagator 
in lowest order
\[
\frac{i}{q^2+i\,\epsilon}\,\left(
-\,g^{\mu\nu} + \xi\,\frac{q^\mu\,q^\nu}{q^2}
\right)
\,.
\]
Obviously the moments and $r_{\Theta,g}^{(2),\overline{\mbox{\tiny
MS}}}$, $\Theta=V,A,S,P$, are not gauge invariant. However,
for $\xi=4$ the combination 
$\alpha_s^2\,R_\infty^{g,\overline{\mbox{\tiny MS}}}/(3\pi)$
coincides with the
complete gluonic and ghost ${\cal{O}}(\alpha_s^2)$ contributions to
the QCD $\beta$-function, and thus 
$r_{\Theta,g}^{(2),\overline{\mbox{\tiny MS}}}|_{\xi=4}$ accounts for the
leading 
logarithmic behavior of the sum of all gluonic ${\cal{O}}(\alpha_s^2)$
contributions (i.e. proportional to $C_A C_F$) in the high energy
limit. It is quite trivial that such a 
$\xi$ can be found, but it is a remarkable fact that 
even in the non-relativistic limit, i.e. keeping only the leading
terms for $\beta=\sqrt{1-4M^2/q^2}\to 0$,
$r_{\Theta,g}^{(2),\overline{\mbox{\tiny MS}}}|_{\xi=4}$ 
accounts for all ${\cal{O}}(\alpha_s^2)$
corrections to the rates proportional to the color factors $C_A
C_F$. This issue has already been pointed out in~\cite{CHKST1} for
the vector current induced rate and has been used in~\cite{Pade} for
the construction of Pad\`e approximants. In
Section~\ref{sectionthreshold} we will explicitly demonstrate this
fact also for the axial-vector, scalar and pseudoscalar current
induced rates. 
\par
\vspace{1cm}
\section{Examining the Region for small $\beta$ -- close to Threshold}
\label{sectionthreshold}
In the previous sections we have determined the
${\cal{O}}(\alpha_s^2)$ light quark non-singlet corrections to the
total heavy quark-antiquark production rates induced by the vector,
axial-vector, scalar and pseudoscalar currents for all energies above
the heavy quark-antiquark threshold, $q^2>4M^2$. Using the method of
the moments we are further able to calculate the
${\cal{O}}(\alpha_s^2)$ corrections to the rates proportional to the
color factors $C_A C_F$ from the (gauge dependent) gluonic and ghost
contributions of the ${\cal{O}}(\alpha_s)$ corrections to the gluon
propagator.\footnote{
The effects of light scalars are
straightforward using the method of the moments and will not be
discussed further in this section.
}
Having these two-loop results at one's disposal it is of special
interest to consider the kinematic regime close to the threshold,
where $\beta=\sqrt{1-4M^2/q^2}\ll 1$. Using the expansions for
$\beta\to 0$ given in Appendix~\ref{appendixexpansions}, the
${\cal{O}}(\alpha_s^2)$ light quark non-singlet corrections to the
rates for all currents read 
\begin{eqnarray}
\Big(
\frac{\alpha_s^{\overline{\mbox{\tiny MS}}}(\mu^2)}{\pi}
\Big)^2\,
r^{(2),\overline{\mbox{\tiny MS}}}_{V, q} 
& \stackrel{\beta\to 0}{=} &
N_c\,C_F\,T\,
\Big(
\frac{\alpha_s^{\overline{\mbox{\tiny MS}}}(\mu^2)}{\pi}
\Big)^2\,
\bigg[\,
\frac{3\,\pi^2}{4}\,\bigg(\,
\frac{1}{3}\,\ln\frac{4\beta^2M^2}{\mu^2}-\frac{5}{9}
\,\bigg) 
\label{vectorlightthresh}\\ & & \quad
- \, 6\,\beta\,\bigg(\,
\frac{1}{3}\,\ln\frac{M^2}{\mu^2}-\frac{11}{36}
\,\bigg)
+ \frac{\pi^2}{2}\,\beta^2\,\bigg(\,
\frac{1}{3}\,\ln\frac{4\beta^2M^2}{\mu^2}-\frac{25}{18}
\,\bigg)
\,\bigg]
+ {\cal{O}}(\beta^3)
\,,
\nonumber\\[2mm]
\Big(
\frac{\alpha_s^{\overline{\mbox{\tiny MS}}}(\mu^2)}{\pi}
\Big)^2\,
r^{(2),\overline{\mbox{\tiny MS}}}_{A, q} 
& \stackrel{\beta\to 0}{=} &
N_c\,C_F\,T\,
\Big(
\frac{\alpha_s^{\overline{\mbox{\tiny MS}}}(\mu^2)}{\pi}
\Big)^2\,
\bigg[\,
\frac{\pi^2}{2}\,\beta^2\,\bigg(\,
\frac{1}{3}\,\ln\frac{4\beta^2M^2}{\mu^2}-\frac{11}{9}
\,\bigg)
\label{axiallightthresh}\\ & & \quad
- \, 2\,\beta^3\,\bigg(\,
\frac{1}{3}\,\ln\frac{M^2}{\mu^2}-\frac{7}{18}
\,\bigg)
+ \frac{\pi^2}{2}\,\beta^4\,\bigg(\,
\frac{1}{3}\,\ln\frac{4\beta^2M^2}{\mu^2}-\frac{5}{9}
\,\bigg)
\,\bigg]
+ {\cal{O}}(\beta^5)
\,,
\nonumber\\[2mm]
\Big(
\frac{\alpha_s^{\overline{\mbox{\tiny MS}}}(\mu^2)}{\pi}
\Big)^2\,
r^{(2),\overline{\mbox{\tiny MS}}}_{S, q} 
& \stackrel{\beta\to 0}{=} &
N_c\,C_F\,T\,
\Big(
\frac{\alpha_s^{\overline{\mbox{\tiny MS}}}(\mu^2)}{\pi}
\Big)^2\,
\bigg[\,
\frac{\pi^2}{2}\,\beta^2\,\bigg(\,
\frac{1}{3}\,\ln\frac{4\beta^2M^2}{\mu^2}-\frac{11}{9}
\,\bigg) 
\label{scalarlightthresh}\\ & & \quad
- \, \beta^3 \,\bigg(\,
\frac{1}{3}\,\ln\frac{M^2}{\mu^2}+\frac{5}{18}
\,\bigg)
+ \frac{\pi^2}{2}\,\beta^4\,\bigg(\,
\frac{1}{3}\,\ln\frac{4\beta^2M^2}{\mu^2}+\frac{4}{9}
\,\bigg)
\,\bigg]
+ {\cal{O}}(\beta^5)
\,,
\nonumber\\[2mm]
\Big(
\frac{\alpha_s^{\overline{\mbox{\tiny MS}}}(\mu^2)}{\pi}
\Big)^2\,
r^{(2),\overline{\mbox{\tiny MS}}}_{P, q} 
& \stackrel{\beta\to 0}{=} &
N_c\,C_F\,T\,
\Big(
\frac{\alpha_s^{\overline{\mbox{\tiny MS}}}(\mu^2)}{\pi}
\Big)^2\,
\bigg[\,
\frac{\pi^2}{2}\,\bigg(\,
\frac{1}{3}\,\ln\frac{4\beta^2M^2}{\mu^2}-\frac{5}{9}
\,\bigg) 
\label{pseudolightthresh}\\ & & \quad
- \, 3\,\beta\,\bigg(\,
\frac{1}{3}\,\ln\frac{M^2}{\mu^2}-\frac{1}{18}
\,\bigg)
+ \frac{\pi^2}{2}\,\bigg(\,
\frac{1}{3}\,\ln\frac{4\beta^2M^2}{\mu^2}-\frac{2}{9}
\,\bigg)\,\beta^2
\,\bigg]
+ {\cal{O}}(\beta^3)
\,,
\nonumber
\end{eqnarray}
where we have displayed an expansion up to next-to-next-to-leading
order in $\beta$. The corresponding ${\cal{O}}(\alpha_s^2)$
contributions from 
the gluon propagator corrections are straightforward using the
gluonic moments given in Section~\ref{subsectiongluon}. We would like
to emphasize that the
expansions~(\ref{vectorlightthresh})--(\ref{pseudolightthresh}) have
to be considered strictly in the context of fixed order multi-loop
perturbation theory which represents primarily an expansion in the
strong coupling $\alpha_s$. Therefore, the
expressions~(\ref{vectorlightthresh})--(\ref{pseudolightthresh})
actually are only good approximations in the kinematic regime where
$\alpha_s\lsim\beta\ll 1$. It is evident that this kinematic regime is
rather small, in particular for the $c\bar c$ and and the $b\bar b$
thresholds, where the strong coupling becomes quite large.
Nevertheless, it is very instructive to examine the expansions because
they allow for some insights into the long- and short-distance
structure of the ${\cal{O}}(\alpha_s^2)$ light quark contributions to
the rates. 
\par
Let us start by comparing the
expansions~(\ref{vectorlightthresh})--(\ref{pseudolightthresh}) with
the expressions of the ${\cal{O}}(\alpha_s)$ corrections in the same
limit,
\begin{eqnarray}
\Big(
\frac{\alpha_s^{\overline{\mbox{\tiny MS}}}(\mu^2)}{\pi}
\Big)\,
r^{(1)}_{V}
& \stackrel{\beta\to 0}{=} &
N_c\,C_F\,\alpha_s^{\overline{\mbox{\tiny MS}}}(\mu^2)\,
\bigg[\,
\frac{3\,\pi}{4} - \frac{6}{\pi}\,\beta + \frac{\pi}{2}\,\beta^2 +
{\cal{O}}(\beta^3)
\,\bigg]
\,,  
\label{vectoronethresh}\\[2mm]
\Big(
\frac{\alpha_s^{\overline{\mbox{\tiny MS}}}(\mu^2)}{\pi} 
\Big)\,
r^{(1)}_{A} & \stackrel{\beta\to 0}{=} &
N_c\,C_F\,\alpha_s^{\overline{\mbox{\tiny MS}}}(\mu^2)\,
\bigg[\,
\frac{\pi}{2}\,\beta^2 - \frac{2}{\pi}\,\beta^3 + \frac{\pi}{2}\,\beta^4 +
{\cal{O}}(\beta^5)
\,\bigg]
\,,
\label{axialonethresh}\\[2mm]
\Big(
\frac{\alpha_s^{\overline{\mbox{\tiny MS}}}(\mu^2)}{\pi} 
\Big)\,
r^{(1)}_{S} & \stackrel{\beta\to 0}{=} &
N_c\,C_F\,\alpha_s^{\overline{\mbox{\tiny MS}}}(\mu^2)\,
\bigg[\,
\frac{\pi}{2}\,\beta^2 - \frac{1}{\pi}\,\beta^3 + \frac{\pi}{2}\,\beta^4 +
{\cal{O}}(\beta^5)
\,\bigg]
\,,
\label{scalaronethresh}\\[2mm]
\Big(
\frac{\alpha_s^{\overline{\mbox{\tiny MS}}}(\mu^2)}{\pi} 
\Big)\,
r^{(1)}_{P} & \stackrel{\beta\to 0}{=} &
N_c\,C_F\,\alpha_s^{\overline{\mbox{\tiny MS}}}(\mu^2)\,
\bigg[\,
\frac{\pi}{2} - \frac{3}{\pi}\,\beta + \frac{\pi}{2}\,\beta^2 +
{\cal{O}}(\beta^3)
\,\bigg]
\,.
\label{pseudoonethresh}
\end{eqnarray} 
Evidently the
expansions~(\ref{vectorlightthresh})--(\ref{pseudolightthresh}) can be
used to carry out an analysis of the respective BLM scales.
This reveals that the leading and next-to-next-to-leading
contributions in the small $\beta$-expansion of the
${\cal{O}}(\alpha_s)$ contributions to the rates are governed by
scales of order $2M\beta$, the relative momentum of the produced heavy
quark-antiquark pair, indicating that they are of long-distance
origin, i.e. generated at non-relativistic momenta. All the
next-to-leading contributions for small $\beta$, on the other hand,
are governed by a hard scale of the order of the heavy quark mass.
This clearly shows that the latter are of short-distance origin,
i.e. generated at relativistic momenta.\footnote{
For the vector and scalar current induced rates these statements
can already be found in~\cite{HKT1,Melnikov}.
}
The most striking feature of the
expansions~(\ref{vectorlightthresh})--(\ref{pseudolightthresh}),
however, is that the leading terms in the small $\beta$-expansion for
the vector and pseudoscalar induced rates, on the one hand, and for
the axial-vector and scalar current induced rates, on the other, are
proportional to each other.
Further, the axial-vector and scalar current induced rates are
suppressed by $\beta^2$ with respect to the vector and pseudoscalar
induced rates. For the rest of this section we will concentrate only
on these dominant contributions for the expansion in $\beta$.
In particular, we will demonstrate explicitly by using
non-relativistic considerations 
that the latter contributions are uniquely determined by the static
color potential\footnote{
Because the heavy quark-antiquark pair is
produced in a color singlet state, we refer only to the color singlet
part of the QCD potential.
}
and the orbital angular momentum states in which the heavy
quark-antiquark pair is produced by the currents. Because the
static potential and the orbital angular momentum represent all the
information needed to completely describe the heavy quark-antiquark
pair in the non-relativistic limit, the calculations in the following
will also illustrate the fact that for small $\beta$ the
${\cal{O}}(\alpha_s^2)$ light quark corrections from secondary heavy
quark production (class(ii)) and the ${\cal{O}}(\alpha_s^2)$ light
quark singlet corrections (class(iii)) are suppressed with respect to
the non-singlet contributions calculated in this work.
Using the same line of reasoning we will also determine the leading
${\cal{O}}(C_A C_F \alpha_s^2)$ corrections in the non-relativistic
limit and demonstrate that they are equal to 
$r_{\Theta,g}^{(2),\overline{\mbox{\tiny MS}}}|_{\xi=4}$, proving
the statement given in Section~\ref{subsectiongluon}.
\par
For illustration let us first discuss the simpler case of the
${\cal{O}}(\alpha_s)$ corrections to the rates in the non-relativistic
limit, Eqs.~(\ref{vectoronethresh})--(\ref{pseudoonethresh}). 
The dominant contributions for
small $\beta$ can be reproduced by the following consideration: the
vector and pseudoscalar currents produce the heavy quark-antiquark
pair in a (relative orbital angular momentum) S-wave state, whereas the
axial-vector and scalar currents lead to a quark-antiquark pair in a
P-wave (see e.g.~\cite{Fadin1}). Via the optical theorem the vector and
pseudoscalar rates are therefore proportional to the absorptive part
of the zero-distance S-wave Green function, and the axial-vector and
scalar rates are proportional to the absorptive part of the zero-distance
P-wave Green function of the non-relativistic Schr\"odinger equation
\begin{equation}
\bigg[\,
-\frac{1}{M_Q}\vec{\nabla}^2_{\vec x} + V_{\mbox{\tiny QCD}}(\vec x) - 
E
\,\bigg]\,G_c(\vec{x},\vec{y}) \, = \,
\delta^{(3)}(\vec{x}-\vec{y})
\,,
\label{eqofmotion}
\end{equation}
where
\begin{equation}
V_{\mbox{\tiny QCD}}(\vec x) \, = \,
-\,C_F\,\frac{\alpha_s^{\overline{\mbox{\tiny MS}}}(\mu^2)}
{|{\vec{x}}|}
\end{equation}
is the leading order static QCD potential, with the strong coupling
fixed at the scale $\mu$
and $E\equiv\sqrt{q^2}-2M$ is the energy relative to the threshold
point. Because we are only interested in the regime where
$\alpha_s\lsim\beta\ll1$ 
we can determine $G_c$ perturbatively using Rayleigh-Schr\"odinger
time-independent perturbation theory (TIPT) starting from the Green
function of the free Schr\"odinger equation,
\begin{equation}
G^{(0)}(\vec{x},\vec{y}) \, = \,
\frac{M}{4\,\pi}\,\frac{\exp(i\,M\,\beta\,|{\vec{x}}-{\vec{y}}|)}
{|{\vec{x}}-{\vec{y}}|}
\,.
\end{equation}
For the ${\cal{O}}(\alpha_s)$ contributions
to the rates it is sufficient to use first order TIPT,
\begin{equation}
G_c^{(1)}(\vec{x},\vec{y}) \, = \,
-\,\int \! d^3{\vec{z}}\,\,\,G^{(0)}(\vec{x},\vec{z})\,\,
V_{\mbox{\tiny QCD}}({\vec{z}})\,\,
G^{(0)}(\vec{z},\vec{y})
\,.
\end{equation}
The S-wave contribution in $G_c^{(1)}$ at zero distances is just
equal to $G_c^{(1)}(\vec{x},\vec{y})|_{|\vec{x}|,|\vec{y}|\to 0}$
and the P-wave contribution at zero-distances reads 
$[\partial_{x^i}\,\partial_{y^i}\,
G_c^{(1)}(\vec{x},\vec{y})]|_{|\vec{x}|,|\vec{y}|\to 0}$, where
$\partial_{x^i}\equiv\partial/\partial x^i$ and summation over the
spatial index $i$ is understood. 
Taking into account the 
normalization of the rates as defined in Eqs.~(\ref{defRVA}) and
(\ref{defRSP}) one finally arrives at  
\begin{eqnarray}
\Big(
\frac{\alpha_s^{\overline{\mbox{\tiny MS}}}(\mu^2)}{\pi}
\Big)\,
r^{(1), \mbox{\tiny nr}}_{V} & = &
N_c\,\frac{6\,\pi}{M^2}\,
\mbox{Im}\,G_c^{(1)}(\vec{x},\vec{y})
\bigg|_{|\vec{x}|,|\vec{y}|\to 0}
\, = \,
N_c\,C_F\,\alpha_s^{\overline{\mbox{\tiny MS}}}(\mu^2)\,
\frac{3\,\pi}{4}
\,,
\label{vectoronethreshnr}\\[2mm]
\Big(
\frac{\alpha_s^{\overline{\mbox{\tiny MS}}}(\mu^2)}{\pi}
\Big)\,
r^{(1), \mbox{\tiny nr}}_{A} & = &
N_c\,\frac{4\,\pi}{M^4}\,
\mbox{Im}\,\bigg[\,
\frac{\partial}{\partial x^i}\frac{\partial}{\partial y^i}
G_c^{(1)}(\vec{x},\vec{y})
\,\bigg]
\bigg|_{|\vec{x}|,|\vec{y}|\to 0}
\, = \,
N_c\,C_F\,\alpha_s^{\overline{\mbox{\tiny MS}}}(\mu^2)\,
\frac{\pi}{2}\,\beta^2
\,,
\label{axialonethreshnr}\\[2mm]
\Big(
\frac{\alpha_s^{\overline{\mbox{\tiny MS}}}(\mu^2)}{\pi}
\Big)\,
r^{(1), \mbox{\tiny nr}}_{S} & = &
N_c\,\frac{4\,\pi}{M^4}\,
\mbox{Im}\,\bigg[\,
\frac{\partial}{\partial x^i}\frac{\partial}{\partial y^i}
G_c^{(1)}(\vec{x},\vec{y})
\,\bigg]
\bigg|_{|\vec{x}|,|\vec{y}|\to 0}
\, = \,
N_c\,C_F\,\alpha_s^{\overline{\mbox{\tiny MS}}}(\mu^2)\,
\frac{\pi}{2}\,\beta^2
\,,\quad
\label{scalaronethreshnr}\\[2mm]
\Big(
\frac{\alpha_s^{\overline{\mbox{\tiny MS}}}(\mu^2)}{\pi}
\Big)\,
r^{(1), \mbox{\tiny nr}}_{P} & = &
N_c\,\frac{4\,\pi}{M^2}\,
\mbox{Im}\,G_c^{(1)}(\vec{x},\vec{y})\bigg|_{|\vec{x}|,|\vec{y}|\to 0}
\, = \,
N_c\,C_F\,\alpha_s^{\overline{\mbox{\tiny MS}}}(\mu^2)\,
\frac{\pi}{2}
\,,
\label{pseudoonethreshnr}
\end{eqnarray} 
which is equivalent to the dominant contributions in
Eqs.~(\ref{vectoronethresh})--(\ref{pseudoonethresh}).
Obviously the $\beta^2$ suppression of the axial-vector and scalar
rates comes from the fact that the latter correspond to heavy
quark-antiquark pairs in a P-wave state.
\par
It is now an easy task to determine all ${\cal{O}}(\alpha_s^2)$
corrections to the rates in the non-relativistic limit coming from
light quarks by taking into account the radiative corrections to the
static QCD potential due to one massless quark
species~\cite{Billoire1} in the configuration space
representation~\cite{Landau1} 
\begin{eqnarray}
\delta V_{\mbox{\tiny QCD}, q}(\vec{x}) & = & 
-\,C_F\,T\,\frac{\alpha_s^{\overline{\mbox{\tiny MS}}}(\mu^2)}
{|{\vec{x}}|}\,
\frac{2\,\alpha_s^{\overline{\mbox{\tiny MS}}}(\mu^2)}{3\,\pi}\,
\bigg[\,
\ln\frac{1}{\mu\,|\vec x|}-\gamma-\frac{5}{6}
\,\bigg]
\nonumber\\ & = &
-\,C_F\,\frac{\alpha_s^{\overline{\mbox{\tiny MS}}}(\mu^2)}
{|{\vec{x}}|}\,
\frac{2\,\alpha_s^{\overline{\mbox{\tiny MS}}}(\mu^2)}{3\,\pi}\,
\bigg[\,
R_\infty^{q,\overline{\mbox{\tiny MS}}}
\bigg(\,
\ln\frac{1}{2\,\mu\,|\vec x|}-\gamma
\,\bigg) +
\frac{1}{2}\,R_0^{q,\overline{\mbox{\tiny MS}}} 
\,\bigg]
\,.
\end{eqnarray}
Because the corrections to the QCD potential from massless quarks
originate completely from the Serber-Uehling light quark vacuum
polarization~\cite{Billoire1}, Eq.~(\ref{vacuumpolx}), $\delta
V_{\mbox{\tiny QCD}, q}$ can be completely expressed in terms of the
moments $R_\infty^{q,\overline{\mbox{\tiny MS}}}$ and
$R_0^{q,\overline{\mbox{\tiny MS}}}$. Using the same line of reasoning
as presented for the ${\cal{O}}(\alpha_s)$ contributions to the rates,
the ${\cal{O}}(\alpha_s^2)$ light quark corrections in the
non-relativistic limit can now be determined by calculating the first
order (TIPT) correction to the Green function due to $\delta
V_{\mbox{\tiny QCD}, q}$,
\begin{equation}
G_{c, q}^{(1)}(\vec{x},\vec{y}) \, = \,
-\,\int \! d^3{\vec{z}}\,\,\,G^{(0)}(\vec{x},\vec{z})\,\,
\delta V_{\mbox{\tiny QCD}, q}({\vec{z}})\,\,
G^{(0)}(\vec{z},\vec{y})
\,.
\end{equation}
Extracting the zero-distance S- and P-wave contributions in
analogy to Eqs.~(\ref{vectoronethreshnr})--(\ref{pseudoonethreshnr})
we arrive at
\begin{eqnarray} 
\Big(
\frac{\alpha_s^{\overline{\mbox{\tiny MS}}}(\mu^2)}{\pi}
\Big)^2\,
r^{(2), \mbox{\tiny nr}}_{V, q} & = &
N_c\,\frac{6\,\pi}{M^2}\,
\mbox{Im}\,G_{c, q}^{(1)}(\vec{x},\vec{y})
\bigg|_{|\vec{x}|,|\vec{y}|\to 0}
\nonumber\\ & = & 
N_c\,C_F\,
\Big(\alpha_s^{\overline{\mbox{\tiny MS}}}(\mu^2)\Big)^2 \,
\frac{1}{4}\,\bigg[\,
R_\infty^{q,\overline{\mbox{\tiny MS}}}\,
\ln\frac{\beta^2M^2}{\mu^2}+R_0^{q,\overline{\mbox{\tiny MS}}}
\,\bigg] 
\nonumber\\ & = &
N_c\,C_F\,T\,
\Big(\alpha_s^{\overline{\mbox{\tiny MS}}}(\mu^2)\Big)^2 \,
\frac{1}{4}\,\bigg[\,
\ln\frac{4\beta^2M^2}{\mu^2}-\frac{5}{3}
\,\bigg] 
\,,
\label{vectorlightthreshnr}\\[2mm]
\Big(
\frac{\alpha_s^{\overline{\mbox{\tiny MS}}}(\mu^2)}{\pi}
\Big)^2\,
r^{(2), \mbox{\tiny nr}}_{A, q} & = &
N_c\,\frac{4\,\pi}{M^4}\,
\mbox{Im}\,\bigg[\,
\frac{\partial}{\partial x^i}\frac{\partial}{\partial y^i}
G_c^{(1)}(\vec{x},\vec{y})
\,\bigg]
\bigg|_{|\vec{x}|,|\vec{y}|\to 0}
\nonumber\\ & = &
N_c\,C_F\,
\Big(\alpha_s^{\overline{\mbox{\tiny MS}}}(\mu^2)\Big)^2 \,
\frac{1}{6}\,\bigg[\,
R_\infty^{q,\overline{\mbox{\tiny MS}}}\,
\bigg(\,
\ln\frac{\beta^2M^2}{\mu^2}-2
\,\bigg) 
+ R_0^{q,\overline{\mbox{\tiny MS}}}
\,\bigg]\,\beta^2
\nonumber\\ & = &
N_c\,C_F\,T\,
\Big(\alpha_s^{\overline{\mbox{\tiny MS}}}(\mu^2)\Big)^2 \,
\frac{1}{6}\,\bigg[\,
\ln\frac{4\beta^2M^2}{\mu^2}-\frac{11}{3}
\,\bigg]\,\beta^2
\,,
\label{axiallightthreshnr}\\[2mm]
\Big(
\frac{\alpha_s^{\overline{\mbox{\tiny MS}}}(\mu^2)}{\pi}
\Big)^2\,
r^{(2), \mbox{\tiny nr}}_{S, q} & = &
\Big(
\frac{\alpha_s^{\overline{\mbox{\tiny MS}}}(\mu^2)}{\pi}
\Big)^2\,
r^{(2), \mbox{\tiny nr}}_{A, q}
\,,
\label{scalarlightthreshnr}\\[2mm]
\Big(
\frac{\alpha_s^{\overline{\mbox{\tiny MS}}}(\mu^2)}{\pi}
\Big)^2\,
r^{(2), \mbox{\tiny nr}}_{P, q} & = &
\frac{2}{3}\,\Big(
\frac{\alpha_s^{\overline{\mbox{\tiny MS}}}(\mu^2)}{\pi}
\Big)^2\,
r^{(2), \mbox{\tiny nr}}_{V, q}
\,,
\label{pseudolightthreshnr}
\end{eqnarray}
which is in agreement with the dominant contributions in
Eqs.~(\ref{vectorlightthresh})--(\ref{pseudolightthresh}). This also
illustrates the statement that the ${\cal{O}}(\alpha_s^2)$ light quark
non-singlet corrections dominate the contributions from classes (ii)
and (iii) in the non-relativistic limit. Further, 
Eqs.~(\ref{vectorlightthreshnr})--(\ref{pseudolightthreshnr}) demonstrate
explicitly why the vector and pseudoscalar rates, and the axial-vector
and scalar rates are proportional to each other in the
non-relativistic limit.
\par 
In the same manner we can now determine all the ${\cal{O}}(C_A C_F
\alpha_s^2)$ contributions to the rate in the non-relativistic limit
by calculating the first order (TIPT) corrections to the Green
function due to the gluonic corrections to the QCD potential
proportional to $C_A C_F$~\cite{Fischler1},
\begin{eqnarray}
\delta V_{\mbox{\tiny QCD}, q}(\vec{x}) & = &
-\,C_F\,C_A\,\frac{\alpha_s^{\overline{\mbox{\tiny MS}}}(\mu^2)}
{|{\vec{x}}|}\,
\frac{2\,\alpha_s^{\overline{\mbox{\tiny MS}}}(\mu^2)}{3\,\pi}\,
\bigg[\,
-\frac{11}{4}\,\bigg(\,
\ln\frac{1}{\mu\,|\vec x|}-\gamma
\,\bigg)+\frac{31}{24}
\,\bigg]
\nonumber\\ & = &
-\,C_F\,\frac{\alpha_s^{\overline{\mbox{\tiny MS}}}(\mu^2)}
{|{\vec{x}}|}\,
\frac{2\,\alpha_s^{\overline{\mbox{\tiny MS}}}(\mu^2)}{3\,\pi}\,
\bigg[\,
R_\infty^{g,\overline{\mbox{\tiny MS}}}
\bigg(\,
\ln\frac{1}{2\,\mu\,|\vec x|}-\gamma
\,\bigg) +
\frac{1}{2}\,R_0^{g,\overline{\mbox{\tiny MS}}} 
\,\bigg]
\,\bigg|_{\xi=4}
\,.
\end{eqnarray}
It is a non-trivial fact that the gluonic corrections to the
QCD-potential can be expressed entirely in terms of the gluonic and
gauge dependent moments $R_\infty^{g,\overline{\mbox{\tiny MS}}}$ and
$R_0^{g,\overline{\mbox{\tiny MS}}}$ for the special choice of the
gauge parameter $\xi=4$ because the $C_A C_F$
corrections to the QCD potential arise from the combination of
one-loop propagator corrections, box diagrams and gluonic vertex
corrections. As already mentioned in Section~\ref{subsectiongluon} it
is quite trivial that such a value of $\xi$ can be found for the
logarithmic contribution in $\delta V_{\mbox{\tiny QCD}, q}$, but it
is a remarkable fact that also the non-logarithmic
term is accounted for completely by $R_0^{g,\overline{\mbox{\tiny
MS}}}$ for the same value of $\xi$. Proceeding along the lines of the
${\cal{O}}(\alpha_s^2)$ light quark corrections we can now easily
determine the ${\cal{O}}(C_A C_F \alpha_s^2)$ corrections to the rates
in the non-relativistic limit,
\begin{eqnarray}
\Big(
\frac{\alpha_s^{\overline{\mbox{\tiny MS}}}(\mu^2)}{\pi}
\Big)^2\,
r^{(2), \mbox{\tiny nr}}_{V, g} & = & 
N_c\,C_F\,
\Big(\alpha_s^{\overline{\mbox{\tiny MS}}}(\mu^2)\Big)^2 \,
\frac{1}{4}\,\bigg[\,
R_\infty^{g,\overline{\mbox{\tiny MS}}}\,
\ln\frac{\beta^2M^2}{\mu^2}+R_0^{g,\overline{\mbox{\tiny MS}}}
\,\bigg]\bigg|_{\xi=4}
\nonumber\\ & = &
N_c\,C_A\,C_F\,
\Big(\alpha_s^{\overline{\mbox{\tiny MS}}}(\mu^2)\Big)^2 \,
\frac{1}{4}\,\bigg[\,
-\frac{11}{4}\,\ln\frac{4\beta^2M^2}{\mu^2}+\frac{31}{12}
\,\bigg] 
\,,
\label{vectorgluonthreshnr}\\[2mm]
\Big(
\frac{\alpha_s^{\overline{\mbox{\tiny MS}}}(\mu^2)}{\pi}
\Big)^2\,
r^{(2), \mbox{\tiny nr}}_{A, g} & = & 
N_c\,C_F\,
\Big(\alpha_s^{\overline{\mbox{\tiny MS}}}(\mu^2)\Big)^2 \,
\frac{1}{6}\,\bigg[\,
R_\infty^{g,\overline{\mbox{\tiny MS}}}\,
\bigg(\,
\ln\frac{\beta^2M^2}{\mu^2}-2
\,\bigg) 
+ R_0^{g,\overline{\mbox{\tiny MS}}}
\,\bigg]\,\beta^2\,\bigg|_{\xi=4}
\nonumber \\ & = &
N_c\,C_A\,C_F\,
\Big(\alpha_s^{\overline{\mbox{\tiny MS}}}(\mu^2)\Big)^2 \,
\frac{1}{6}\,\bigg[\,
-\frac{11}{4}\,\ln\frac{4\beta^2M^2}{\mu^2}+\frac{97}{12}
\,\bigg]\,\beta^2
\,,
\label{axialgluonthreshnr}\\[2mm]
\Big(
\frac{\alpha_s^{\overline{\mbox{\tiny MS}}}(\mu^2)}{\pi}
\Big)^2\,
r^{(2), \mbox{\tiny nr}}_{S, g} & = &
\Big(
\frac{\alpha_s^{\overline{\mbox{\tiny MS}}}(\mu^2)}{\pi}
\Big)^2\,
r^{(2), \mbox{\tiny nr}}_{A, g}
\,,
\label{scalargluonthreshnr}\\[2mm]
\Big(
\frac{\alpha_s^{\overline{\mbox{\tiny MS}}}(\mu^2)}{\pi}
\Big)^2\,
r^{(2), \mbox{\tiny nr}}_{P, g} & = &
\frac{2}{3}\,\Big(
\frac{\alpha_s^{\overline{\mbox{\tiny MS}}}(\mu^2)}{\pi}
\Big)^2\,
r^{(2), \mbox{\tiny nr}}_{V, g}
\,,
\label{pseudogluonthreshnr}
\end{eqnarray}
which proves our statement given in Section~\ref{subsectiongluon}.
\par
To conclude, we would like to note that although the results in 
Eqs.~(\ref{vectorlightthresh})--(\ref{pseudolightthresh}) (and also
Eqs.~(\ref{vectorlightthreshnr})--(\ref{pseudolightthreshnr})) are
only applicable for $\alpha_s\lsim\beta\ll 1$, they represent very 
important results for predictions of the rates for
$|\beta|\lsim\alpha_s$ because they allow for an extraction of the
short distance effects coming from the light quark dynamics. These
short-distance effects are universal for $|\beta|\ll 1$, regardless
whether $\beta$ is smaller or larger than $\alpha_s$. Therefore, in an
effective field theory approach like NRQCD~\cite{Caswell1},
results~(\ref{vectorlightthresh})--(\ref{pseudolightthresh}) are 
an important input for the so called ``matching calculations'' which
allow for a systematic extraction of short-distance contributions.
Such a treatment, however, is beyond the scope of this paper and shall
be carried out elsewhere.
\par\vspace{1cm}
\section{Summary}
\label{sectionsummary}
The non-singlet ${\cal{O}}(\alpha_s^2)$ corrections from light quarks
to the rates for primary production of massive quark pairs have been
calculated analytically for the production through the vector,
axial-vector, scalar and pseudoscalar currents. The results have been
expressed in terms of moments of the vacuum polarization from the
light quarks which allows for an immediate determination of
${\cal{O}}(\alpha_s^2)$ corrections from light colored scalar
particles and from the gluonic vacuum polarization insertion. We have
analyzed the results in the kinematic regime close to the massive
quark-antiquark threshold and have shown that the dominant
contributions in the non-relativistic limit can be reproduced from
non-relativistic considerations using only the radiative corrections
to the QCD potential and the information on the orbital angular
momentum state in which the massive quark pair is produced.
\par
\vspace{.5cm}
\section*{Acknowledgement}
The authors are grateful to J.H. K\"uhn and
M. Steinhauser for useful conversation and to J.H. K\"uhn for his
support. In particular, we thank the
Graduiertenkolleg Elementarteilchenphysik for the financial support
during our time at the Institut f\"ur Theoretische Teilchenphysik at
the University of Karlsruhe, where a large part of
the work in this paper has been carried out. T.T. also thanks the
British PPARC.
A.H.H. is supported in part by the U.S. Department of Energy under
Contract No.~DOE~DE-FG03-90ER40546.

\vspace{1.0cm}
\begin{appendix}
\par\vspace{1cm}
\section{First Order Corrections from Virtual and Real 
Radiation of a Massive Vector Boson }
\label{appendixfirstorder}
In this appendix we present the ${\cal{O}}(\alpha)$ QED corrections to the heavy
quark production vertices from virtual and real radiation of a vector
boson of mass $\lambda$. Virtual and real 
radiation corrections are displayed separately. The virtual corrections
are presented in terms of the functions
\begin{equation}
 \Phi(\xi) \, \equiv \, \frac{1}{2}\sqrt{\xi^2-4\,\xi}
  \ln\left(\frac{\xi-\sqrt{\xi^2-4\,\xi}}
                {\xi+\sqrt{\xi^2-4\,\xi}}\right) 
  \sqrt{4\,\xi-\xi^2}\,\arctan\left(
  \frac{\sqrt{4\,\xi-\xi^2}}{\xi} \right)
\,,\qquad \xi > 0
\,,\label{Phi}
\end{equation}
and
\begin{eqnarray}
\lefteqn{
\Psi(p,\xi) \, \equiv \, \frac{1}{2}\ln^2\left(
  \frac{1}{2}\left[\xi-2+\sqrt{\xi^2-4\,\xi}\right]\right) 
}\nonumber\\& &+
  \mbox{Li}_2\left(1+\frac{p}{2}
    \left[-2+\xi+\sqrt{\xi^2-4\,\xi}\right]\right) +
  \mbox{Li}_2\left(1+\frac{p}{2}
    \left[-2+\xi-\sqrt{\xi^2-4\,\xi}\right]\right) \\& &+
  \left\{
    \begin{array}{ll}
     \displaystyle{
     -\frac{3}{2}\,\pi^2 
     + 4\,\pi\arctan\left(\frac{\sqrt{4\,\xi-\xi^2}}{\xi}\right)
     + 2\,\pi\arctan\left(\frac{2\,p+\xi-2}
         {\sqrt{4\,\xi-\xi^2}}\right)\,,}
      & 0<\xi<4 \\
     \displaystyle{
     -\frac{1}{2}\,\pi^2 + i\,\pi\,\ln\Big((1-p)^2+p\,\xi\Big)}\,,
      & \xi>4
    \end{array} \right.\,,\nonumber
\label{Psi}
\end{eqnarray}
whereas the real corrections are expressed in terms of an one-dimensional
integral representation. The parameter $\ell$ is defined as
\[
\ell \, \equiv \, \frac{\lambda^2}{M^2}\,,
\]
and
\[ \beta \, = \, \sqrt{1-\frac{4\,M^2}{q^2}}\,,
\qquad
p \, = \, \frac{1-\beta}{1+\beta}\,.
\]
The real parts of the vector current form factors $F_{1/2}^{(1)}$ have
already been given in~\cite{HKT1, andrediss}.
\subsection{Vector Current Vertex}
\label{appendixfirstordervector}
\begin{eqnarray}
\lefteqn{
F_{1}^{(1)}(M^2, q^2, \lambda^2) \, = \,
- \frac{1}{4\,\beta}\,\bigg[\, 1 + {{\beta}^2} + 
       \frac{1 - {{\beta}^2}}{{{\beta}^2}}\,\ell\,
        \bigg( 1 + \frac{\left( 3 - 2\,{{\beta}^2} \right) \,
             \left( 1 - {{\beta}^2} \right) }{8\,{{\beta}^2}}\,\ell \bigg) \
        \,\bigg] \,\Psi(p,\ell)\,
}
\nonumber\,\\ 
 & & \mbox{}  + 
  \bigg[\, \frac{-1 + 2\,{{\beta}^2}}{2\,{{\beta}^2}} + 
     \frac{-3 + 7\,{{\beta}^2} + 2\,{{\beta}^4}}{8\,{{\beta}^4}}\,\ell + 
     \frac{3}{\ell - 4} \,\bigg] \,\Phi(\ell)\,\nonumber\,\\ 
 & & \mbox{} + 
  \frac{1}{16\,{{\beta}^4}}\,\bigg[\, -3 + 7\,{{\beta}^2} + 2\,{{\beta}^4} + 
     \frac{{{\left( 1 - {{\beta}^2} \right) }^2}\,
        \left( -3 + 2\,{{\beta}^2} \right) }{2\,\beta}\,\ln p \,\bigg] \,
   {{\ell}^2}\,\ln \ell\,\nonumber\,\\ 
 & & \mbox{} - 
  \frac{1}{2\,{{\beta}^2}}\,\bigg[\, 1 + 
     \frac{1 - {{\beta}^2}}{2\,\beta}\,\ln p \,\bigg] \,\ell\,\ln \ell - 
  \bigg[\, \frac{1}{2} + \frac{1 + {{\beta}^2}}{4\,\beta}\,\ln p \,\bigg] \,
   \ln \ell\,\nonumber\,\\ 
 & & \mbox{} + 
  \bigg[\,  \frac{{{\left( 1 - {{\beta}^2} \right) }^2}\,
     \left( -3 + 2\,{{\beta}^2} \right) }{64\,{{\beta}^5}}\,\ln^2 p
  \,\bigg]\,\ell^2\,
   \nonumber\,\\ 
 & & \mbox{} + 
  \frac{1}{4\,{{\beta}^2}}\,\bigg[\, -\left( 1 + 2\,{{\beta}^2} \right)  + 
     \frac{1 - {{\beta}^2}}{2\,\beta}\,\ln p\,
      \left( -3 + 2\,{{\beta}^2} - \ln p \right)  \,\bigg] \,\ell\,\nonumber\,
   \\ 
 & & \mbox{} - \frac{1}{8\,\beta}\,\ln p\,
   \bigg( 2\,\left( 1 + 2\,{{\beta}^2} \right)  + 
     \left( 1 + {{\beta}^2} \right) \,\ln p \bigg)  - 
  1\,\nonumber\,\\ 
 & & \mbox{} + 
  i\,\pi \,\left\{ -\frac{1 + 2\,{{\beta}^2}}{4\,\beta} + 
     \frac{\left( 1 - {{\beta}^2} \right) \,
        \left( -3 + 2\,{{\beta}^2} \right) }{8\,{{\beta}^3}}\,\ell\,\right.\,
      \nonumber\,\\ 
 & & \mbox{}\,\qquad\,\left. + 
     \frac{1}{4\,\beta}\,\bigg[\, 1 + {{\beta}^2} + 
        \frac{1 - {{\beta}^2}}{{{\beta}^2}}\,\ell + 
        \frac{\left( 3 - 2\,{{\beta}^2} \right) \,
           {{\left( 1 - {{\beta}^2} \right) }^2}}{8\,{{\beta}^4}}\,\ell^2 
   \,\bigg] \,\ln(1 + \frac{{{\left( 1 - p \right) }^2}}{\ell\,p})
\right\}
\,,
\label{F1oneloopfinitelambda}
\\[2mm]
\lefteqn{
F_{2}^{(1)}(M^2, q^2, \lambda^2) \, = \,
\frac{{{\left( 1 - {{\beta}^2} \right) }^2}}{4\,{{\beta}^3}}\,
   \bigg[\, 1 + \frac{3\,\left( 1 - {{\beta}^2} \right) }{8\,{{\beta}^2}}\,
      \ell \,\bigg] \,\ell\,\Psi(p,\ell)
 +\frac{1 - {{\beta}^2}}{2\,{{\beta}^2}}\,
   \bigg[\, 1 + \frac{3 - 5\,{{\beta}^2}}{4\,{{\beta}^2}}\,\ell \,\bigg] \,
   \Phi(\ell)\,
}
\nonumber\,\\ 
 & & \mbox{} + 
  \frac{{{\left( 1 + \beta \right) }^2}\,\left( 1 - {{\beta}^2} \right) }{
    16\,{{\beta}^4}}\,\bigg[\, \frac{3 - 5\,{{\beta}^2}}{
      {{\left( 1 + \beta \right) }^2}} + 
     \frac{3\,{{\left( 1 - \beta \right) }^2}}{2\,\beta}\,\ln p \,\bigg] \,
   {{\ell}^2}\,\ln \ell\,\nonumber\,\\ 
 & & \mbox{} + 
  \frac{1 - {{\beta}^2}}{2\,{{\beta}^2}}\,
   \bigg[\, 1 + \frac{1 - {{\beta}^2}}{2\,\beta}\,\ln p \,\bigg] \,\ell\,
   \ln \ell + \frac{3\,{{\left( 1 - {{\beta}^2} \right) }^3}}{64\,{{\beta}^5}
    }\,\bigg[\,\ln^2 p\,\bigg]\,{{\ell}^2}\,\nonumber\,\\ 
 & & \mbox{} + 
  \frac{1 - {{\beta}^2}}{4\,{{\beta}^2}}\,
   \bigg[\, 1 + \frac{1 - {{\beta}^2}}{2\,\beta}\,\ln p\,
      \left( 3 + \ln p \right)  \,\bigg] \,\ell + 
  \frac{1 - {{\beta}^2}}{4\,\beta}\,\ln p\,\nonumber\,\\ 
 & & \mbox{} + 
  i\,\pi \,\left\{\, \frac{1 - {{\beta}^2}}{4\,\beta} + 
     \frac{3\,{{\left( 1 - {{\beta}^2} \right) }^2}}{8\,{{\beta}^3}}\,\ell\,
      \right.\,\nonumber\,\\ 
 & & \mbox{}\,\qquad\,\left. - 
     \frac{{{\left( 1 - {{\beta}^2} \right) }^2}}{4\,{{\beta}^3}}\,
      \bigg[\, 1 + \frac{3\,\left( 1 - {{\beta}^2} \right) }{8\,{{\beta}^2}}\,
         \ell \,\bigg] \,\ell\,\ln(1 + 
        \frac{{{\left( 1 - p \right) }^2}}{\ell\,p}) \,\right\}\,,
\label{F2oneloopfinitelambda}
\end{eqnarray}
\begin{eqnarray}
\lefteqn{
r_V^{(1),real}(M^2,q^2,\lambda^2) \, = \,
}
\nonumber\\ & & 
\int_{4M^2/q^2}^{(1-\sqrt{z})^2} {\rm d}y \, 
 \Bigg\{ -\sqrt{1-\frac{4M^2}{q^2 y}}\,\Lambda^{1/2}(y,z)\, 
 \Bigg[ 1 + \Bigg( \frac{16M^4}{q^4}+\frac{8M^2}{q^2} + 
        4\bigg(1+\frac{2M^2}{q^2}\bigg)z \, \Bigg) \, D \, \Bigg] 
\nonumber\\[1mm]
& & \mbox{} \qquad \qquad \qquad + 
 \bigg[ \frac{8M^4}{q^4} + \frac{4M^2}{q^2}(1-y+z) - (1-y+z)^2 - 2(1+z)y
\, \bigg] \, L \, \Bigg\} \,,
\end{eqnarray}
where 
\begin{eqnarray}
z & = & \frac{\lambda^2}{q^2} \\
L & \equiv & \frac{1}{1-y+z}\,\ln
 \frac{1-y+z-\sqrt{1-\frac{4M^2}{q^2 y}}\,\Lambda^{1/2}(y,z)}
      {1-y+z+\sqrt{1-\frac{4M^2}{q^2 y}}\,\Lambda^{1/2}(y,z)}\,,
\label{defL}\\[2mm]
D & \equiv & \bigg[ (1-y+z)^2-\bigg(1-\frac{4M^2}{q^2
     y}\bigg)\,\Lambda(y,z)\, 
       \bigg]^{-1}\,,
\label{defD}\\[2mm]
\Lambda(y,z) & \equiv & 1+y^2+z^2-2(y+z+y z)
\,.
\end{eqnarray}
\subsection{Axial-Vector Current Vertex}
\label{appendixfirstorderaxialvector}
\begin{eqnarray}
\lefteqn{
F_{3}^{(1)}(M^2, q^2, \lambda^2) \, = \,
- \frac{1 + {{\beta}^2}}{4\,\beta}\,
     \bigg[\, 1 + \frac{1 - {{\beta}^2}}{2\,{{\beta}^2}}\,\ell\,
        \bigg( 1 + \frac{1 - {{\beta}^2}}{4\,\left( 1 + {{\beta}^2} \right) 
            }\,\ell \bigg)  \,\bigg] \,\Psi(p,\ell)\,
}
\nonumber\\ 
 & & \mbox{}  - 
  \bigg[\, \frac{1 - 2\,{{\beta}^2}}{2\,{{\beta}^2}} + 
     \frac{1 - 7\,{{\beta}^2}}{8\,{{\beta}^2}}\,\ell - \frac{3}{\ell - 4} 
  \,\bigg] \,\Phi(\ell)\,\nonumber\,\\ 
 & & \mbox{} - 
  \frac{1}{16\,{{\beta}^2}}\,\bigg[\, 1 - 7\,{{\beta}^2} + 
     \frac{{{\left( 1 - {{\beta}^2} \right) }^2}}{2\,\beta}\,\ln p \,\bigg] \,
   {{\ell}^2}\,\ln \ell\,\nonumber\,\\ 
 & & \mbox{} - 
  \frac{1 + {{\beta}^2}}{4\,{{\beta}^2}}\,
   \bigg[\, 1 + \frac{1 - {{\beta}^2}}{2\,\beta}\,\ln p \,\bigg] \,\ell\,
   \ln \ell - \frac{1}{2}\,\bigg[\, 1 + 
     \frac{1 + {{\beta}^2}}{2\,\beta}\,\ln p \,\bigg]\,\ln \ell\,\nonumber\,
   \\ 
 & & \mbox{} - \frac{{{\left( 1 - {{\beta}^2} \right) }^2}}{
    64\,{{\beta}^3}}\,\bigg[\,\ln^2 p\,\bigg]\,\ell^2
 -\frac{1}{4}\,\bigg[\, 3 + \frac{1 - {{\beta}^2}}{2\,\beta}\,\ln p + 
     \frac{1 - {{\beta}^4}}{4\,{{\beta}^3}}\,\ln^2 p \,\bigg] \,\ell\,
   \nonumber\,\\ 
 & & \mbox{} 
  -\frac{1}{4\,\beta}\,\bigg[\, 2 + {{\beta}^2} + 
     \frac{1 + {{\beta}^2}}{2}\,\ln p \,\bigg] \,\ln p - 
  1\,\nonumber\,\\ 
 & & \mbox{} + 
  i\,\pi \,\left\{\, -\frac{2 + {{\beta}^2}}{4\,\beta} - 
     \frac{1 - {{\beta}^2}}{8\,\beta}\,\ell\,\right.\,\nonumber\,
      \\ 
 & & \mbox{}\,\qquad\,\left. + 
     \frac{1}{4\,\beta}\,\bigg[\, 1 + {{\beta}^2} + 
        \frac{1 - {{\beta}^4}}{2\,{{\beta}^2}}\,\ell + 
        \frac{{{\left( 1 - {{\beta}^2} \right) }^2}}{8\,{{\beta}^2}}\,
         {{\ell}^2} \,\bigg] \,\ln(1 + 
        \frac{{{\left( 1 - p \right) }^2}}{\ell\,p}) \,\right\}\,,
\label{F3oneloopfinitelambda}
\\[2mm]
\lefteqn{
F_{4}^{(1)}(M^2, q^2, \lambda^2) \, = \,
-\frac{{{\left( 1 - {{\beta}^2} \right) }^2}}{8\,{{\beta}^3}}\,
     \bigg[\, 1 + {{\beta}^2} + \frac{1 - {{\beta}^2}}{4}\,\ell \,\bigg] \,
     \ell\,\Psi(p,\ell)  
}
\nonumber\\ & &  - 
  \frac{1 - {{\beta}^2}}{8\,{{\beta}^2}}\,
   \bigg[\, 4 + \left( 1 + {{\beta}^2} \right) \,\ell \,\bigg]
\,\Phi(\ell)\,
 -  \frac{1 - {{\beta}^2}}{16\,{{\beta}^2}}\,
   \bigg[\, 1 + {{\beta}^2} + \frac{{{\left( 1 - {{\beta}^2} \right) }^2}}{
       2\,\beta}\,\ln p \,\bigg] \,{{\ell}^2}\,\ln \ell\,\nonumber\,
   \\ 
 & & \mbox{} - \frac{{{\left( 1 - {{\beta}^2} \right) }^2}}{
    4\,{{\beta}^2}}\,\bigg[\, 1 + \frac{1 + {{\beta}^2}}{2\,\beta}\,\ln p 
 \,\bigg] \,\ell\,\ln \ell - \frac{{{\left( 1 - {{\beta}^2} \right) }^3}}{
    64\,{{\beta}^3}}\,\bigg[\,\ln^2 p\,\bigg]\,{{\ell}^2}\,\nonumber\,\\ 
 & & \mbox{} + 
  \frac{1 - {{\beta}^2}}{4}\,\bigg[\, 1 - 
     \frac{1 - {{\beta}^2}}{2\,\beta}\,\ln p - 
     \frac{1 - {{\beta}^4}}{4\,{{\beta}^3}}\,\ln^2 p \,\bigg] \,\ell - 
  \frac{1 - {{\beta}^2}}{2}\,\bigg[\, 1 + 
     \frac{2 + {{\beta}^2}}{2\,\beta}\,\ln p \,\bigg] \,\nonumber\,
   \\ 
 & & \mbox{} + i\,\pi \,
   \left\{\, -\frac{\left( 1 - {{\beta}^2} \right) \,
        \left( 2 + {{\beta}^2} \right) }{4\,\beta} - 
     \frac{{{\left( 1 - {{\beta}^2} \right) }^2}}{8\,\beta}\,\ell\,\right.\,
      \nonumber\,\\ 
 & & \mbox{}\,\qquad\,\left. + 
     \frac{{{\left( 1 - {{\beta}^2} \right) }^2}}{8\,{{\beta}^3}}\,
      \bigg[\, 1 + {{\beta}^2} + \frac{1 - {{\beta}^2}}{4}\,\ell \,\bigg] \,
      \ell\,\ln(1 + \frac{{{\left( 1 - p \right) }^2}}{\ell\,p})
\,\right\}
\,,  
\label{F4oneloopfinitelambda}
\end{eqnarray}
\begin{eqnarray}
\lefteqn{
r_A^{(1),real}(M^2, q^2, \lambda^2) \, = \,
}
\nonumber\\ & &
\int_{4M^2/q^2}^{(1-\sqrt{z})^2} {\rm d}y \, 
 \Bigg\{ -\sqrt{1-\frac{4M^2}{q^2 y}}\,\Lambda^{1/2}(y,z)\, 
 \Bigg[ 1 - \Bigg( \frac{32M^4}{q^4}-\frac{8M^2}{q^2} - 
        4\bigg(1-\frac{4M^2}{q^2}\bigg)z \, \Bigg) \, D \, \Bigg] 
\nonumber\\[1mm]
& & \mbox{} \quad + 
 \bigg[ -\frac{16M^4}{q^4} + \frac{4M^2}{q^2}(1+2y) - 
 \bigg(1+\frac{2M^2}{q^2}\bigg)(1-y+z)^2 - 2(1+z)y
\, \bigg] \, L \, \Bigg\} \,,
\end{eqnarray}
where the functions $L$ and $D$ are given in Eqs.~(\ref{defL}) and  
(\ref{defD}). 
\subsection{Scalar Current Vertex}
\label{appendixfirstorderscalar}
\begin{eqnarray}
\lefteqn{
S^{(1)}(M^2, q^2, \lambda^2) \, = \,
-\frac{1}{4\,\beta}\,\bigg[\, 1 + {{\beta}^2} + 
       \frac{{{\left( 1 - {{\beta}^2} \right) }^2}}{2\,{{\beta}^2}}\,\ell 
  \,\bigg] \,\Psi(p,\ell)   - 
  \bigg[\, \frac{1 - 3\,{{\beta}^2}}{2\,{{\beta}^2}} - \ell - 
     \frac{3}{\ell - 4} \,\bigg] \,\Phi(\ell)\,
}
\nonumber\,\\ 
 & & \mbox{} + 
  \frac{1}{2}\,{{\ell}^2}\,\ln \ell - 
  \bigg[\, \frac{1}{2} + \frac{1 + {{\beta}^2}}{4\,{{\beta}^2}}\,\ell + 
     \frac{1}{4\,\beta}\,\bigg( 1 + {{\beta}^2} + 
        \frac{{{\left( 1 - {{\beta}^2} \right) }^2}}{2\,{{\beta}^2}}\,\ell 
  \bigg) \,\ln p \,\bigg] \,\ln \ell\,\nonumber\,\\ 
 & & \mbox{} - 
  \ell - \frac{1}{8\,\beta}\,\bigg[\, 1 + {{\beta}^2} + 
     \frac{{{\left( 1 - {{\beta}^2} \right) }^2}}{2\,{{\beta}^2}}\,\ell 
 \,\bigg] \,\ln^2 p - \frac{1 - {{\beta}^2}}{2\,\beta}\,\ln p - 
  \frac{1}{2}\,\nonumber\,\\ 
 & & \mbox{} + 
  i\,\pi \,\left\{\, -\frac{1 - {{\beta}^2}}{2\,\beta} + 
     \frac{1}{4\,\beta}\,\bigg[\, 1 + {{\beta}^2} + 
        \frac{{{\left( 1 - {{\beta}^2} \right) }^2}}{2\,{{\beta}^2}}\,\ell 
  \,\bigg] \,\ln(1 + \frac{{{\left( 1 - p \right) }^2}}{\ell\,p}) 
  \,\right\} 
\,,
\label{Soneloopfinitelambda}
\end{eqnarray}
\begin{eqnarray}
\lefteqn{
r_S^{(1),real}(M^2, q^2,\lambda^2) \, = \,
}
\nonumber\\ & &
\int_{4M^2/q^2}^{(1-\sqrt{z})^2} {\rm d}y \, 
 \Bigg\{ \sqrt{1-\frac{4M^2}{q^2 y}}\,\Lambda^{1/2}(y,z)\, 
 \bigg[ \frac{32M^4}{q^4}-\frac{8M^2}{q^2} - 
        4\bigg(1-\frac{4M^2}{q^2}\bigg)z \, \bigg] \, D 
\nonumber\\[1mm]
& & \mbox{} \qquad \qquad \qquad + 
 \bigg[ -\frac{16M^4}{q^4} + \frac{4M^2}{q^2}(1+2y-4z) - 1 - (z-y)^2
\, \bigg] \, L \, \Bigg\}
\,. 
\end{eqnarray}
\subsection{Pseudoscalar Current Vertex}
\label{appendixfirstorderpseudoscalar}
\begin{eqnarray}
\lefteqn{
P^{(1)}(M^2, q^2, \lambda^2) \, = \,
-\frac{1 + {{\beta}^2}}{4\,\beta}\,\Psi(p,\ell)   + 
  \bigg[\, 1 + \ell + \frac{3}{\ell - 4} \,\bigg]\,\Phi(\ell)\,
}
\nonumber\\ 
 & & \mbox{} - \frac{1}{2}\,\ell\,\left( 1 - \ell \right) \,
   \ln \ell - \frac{1}{2}\,\bigg[\, 1 + 
     \frac{1 + {{\beta}^2}}{2\,\beta}\,\ln p \,\bigg] \,\ln \ell - \ell - 
  \frac{1 + {{\beta}^2}}{8\,\beta}\,\ln^2 p - 
  \frac{1}{2}\,\nonumber\,\\ 
 & & \mbox{} + 
  i\,\pi \,\left\{\,\frac{1 + {{\beta}^2}}{4\,\beta}\,
   \ln(1 + \frac{{{\left( 1 - p \right) }^2}}{\ell\,p})\,\right\}
\,,
\label{Poneloopfinitelambda}
\end{eqnarray}
\begin{eqnarray}
\lefteqn{
r_P^{(1),real}(M^2, q^2, \lambda^2) \, = \,
}
\nonumber\\ & &
\int_{4M^2/q^2}^{(1-\sqrt{z})^2} {\rm d}y \, 
 \Bigg\{ -4\sqrt{1-\frac{4M^2}{q^2 y}}\,\Lambda^{1/2}(y,z)\, 
 \bigg( z+\frac{2M^2}{q^2}\bigg) \, D 
\nonumber\\[1mm]
& & \mbox{} \qquad \qquad \qquad + 
 \bigg[ \frac{4M^2}{q^2} - 1 - (z-y)^2
\, \bigg] \, L \, \Bigg\}
\,. 
\end{eqnarray}
\par\vspace{1cm}
\section{Second Order Corrections from Virtual Massive
Quarks }
\label{appendixequalmass}
The main focus of this paper is the presentation and discussion of the
effects of light quark vacuum polarization insertions into the gluon
exchange diagrams for massive quark-antiquark production. This is not
an easy task because for the mass assignment $m\ll M$ infrared
singularities 
in the form of logarithms of the ratio $m^2/q^2$ have to be handled.
For the mass constellation $m=M$, however, it has been observed 
in~\cite{HKT1} that the integrations 
in Eq.~(\ref{masterformulavectorvirtual}) become trivial because the
square root  
connected with the born cross-section $\tilde R_{q\bar q}$ can be 
substituted away. The results can then be
expressed entirely in terms of logarithms.
In this appendix we present the results
for the axial-vector and pseudoscalar second order form
factors for the massive quark pair production vertices coming from the
the massive quark-antiquark vacuum polarization insertion into the
gluon line, i.e. for $m=M$. For
completeness and convenience of the reader we also review the older
results for the real parts of the vector and scalar form factors
of~\cite{HKT1} and \cite{Melnikov}. All the imaginary parts of the
form factors are new and have to our knowledge never been presented in
the literature before. For simplicity all formulae in this appendix are
presented in the framework of on-shell QED. 
\subsection{Vector Current Vertex}
\label{appendixsecondordervectormassive}
\begin{eqnarray}
F_{1,Q}^{(2)} & = &
\frac{1}{12\,{{\beta}^2}}\,\bigg[\, - \frac{
         \left( 1 - 2\,{{\beta}^2} \right) \,
          \left( 9 - 6\,{{\beta}^2} + 5\,{{\beta}^4} \right) }{
         24\,{{\beta}^3}}\,\ln^3 p   + 
     \frac{31 - 23\,{{\beta}^2} + 30\,{{\beta}^4}}{12\,{{\beta}^2}}\,
      \ln^2 p\,\nonumber\,\\ 
 & & \mbox{}\,\qquad + 
     \bigg( \frac{267 - 238\,{{\beta}^2} + 236\,{{\beta}^4}}{18\,\beta} + 
        \frac{\left( 1 - 2\,{{\beta}^2} \right) \,
           \left( 9 - 6\,{{\beta}^2} + 5\,{{\beta}^4} \right) }{
          2\,{{\beta}^3}}\,\zeta(2) \bigg) \,\ln p\,\nonumber\,
      \\ 
 & & \mbox{}\, + 
     \frac{9 - 21\,{{\beta}^2} - 10\,{{\beta}^4}}{{{\beta}^2}}\,\zeta(2) + 
     \frac{147 + 236\,{{\beta}^2}}{9} \,\bigg] \,\nonumber\,
      \\ 
 & & \mbox{}\, + 
  i\,\frac{\pi }{16\,{{\beta}^3}}\,
   \bigg\{\, -\frac{\left( 1 - 2\,{{\beta}^2} \right) \,
          \left( 9 - 6\,{{\beta}^2} + 5\,{{\beta}^4} \right) }{6\,{{\beta}^2}
         }\,\ln^2 p   + 
     \frac{2\,\left( 31 - 23\,{{\beta}^2} + 30\,{{\beta}^4} \right) }{
       9\,\beta}\,\ln p\,\nonumber\,\\ 
 & & \mbox{}\, \qquad + 
     \frac{2\,\left( 267 - 238\,{{\beta}^2} + 236\,{{\beta}^4} \right) }{27}\
      \,\bigg\} \,,
\\[2mm]
F_{2,Q}^{(2)} & = &
\frac{1 - {{\beta}^2}}{4\,{{\beta}^2}}\,
   \bigg[\, \frac{{{\left( 1 - {{\beta}^2} \right) }^2}}{8\,{{\beta}^3}}\,
      \ln^3 p - \frac{11 - 9\,{{\beta}^2}}{12\,{{\beta}^2}}\,\ln^2 p\,
      \nonumber\,\\ 
 & & \mbox{}\, - 
     \bigg( \frac{93 - 68\,{{\beta}^2}}{18\,\beta} + 
        \frac{3\,{{\left( 1 - {{\beta}^2} \right) }^2}}{2\,{{\beta}^3}}\,
         \zeta(2) \bigg) \,\ln p - 
     \frac{3 - 5\,{{\beta}^2}}{{{\beta}^2}}\,\zeta(2) - \frac{17}{3} \,\bigg] 
    \,\nonumber\,\\ 
 & & \mbox{} + 
  i\,\pi \,\frac{1 - {{\beta}^2}}{{{\beta}^3}}\,
   \bigg\{\, 
      \frac{3\,{{\left( 1 - {{\beta}^2} \right) }^2}}{32\,{{\beta}^2}}\,
      \ln^2 p - \frac{11 - 9\,{{\beta}^2}}{24\,\beta}\,\ln p - 
     \frac{93 - 68\,{{\beta}^2}}{72} \,\bigg\} \,.
\end{eqnarray}
\subsection{Axial-Vector Current Vertex}
\label{appendixsecondorderaxialvectormassive}
\begin{eqnarray}
F_{3,Q}^{(2)} & = &
\frac{1}{12\,{{\beta}^2}}\,\bigg[\, - \frac{
         3 - 10\,{{\beta}^2} - {{\beta}^4}}{24\,\beta}\,\ln^3 p   + 
     \frac{10 + 25\,{{\beta}^2} + 3\,{{\beta}^4}}{12\,{{\beta}^2}}\,\ln^2 p\,
      \nonumber\,\\ 
 & & \mbox{} + 
     \bigg( \frac{60 + 197\,{{\beta}^2} + 8\,{{\beta}^4}}{18\,\beta} + 
        \frac{3 - 10\,{{\beta}^2} - {{\beta}^4}}{2\,\beta}\,\zeta(2) \bigg) 
       \,\ln p\,\nonumber\,\\ 
 & & \mbox{}\, + 
     ( 3 - 25\,{{\beta}^2} ) \,\zeta(2) + 
     \frac{30 + 353\,{{\beta}^2}}{9} \,\bigg] \,\nonumber\,\\ 
 & & \mbox{} + 
  i\,\frac{\pi }{24\,{{\beta}^3}}\,
   \bigg\{\, - \frac{3 - 10\,{{\beta}^2} - {{\beta}^4}}{4}\,\ln^2 p 
     + \frac{10 + 25\,{{\beta}^2} + 3\,{{\beta}^4}}{3\,\beta}\,\ln p\,
      \nonumber\,\\ 
 & & \mbox{}\, \qquad + 
     \frac{60 + 197\,{{\beta}^2} + 8\,{{\beta}^4}}{9} \,\bigg\} \,,
\\[2mm]
F_{4,Q}^{(2)} & = &
\frac{1 - {{\beta}^2}}{{{\beta}^2}}\,
   \bigg[\, - \frac{{{\left( 1 - {{\beta}^2} \right) }^2}}{96\,\beta}\,
        \ln^3 p   + \frac{4 + 5\,{{\beta}^2} - 3\,{{\beta}^4}}{
       48\,{{\beta}^2}}\,\ln^2 p\,\nonumber\,\\ 
 & & \mbox{} + 
     \bigg( \frac{24 + 47\,{{\beta}^2} - 20\,{{\beta}^4}}{72\,\beta} + 
        \frac{{{\left( 1 - {{\beta}^2} \right) }^2}}{8\,\beta}\,\zeta(2) 
   \bigg) \,\ln p\,\nonumber\,\\ 
 & & \mbox{}\, + 
     \frac{1 + {{\beta}^2}}{4}\,\zeta(2) + \frac{3 - 5\,{{\beta}^2}}{9} 
  \,\bigg] \,\nonumber\,\\ 
 & & \mbox{} + 
  i\,\pi \,\frac{1 - {{\beta}^2}}{8\,{{\beta}^3}}\,
   \bigg\{\, -\frac{{{\left( 1 - {{\beta}^2} \right) }^2}}{4}\,
        \ln^2 p   + \frac{4 + 5\,{{\beta}^2} - 3\,{{\beta}^4}}{
       3\,\beta}\,\ln p \,\nonumber\,\\ 
 & & \mbox{} \,\qquad
   + \frac{24 + 47\,{{\beta}^2} - 20\,{{\beta}^4}}{9} 
     \,\bigg\} \,.
\end{eqnarray}
\subsection{Scalar Current Vertex}
\label{appendixsecondorderscalarmassive}
\begin{eqnarray}
S_{Q}^{(2)} & = &
\frac{1 + {{\beta}^2}}{72\,\beta}\,\ln^3 p + 
  \frac{5 - {{\beta}^2} + 6\,{{\beta}^4}}{72\,{{\beta}^4}}\,\ln^2 p
 + \bigg( \frac{15 + 2\,{{\beta}^2} + 11\,{{\beta}^4}}{54\,{{\beta}^3}} - 
     \frac{1 + {{\beta}^2}}{6\,\beta}\,\zeta(2) \bigg) \,\ln p\,\nonumber\,
   \\ 
 & & \mbox{}\,\quad - \frac{7}{3}\,\zeta(2) + 
  \frac{30 + 377\,{{\beta}^2}}{108\,{{\beta}^2}}\,\nonumber\,
   \\ 
 & & \mbox{} + i\,\frac{\pi }{6\,\beta}\,
   \bigg\{\, \frac{1 + {{\beta}^2}}{4}\,\ln^2 p + 
     \frac{5 - {{\beta}^2} + 6\,{{\beta}^4}}{6\,{{\beta}^3}}\,\ln p
  + \frac{15 + 2\,{{\beta}^2} + 11\,{{\beta}^4}}{9\,{{\beta}^2}} \,\bigg\} 
\end{eqnarray}
\subsection{Pseudoscalar Current Vertex}
\label{appendixsecondorderpseudoscalarmassive}
\begin{eqnarray}
P_{Q}^{(2)} & = &
\frac{1 + {{\beta}^2}}{6\,\beta}\,
   \bigg[\, \frac{1}{12}\,\ln^3 p - 
     \frac{1 - 6\,{{\beta}^2}}{12\,{{\beta}^3}}\,\ln^2 p - 
     \bigg( \frac{3 - 17\,{{\beta}^2}}{9\,{{\beta}^2}} + \zeta(2) \bigg) \,
      \ln p \,\bigg] \,\nonumber\,\\ 
 & & \mbox{} \,\quad - \frac{7}{3}\,\zeta(2) - 
  \frac{6 - 413\,{{\beta}^2}}{108\,{{\beta}^2}}\,\nonumber\,
   \\ 
 & & \mbox{} + i\,\pi \,\frac{1 + {{\beta}^2}}{6\,\beta}\,
   \bigg\{\, \frac{1}{4}\,\ln^2 p - 
     \frac{1 - 6\,{{\beta}^2}}{6\,{{\beta}^3}}\,\ln p - 
     \frac{3 - 17\,{{\beta}^2}}{9\,{{\beta}^2}} \,\bigg\}
\end{eqnarray}
\par\vspace{1cm}
\section{Expansions}
\label{appendixexpansions}
Below we list the expansions of the quantities $\tilde r^{(0,1)}_{\Theta}$
and $\delta^{(2)}_{\Theta}$, $\Theta = V, A, S, P$, for energies close
to the heavy quark-antiquark production threshold, 
$\beta = \sqrt{1-4M^2/q^2} \to 0$, as well as in the high 
energy limit, $x \equiv M^2/q^2 \to 0$. 
\subsection{Vector Current Vertex}
\begin{eqnarray}
\tilde r^{(0)}_V & = & \frac{1}{2} \beta \big( 3 - \beta^2 \big) 
\quad\stackrel{x \to 0}{\longrightarrow}\quad 
1 - 6 x^2 - 8 x^3 + {\cal O}(x^4)\,,
\\[2mm]
\tilde r_V^{(1)} & \stackrel{\beta \to 0}{\longrightarrow} &
\frac{9}{2} \zeta(2) - 6 \beta + 3 \zeta(2) \beta^2 + 
\Big( 8 \ln \beta + 12 \ln 2 - \frac{37}{3} \Big) \beta^3 + 
{\cal O}(\beta^4)\,,
\label{exprho1vlow}
\\[2mm]
\delta^{(2)}_V & \stackrel{\beta \to 0}{\longrightarrow} &
3 \zeta(2) \ln\frac{\beta}{2} + 
\Big( 8 \ln 2 - \frac{3}{2} \Big) \beta + 
2 \zeta(2) \Big( \ln\frac{\beta}{2} - 2 \Big) \beta^2 
\\ & & 
+\frac{1}{18} \Big( 96 \ln^2 \beta + 96 \ln 2 \ln \beta - 336 \ln
\beta - 96 \zeta(2) - 72 \ln^2 2 - 208 \ln 2 + 437 \Big) \beta^3 + 
{\cal O}(\beta^4)\,,
\label{expdelta2vlow}
\nonumber\\[2mm]
\tilde r_V^{(1)} & \stackrel{x \to 0}{\longrightarrow} &
\frac{3}{4} + 9 x + \Big( - 18 \ln x + \frac{15}{2} \Big) x^2 - 
\frac{4}{9} \Big( 87 \ln x + 47 \Big) x^3 + {\cal O}(x^4)\,,
\label{exprho1vhigh}
\\[2mm]
\delta^{(2)}_V & \stackrel{x \to 0}{\longrightarrow} &
\zeta(3) - \frac{1}{2} \ln 2 - \frac{23}{24} - 
3 \left[ 2 \ln 2 + \frac{1}{2} \right] x 
\nonumber\\ & & 
+\left[ - 3 \ln^2 x + \ln x \left( 12 \ln 2 + \frac{7}{2} \right) 
       - 4 \zeta(3) - 18 \zeta(2) -5 \ln 2 - \frac{5}{3} \right] x^2 
\nonumber\\ & & 
+\frac{2}{9}
\left[ - 36 \ln^2 x + 2 \ln x \left( 58 \ln 2 + \frac{43}{3} \right) 
       - 152 \zeta(2) + \frac{188}{3} \ln 2 + 19 \right] x^3 + {\cal O}(x^4) 
\,.\label{expdelta2vhigh}
\end{eqnarray}
\subsection{Axial-Vector Current Vertex}
\begin{eqnarray}
\tilde r^{(0)}_A & = & \beta^3 
\quad\stackrel{x \to 0}{\longrightarrow}\quad
1 - 6 x + 6 x^2 + 4 x^3 + {\cal O}(x^4)\,,
\\[2mm]
\tilde r_A^{(1)} & \stackrel{\beta \to 0}{\longrightarrow} &
3 \zeta(2) \beta^2 - 2 \beta^3 + 3 \zeta(2) \beta^4 + 
\left( -\frac{451}{45} + 8 \ln 2 + \frac{16}{3} \ln\beta \right)
\beta^5 + {\cal O}(\beta^7)\,,
\label{exprho1alow}
\\[2mm]
\delta^{(2)}_A & \stackrel{\beta \to 0}{\longrightarrow} &
2 \zeta(2) \Big( \ln\frac{\beta}{2} - 1 \Big) \beta^2 +
\frac{1}{3} \Big( 8 \ln 2 - 1 \Big) \beta^3 + 
\zeta(2)\left(-1+2\,\ln\frac{\beta}{2}\right) \beta^4 
\label{expdelta2alow}\\ & & + 
\left( \frac{6211}{450} - \frac{32}{9} \zeta(2) - \frac{572}{135}
\ln 2 - \frac{8}{3} \ln^2 2 - \frac{176}{15} \ln\beta + \frac{32}{9}
\ln(2\beta) \ln\beta\right) \beta^5 + {\cal O}(\beta^7)\,,
\nonumber
\\[2mm]
\tilde r_A^{(1)} & \stackrel{x \to 0}{\longrightarrow} &
\frac{3}{4} - 9 \Big( \ln x + \frac{1}{2} \Big) x + 
\Big( 18 \ln x - \frac{33}{2} \Big) x^2 + 
\frac{2}{9} \Big( 42 \ln x + 41 \Big) x^3 + {\cal O}(x^4)\,,
\label{exprho1ahigh}
\\[2mm]
\delta^{(2)}_A & \stackrel{x \to 0}{\longrightarrow} &
 \zeta(3) - \frac{1}{2} \ln 2 - \frac{23}{24} 
\nonumber\\ & & 
 +\frac{3}{2} \left[ - \ln^2 x + \ln x \left( 4 \ln 2 + 1 \right) 
                    - 4 \zeta(3) - 6 \zeta(2) + 2 \ln 2 + 5 \right] x 
\nonumber\\ & & 
 +\left[ 3 \ln^2 x - \ln x \left( 12 \ln 2 + \frac{13}{2} \right)
        + 8 \zeta(3) + 18 \zeta(2) + 11 \ln 2 - \frac{5}{3} \right] x^2 
\nonumber\\ & & 
 +\frac{1}{9} \left[ 45 \ln^2 x 
                    - \ln x \left( 56 \ln 2 + \frac{179}{3} \right)
                    + 74 \zeta(2) - \frac{164}{3} \ln 2 +
                      \frac{121}{2} \right] x^3 + {\cal O}(x^4)\,.
\label{expdelta2ahigh}
\end{eqnarray}
\subsection{Scalar Current Vertex}
\begin{eqnarray}
\tilde r^{(0)}_S & = & \beta^3 
\quad\stackrel{x \to 0}{\longrightarrow}\quad 
1 - 6 x + 6 x^2 + 4 x^3 + {\cal O}(x^4)\,,
\\[2mm]
\tilde r_S^{(1)} & \stackrel{\beta \to 0}{\longrightarrow} &
3 \zeta(2) \beta^2 - \beta^3 + 3 \zeta(2) \beta^4 +
\left(- \frac{586}{45} + 8 \ln 2 + \frac{16}{3} \ln\beta\right) 
\beta^5 + {\cal O}(\beta^7)\,,
\label{exprho1slow}
\\[2mm]
\delta^{(2)}_S & \stackrel{\beta \to 0}{\longrightarrow} &
2 \zeta(2) \Big( \ln\frac{\beta}{2} - 1 \Big) \beta^2 +
\frac{1}{6} \Big( 8 \ln 2 - 5 \Big) \beta^3 
+ 2 \zeta(2) \left( 1 + \ln\frac{\beta}{2} \right) \beta^4
\label{expdelta2slow}\\
& & + \left(
\frac{2843}{225} - \frac{32}{9} \zeta(2) - \frac{32}{135}
\ln 2 - \frac{8}{3} \ln^2 2 - \frac{176}{15} \ln\beta + \frac{32}{9}
\ln(2\beta) \ln\beta
 \right) \beta^5
+ {\cal O}(\beta^7)\,,
\nonumber
\\[2mm]
\tilde r_S^{(1)} & \stackrel{x \to 0}{\longrightarrow} &
\frac{3}{2} \ln x + \frac{9}{4} - 6 \Big( 3 \ln x + 1 \Big) x + 
3 \Big( 12 \ln x - 11 \Big) x^2 + 
\frac{2}{9} \Big( 69 \ln x + 140 \Big) x^3 + {\cal O}(x^4)\,,
\label{exprho1shigh}
\\[2mm]
\delta^{(2)}_S & \stackrel{x \to 0}{\longrightarrow} &
\frac{1}{4} \ln^2 x - \ln x \left( \ln 2 + \frac{1}{4} \right)
+ \zeta(3) + \frac{3}{2} \zeta(2) - \frac{3}{2} \ln 2 - \frac{4}{3} 
\nonumber\\ & & 
 +\left[ - 3 \ln^2 x + 3 \ln x \left( 4 \ln 2 + 1 \right) - 6 \zeta(3)
- 18 \zeta(2) + 4 \ln 2 + 5 \right] x 
\nonumber\\ & & 
 +\left[ \frac{9}{2} \ln^2 x - \ln x \left( 24 \ln 2 + \frac{17}{2}
\right) + 8 \zeta(3) + 33 \zeta(2) + 22 \ln 2 - \frac{5}{12} \right] x^2 
\nonumber\\ & & 
 +\frac{1}{9} \left[ 54 \ln^2 x - \ln x \left( 92 \ln 2 + \frac{449}{3}
\right) + 128 \zeta(2) - \frac{560}{3} \ln 2 + \frac{337}{2} \right]
x^3 + {\cal O}(x^4)\,.
\label{expdelta2shigh}
\end{eqnarray}
\subsection{Pseudoscalar Current Vertex}
\begin{eqnarray}
\tilde r^{(0)}_P & = & \beta 
\quad\stackrel{x \to 0}{\longrightarrow}\quad 
1 - 2 x - 2 x^2 - 4 x^3 + {\cal O}(x^4)\,,
\\[2mm]
\tilde r_P^{(1)} & \stackrel{\beta \to 0}{\longrightarrow} &
3 \zeta(2) - 3 \beta + 3 \zeta(2) \beta^2 + 
\frac{1}{9} \Big( 48 \ln \beta + 72 \ln 2 - 110 \Big) \beta^3 +
{\cal O}(\beta^5)\,,
\label{exprho1plow}
\\[2mm]
\delta^{(2)}_P & \stackrel{\beta \to 0}{\longrightarrow} &
2 \zeta(2) \ln\frac{\beta}{2} + 
\Big( 4 \ln 2 - \frac{3}{2} \Big) \beta + 
2 \zeta(2) \ln\frac{\beta}{2} \beta^2 
\\ & & 
+\frac{1}{27} \Big( 96 \ln^2 \beta + 96 \ln 2 \ln \beta - 336 \ln
\beta - 96 \zeta(2) - 72 \ln^2 2 - 64 \ln 2 + 395 \Big) \beta^3 + 
{\cal O}(\beta^5)\,,
\label{expdelta2plow}
\nonumber\\[2mm]
\tilde r_P^{(1)} & \stackrel{x \to 0}{\longrightarrow} &
\frac{3}{2} \ln x + \frac{9}{4} - 6 \Big( \ln x - 1 \Big) x - 
3 \Big( 4 \ln x + 3 \Big) x^2 - 
\frac{2}{9} \Big( 93 \ln x + 130 \Big) x^3 + {\cal O}(x^4)\,,
\label{exprho1phigh}
\\[2mm]
\delta^{(2)}_P & \stackrel{x \to 0}{\longrightarrow} &
\frac{1}{4} \ln^2 x - \ln x \left( \ln 2 + \frac{1}{4} \right)
+ \zeta(3) + \frac{3}{2} \zeta(2) - \frac{3}{2} \ln 2 - \frac{4}{3} 
\nonumber\\ & & 
 +\left[ - \ln^2 x + \ln x \left( 4 \ln 2 + 1 \right) - 2 \zeta(3)
- 6 \zeta(2) - 4 \ln 2 - \frac{1}{3} \right] x 
\nonumber\\ & & 
 +\left[ - \frac{3}{2} \ln^2 x + \ln x \left( 8 \ln 2 + \frac{7}{2}
\right) - 11 \zeta(2) + 6 \ln 2 - \frac{7}{4} \right] x^2 
\nonumber\\ & & 
 +\frac{1}{9} \left[ - 54 \ln^2 x + \ln x \left( 124 \ln 2 + \frac{253}{3}
\right) - 160 \zeta(2) + \frac{520}{3} \ln 2 + \frac{139}{2} \right]
x^3 + {\cal O}(x^4)\,.
\label{expdelta2phigh}
\end{eqnarray}
\end{appendix}
%
%
\sloppy
\raggedright
\def\app#1#2#3{{\it Act. Phys. Pol. }{\bf B #1} (#2) #3}
\def\apa#1#2#3{{\it Act. Phys. Austr.}{\bf #1} (#2) #3}
\def\lhc{Proc. LHC Workshop, CERN 90-10}
\def\npb#1#2#3{{\it Nucl. Phys. }{\bf B #1} (#2) #3}
\def\nP#1#2#3{{\it Nucl. Phys. }{\bf #1} (#2) #3}
\def\plb#1#2#3{{\it Phys. Lett. }{\bf B #1} (#2) #3}
\def\prd#1#2#3{{\it Phys. Rev. }{\bf D #1} (#2) #3}
\def\pra#1#2#3{{\it Phys. Rev. }{\bf A #1} (#2) #3}
\def\pR#1#2#3{{\it Phys. Rev. }{\bf #1} (#2) #3}
\def\prl#1#2#3{{\it Phys. Rev. Lett. }{\bf #1} (#2) #3}
\def\prc#1#2#3{{\it Phys. Reports }{\bf #1} (#2) #3}
\def\cpc#1#2#3{{\it Comp. Phys. Commun. }{\bf #1} (#2) #3}
\def\nim#1#2#3{{\it Nucl. Inst. Meth. }{\bf #1} (#2) #3}
\def\pr#1#2#3{{\it Phys. Reports }{\bf #1} (#2) #3}
\def\sovnp#1#2#3{{\it Sov. J. Nucl. Phys. }{\bf #1} (#2) #3}
\def\sovpJ#1#2#3{{\it Sov. Phys. LETP Lett. }{\bf #1} (#2) #3}
\def\jl#1#2#3{{\it JETP Lett. }{\bf #1} (#2) #3}
\def\jet#1#2#3{{\it JETP Lett. }{\bf #1} (#2) #3}
\def\zpc#1#2#3{{\it Z. Phys. }{\bf C #1} (#2) #3}
\def\ptp#1#2#3{{\it Prog.~Theor.~Phys.~}{\bf #1} (#2) #3}
\def\nca#1#2#3{{\it Nuovo~Cim.~}{\bf #1A} (#2) #3}
\def\ap#1#2#3{{\it Ann. Phys. }{\bf #1} (#2) #3}
\def\hpa#1#2#3{{\it Helv. Phys. Acta }{\bf #1} (#2) #3}
\def\ijmpA#1#2#3{{\it Int. J. Mod. Phys. }{\bf A #1} (#2) #3}
\def\ZETF#1#2#3{{\it Zh. Eksp. Teor. Fiz. }{\bf #1} (#2) #3}
\def\jmp#1#2#3{{\it J. Math. Phys. }{\bf #1} (#2) #3}
\def\yf#1#2#3{{\it Yad. Fiz. }{\bf #1} (#2) #3}
\vglue 0.4cm

\end{document}